\documentclass{aa}
\usepackage{graphicx,natbib}
\usepackage{txfonts}

\newcommand{\be}{\begin{equation}}

\newcommand{\emm}[1]{\ensuremath{#1}}   
\newcommand{\emr}[1]{\emm{\mathrm{#1}}} 
\newcommand{\unit}[1]{\emm{\, \emr{#1}}}
\newcommand{\K}   {\unit{K}}

\newcommand{\pscm}{\unit{cm^{-2}}}
\newcommand{\pccm}{\unit{cm^{-3}}}
\newcommand{\Kpccm}{\unit{K\,cm^{-3}}}

\newcommand{\kms}   {\unit{km\,s^{-1}}}
\newcommand{\Kkms}{\unit{K\,km\,s^{-1}}}

\newcommand{\kpc}   {\unit{kpc}}

\def\ltsimeq{\,\raise 0.3 ex\hbox{$ < $}\kern -0.75 em \lower 0.7 ex\hbox{$\sim$}\,}
\def\gtsimeq{\,\raise 0.3 ex\hbox{$ > $}\kern -0.75 em \lower 0.7
  ex\hbox{$\sim$}\,}

\newcommand{\CI}{C\,{\sc i}}
\newcommand{\OI}{O\,{\sc i}}
\newcommand{\CII}{C\,{\sc ii}}

\begin{document}

\title{[\CII] absorption and emission in the diffuse interstellar medium across the Galactic Plane\thanks{{\sl Herschel}  is an ESA space observatory with science instruments provided by European-led Principal Investigator consortia and with important participation from NASA.} }

\author{
M. Gerin \inst{1,2} 
\and M. Ruaud \inst{3}
\and  J. R. Goicoechea \inst{4}
\and A. Gusdorf \inst{1,2}
\and B. Godard \inst{2,1}
\and  M. de Luca \inst{1,2}
\and E. Falgarone \inst{1,2}
\and P. Goldsmith \inst{5}
\and D.C. Lis \inst{1,2,6}
\and K. M. Menten \inst{7}
\and D. Neufeld \inst{8}
\and T.G. Phillips \inst{6}
\and H. Liszt \inst{9}
}

\institute{
CNRS UMR8112, LERMA, Observatoire de Paris and Ecole Normale
  Sup\'erieure.
24 Rue Lhomond, 75231 Paris cedex 05, France.\\
\email{maryvonne.gerin@ens.fr} 
\and
Sorbonne Universit\'es, UPMC Univ Paris 06, UMR 8112, LERMA,
  F-75005, Paris, France.
\and
CNRS UMR5804, Laboratoire d'Astrophysique de Bordeaux, 2 rue de
l'Observatoire, BP89, 33271 Floirac cedex, France.\\ 
\email{maxime.ruaud@obs.u-bordeaux1.fr}
\and
Instituto de Ciencia de Materiales de Madrid (ICMM-CSIC). E-28049, Cantoblanco,
Madrid, Spain.
\email{jr.goicoechea@icmm.csic.es}
\and 
Jet Propulsion Laboratory, California Institute of Technology, 4800 Oak Grove Drive, Pasadena, CA 91109-8099, USA.
\email{pauf.f.goldsmith@jpl.nasa.gov}
\and 
California Institute of Technology, Cahill Center for Astronomy and Astrophysics 301-17, Pasadena, CA 91125, USA.
\email{darek.lis@obspm.fr,tgp@caltech.edu}
\and 
Max-Planck Institute f\"ur Radioastronomie, Auf dem H\"ugel 69, D-53121 Bonn, Germany.
\email{kmenten@mpifr-bonn.mpg.de}
\and 
Department of Physics and Astronomy, The John Hopkins University, Baltimore, MD, 21218, USA.
\email{neufeld@pha.jhu.edu}
\and 
National Radio Astronomy Observatory, 520 Edgemont Road
Charlottesville, VA  22903, USA.
\email{hliszt@nrao.edu}
}

\date{Received ; accepted}

\abstract
{}
{Ionized carbon is the main  gas-phase reservoir of carbon in the  neutral diffuse interstellar
  medium and its 158~$\mu$m fine structure transition [\CII] is the most
  important cooling line of the diffuse interstellar medium (ISM).  We combine
  [\CII] absorption and emission spectroscopy to gain an improved understanding of physical conditions in the different phases of the ISM.}
{We present high resolution [\CII]  spectra obtained
with the {\it Herschel}/HIFI instrument towards  bright dust continuum
 sources regions in the Galactic plane, probing simultaneously the diffuse gas
along the line of sight and the background  high-mass star forming regions.
These data are complemented by single pointings in the 492 and 809~GHz fine
structure lines of atomic carbon and by medium spectral resolution  spectral maps  of the fine structure lines of atomic oxygen at 63
and 145 $\mu$m with {\it Herschel}/PACS.}
{We show that the presence of foreground absorption may completely cancel the
  emission from the background source in medium spectral resolution PACS data and
  that high spectral resolution spectra are needed to interpret the [\CII] and [\OI]  emission and the [\CII]/FIR ratio. This phenomenon may explain part of
  the [\CII]/FIR deficit seen in external
  luminous infrared galaxies where the bright emission from the 
  nuclear regions may be partially canceled by absorption from diffuse gas in the
  foreground. 
The C$^+$ and C excitation in the diffuse gas is
consistent with a median pressure of  $\sim 5900$~\Kpccm \  for
a mean kinetic temperature of $\sim 100$~K. A few higher pressure regions 
 are detected along the lines of sight, as emission features in both fine
structure lines of atomic carbon.  The knowledge of the  gas density
 allows us to determine the  filling factor of the absorbing gas along
the selected lines of sight. {\bf The derived median value of the 
filling factor  is 2.4 \%, in good agreement with the properties of the Galactic Cold Neutral Medium.} The mean excitation temperature is used to
derive the average cooling due to C$^+$ in the Galactic plane :
 $9.5 \times 10^{-26}$ ergs$^{-1}$H$^{-1}$.  Along the observed lines of sight, the gas phase carbon abundance does not
 exhibit a strong gradient  as a function of  Galacto-centric radius and has a
 weighted average of  C/H = $1.5 \pm 0.4 \times 10^{-4} $.}
{}

\maketitle

\section{Introduction}

The diffuse interstellar medium presents a complex structure due to the
coexistence of two stable phases of neutral gas, the Cold Neutral Medium (CNM)
with moderate densities and temperatures 
($n = n({\rm\ion{H}{I}})+n({\rm H_2}) \sim 30$~\pccm \ and  $T \sim 100$~\K)  and the Warm Neutral Medium
(WNM) with low densities and high temperatures 
($n \sim 0.4$~\pccm \ and  $T \sim 8000$~\K), the two phases being
in pressure equilibrium \citep[see][]{draine}. Higher pressure phases are associated either with
HII regions created by massive stars, and with molecular clouds where the
interplay of gravity and magnetic fields contributes to the creation of 
denser structures, in which thermal pressure does not balance other forces present. 
The classical tracer of both diffuse ISM phases is the hyperfine
transition of neutral hydrogen at 21~cm (HI). Both phases contribute to the 
widespread emission, while the CNM can also be detected in absorption
against either continuum sources or background HI emission. However it is
still fairly difficult to estimate the relative contributions of these
phases to the total amount of atomic hydrogen, and how this varies
with the physical conditions in the Galaxy  \citep{kalberla}. 

The  $^{2}P_{3/2} - ^{2}P_{1/2}$ fine structure line
of ionized carbon at 158~$\mu$m, [\CII], is the main cooling line of the CNM and
as such is a complementary probe of this diffuse phase. The 
{\it Herschel} satellite has performed the first [\CII] survey of the
Galactic plane \citep{langer,pineda,langer:14}. 
The measured [\CII] emission has been compared
with HI and CO data to identify the contributions from the different ISM
phases, leading to the conclusion that the CNM gas represents $\sim 40$~\%
 of the atomic gas in the Galactic Plane. In addition to the molecular gas
 traced by CO emission, an additional 30 \% is traced by the [\CII]
 emission. This ``CO dark'' molecular gas is also detected as an excess in thermal dust emission in the
sub-millimeter regime as discussed by \citet{planck1} using the {\it Planck Surveyor} data.

Detection of [\CII] emission is limited by the  sensitivity
of the HIFI instrument. Absorption spectroscopy toward strong far infrared
continuum sources can improve the ability to study low column density regions.
This method has been 
pioneered for the fine structure lines of ionized carbon and  neutral oxygen by
\citet{vastel:00,vastel:02} and by \citet{keene} using the {\it Infrared Space
  Observatory} (ISO) long wavelength spectrograph (LWS)  at moderate
spectral resolution. The HIFI instrument on board the {\it Herschel} satellite
has allowed for the first time high spectral resolution observations of both
neutral and ionized carbon fine  structure lines, with high
sensitivity. It has also enabled the detection of ground state hydride lines from the diffuse interstellar medium \citep{gerin:12}.  These observations have revealed that neutral
species like HF, CH and possibly water vapor, are reliable
tracers of molecular hydrogen \citep{gerin:10b,neufeld:10a,flagey}. Given
the very large critical densities of the  ground state hydride lines compared to the moderate densities in the diffuse ISM, the
absorption data do not provide information on the physical conditions
of the absorbing gas. 

In this paper we present complementary data
on the fine structure lines of ionized and neutral carbon aimed at 
better understanding the properties of the diffuse gas in the
Galactic Plane. The high spectral resolution HIFI  data are complemented by
PACS  spectral maps for the [\CII] and [\OI] fine structure lines, at medium spectral
resolution. We study the density and pressure structure, as well as the
 the volume filling factor of the colder and denser 
phases (CNM and diffuse molecular
 gas) from the comparison of the emission and absorption data.
Section \ref{sect:obs} presents the observations. Section \ref{sect:results}
discusses the analysis methods and gives the main results, while the
conclusions are presented in Section \ref{sect:discussion} and summarized in
Section \ref{sect:conclusion}.  Appendices A and B give additional information
on the observations and present the data in the complete set of sources observed.

\section{Observations}
\label{sect:obs}

\begin{figure}
\resizebox{8cm}{!}{
\includegraphics{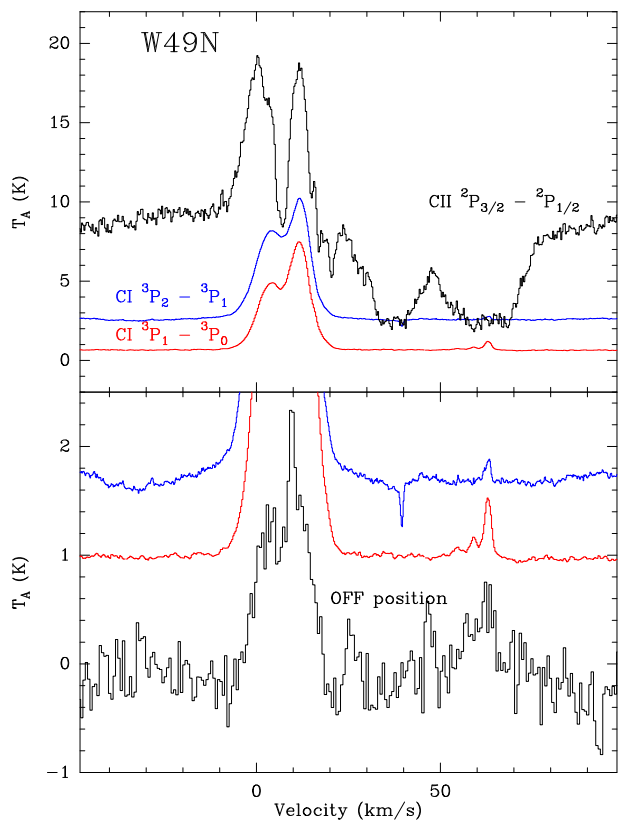}}
\caption{\label{fig:specw49n} Top : {\sl Herschel}/HIFI spectra towards W49N.
The red line shows the [\CI]$^3P_1 - ^3P_0$ line at 492~GHz, the blue line shows
the [\CI]$^3P_2 - ^3P_1$ line at 809~GHz and the black line the
[\CII]$^2P_{3/2}-^2P_{1/2}$ line at 1.9~THz. The horizontal axis is the LSR
velocity in \kms \ and the vertical axis is the  SSB antenna temperature in
Kelvin.
Bottom : zoom on the [\CI] lines (red, blue as above) and the average [\CII] 
spectrum of the OFF positions (black). The continuum levels have been shifted
for clarity in the bottom panel.  }
\end{figure}

\subsection{Presentation of the HIFI observations}

\begin{table*}
\caption{\label{tab:sources} Summary of source properties}
\begin{tabular}{l|ccccc|cccc}
\hline
\hline
Name & $RA$  & $Dec$ & $V_{LSR}$ & $D$ & Ref.$^a$ & Velocity& $N_{tot}(\rm{H})$&
$N(\rm{H}\ion{I})$&$N(\rm{H}_2)$\\
     &           &           &  &  & &interval &($10^{21}$&
     ($10^{21}$& ($10^{21}$\\
    &(J2000)    & (J2000)     &  (\kms)        & (\kpc) &  &  (\kms) &  
\pscm)& \pscm)& \pscm)\\
\hline
W28A(G5.89-0.4) & 18:00:30.4 & -24:04:00 & 10 & 1.3& 1 &  [15,25]       & $6.5
\pm  1.2$ & 3.2$^f$ & $1.7\pm 0.4$$^e$ \\
W31C(G10.6-0.4) & 18:10:28.7 & -19:55:50 & -2 & 5.0 & 2& [10,23] & $ 15.6 \pm
1.1$& 5.0$^b$ & $5.3 \pm 0.5$$^c$\\
                &            &           &    &     &  & [23,34] & 
$ 21.6 \pm 2.0  $ &5.4$^b$& $8.1 \pm 0.7$$^c$\\
                &          &             &    &     &   & [34,61] & 
$19.3 \pm 5.0$& 6.9$^b$ & $6.2 \pm 0.6$$^c$\\
W33A(G12.9-0.3) & 18:14:39.4 & -17:52:00 & 37 & 2.4 & 3 &[14,25] 
& $2.4 \pm 0.5$ & $2.0$$^c$ & $0.2 \pm 0.04$$^c$   \\
G34.3+0.15 & 18:53:18.7 & +01:14:58 & 58 & 3.3 & 4& [8,20]& $4.6 \pm 0.5$&
2.2$^b$&$1.2 \pm 0.1$$^c$\\
           &            &           &    &     &   & [20,35]& $7.6 \pm 4.0$&
           5.0$^b$ & $1.3 \pm 0.1$$^c$\\
W49N(G43.2-0.1) & 19:10:13.2 & +09:06:12 & 11 & 11.4 & 5& [25,50]&$14.3 \pm 0.7$
&7.0$^b$&$3.7 \pm 0.4$$^c$\\
                &             &         &    &      &  & [50,80]& $22.6 \pm
                1.1$& 9.3$^b$&$6.7 \pm 0.7$$^c$\\ 
W51(G49.5-0.4) & 19:23:43.9 & +14:30:30.5 & 57 & 5.4 & 6 & [1,10] & 
$3.3 \pm 0.2$&1.5$^b$ & $0.9 \pm 0.1$$^c$\\
               &            &             &    &     &   & [10,16]& 
$1.2 \pm 0.1$& 1.0$^b$ &$0.12 \pm  0.08$$^c$ \\
              &             &             &    &      &  & [16,35] &
$2.5 \pm 0.2$& 2.4$^b$ &$0.07 \pm 0.02$$^c$\\
DR21OH(G81.7+0.6) & 20:39:00.7 & +42:22:46.7 & -2 & 1.5 & 7 & [1,20] 
& $12.8\pm 4.0$ &  5.0$^b$ & $3.9 \pm 0.4$$^c$\\
W3-IRS5 &  02:25:40.6  & +62:05:51   & -37    & 2.0     & 8 & [-25,-15] & $1.6
\pm 0.3$& 0.7$^{b,d}$ & $0.41 \pm 0.1$$^e$ \\
        &              &            &        &         &   & [-15,10] & $1.7
        \pm 0.3$& 0.9$^{b,d}$ & $0.41 \pm 0.1$$^e$ \\
\hline
\end{tabular}

$^a$ References:  (1) \citet{motogi},(2) \citet{sanna}, (3) \citet{immer}, (4) \citet{kuchar}, 
(5) \citet{gwinn}, (6) \citet{sato:10},  (7) \citet{rygl}, (8) \citet{xu:06}\\
$^b$ from \citep{winkel}, $^c$ From \citet{godard:12} ; $^d$ using data from
\citet{normandeau:99} ; $^e$ this paper using [CH]/[H$_2$] = $3.6 \times 10^{-8}$\citep{sheffer:08} and [HF]/[H$_2$] = $1.3 \times 10^{-8}$ \citep{sonnentrucker:10} ; $^f$ from \citet{fish}.\\
\end{table*}

The source sample is based on  the PRISMAS sources, as presented in e.g.
\citet{godard:10,flagey}.
W3-IRS5 was targeted as an additional sight-line to probe the outer Galaxy as
this source is located in the Perseus arm. 
The source positions, LSR velocities and distances are
listed in Table \ref{tab:sources}  as well as the velocity intervals corresponding to the diffuse matter absorption and the associated atomic, molecular and total hydrogen  column densities.  Table A.1 in Appendix
\ref{app:obs} summarizes
the observing modes and provides the ObsID analyzed in this paper.

Analysis of absorption spectra requires accurate knowledge
of the continuum level, which is best measured in DBS (Double Beam Switching)
mode with heterodyne receivers like HIFI. However, since 
the [\CII] emission is very extended in the Galaxy, this mode produces
spectra that are contaminated by the presence of signals in the OFF beam.
To avoid this problem, HIFI could be operated using the internal cold
load (employed for  calibration), as a reference for monitoring the
system gain. This observing mode called ``Load Chop'' (LC) minimizes
the contamination at the expense of  a lower accuracy measurement of the continuum
\citep{degraauw,roelfsema}.
We therefore performed the [\CII] and [\CI] observations in two steps :
 we obtained a single pointing at the source position in DBS mode, followed
by either a small map covering $\sim 50'' x\ 50''$ (for [\CII]) or a single pointing (for both [\CI] lines) in LC mode. This
combination provides the continuum at the central position and an accurate
spectrum free of contamination, including both
the background source emission and the foreground diffuse medium along the
line of sight.
As the continuum emission of the background source is semi-extended, we
also used the PACS spectrometer \citep{Pog10} to determine the 158$\mu$m continuum
over the area mapped with HIFI.

The spectroscopic parameters of the targeted fine structure lines are
given in Table \ref{tab:lines}. At these frequencies, the angular resolution
of the {\it Herschel} telescope is limited by diffraction, and the FWHM beam size is
44'', 26'' and 11'', at 492, 809, and 1900~GHz, respectively. As we are
focusing on the foreground absorption in this paper, we have not attempted
to correct for the difference in angular resolution among the different lines.

{\small
\begin{table}
\caption{\label{tab:lines}Observed  fine structure lines}
\begin{tabular}{lrrrcc}
\hline
\hline
  Transition &  $\nu$ \ \ & $\lambda$\ \ \ & $E_{u}$ & $g_u$ & $A$ \\
                   &  (GHz)  & ($\mu$m) &  (\K)   &       & (s$^{-1}$)\\
\hline
C$^+$ ($^2P_{3/2} - ^2P_{1/2}$) &  1900.537 & 157.7 & 91.2 & 4 & $2.32 \times
10^{-6}$ \\ 
C  ($^3P_1 - ^3P_0$) & 492.160 & 609.1 &23.6 & 3 & $7.93 \times 10^{-8}$ \\
C  ($^3P_2 - ^3P_1$) & 809.341 & 370.4 &62.5 & 5 & $2.68 \times 10^{-7}$ \\
O  ($^3P_1 - ^3P_2$) & 4744.777  & 63.2 &227.7 & 3 & $8.54 \times 10^{-5}$\\
O  ($^3P_0 - ^3P_1$) &  2060.069 & 145.5 &326.6 & 1 & $1.64 \times 10^{-5}$ \\
\hline

\end{tabular}

From CDMS \citep{muller:01,muller:05}
\end{table}
}

\begin{figure}
\rotatebox{0}{
\resizebox{8.5cm}{!}{
\includegraphics{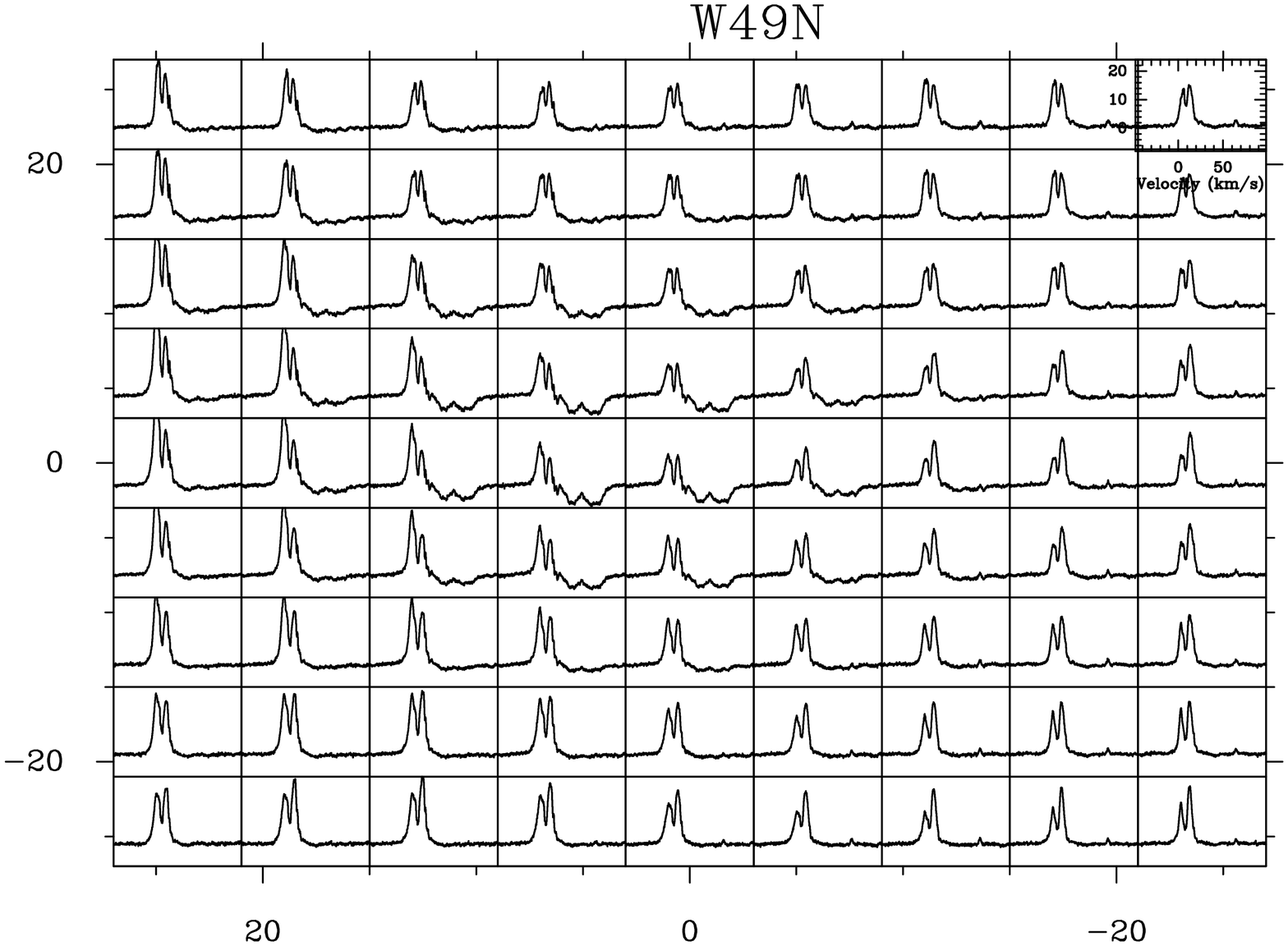}}}
\caption{\label{fig:mapw49n} Montage of [\CII] HIFI spectra towards W49N. A 
  baseline has been subtracted from each spectrum. In each box, the horizontal axis is 
the LSR   velocity in  \kms, which runs  from -48 \kms \  to 98  \kms and the 
 vertical axis is the antenna temperature in Kelvin, which runs from -8
  to 24 \K.   The x--axis shows the right ascension offset in arc-sec and the
  y--axis the declination offset in arc-sec, relative to the source position given
in Table \ref{tab:sources}.}
\end{figure}

\subsection{HIFI Data processing}
The observations have been analyzed with the Herschel Interactive
Processing Environment (HIPE version 9 \footnote{available at http://herschel.esac.esa.int/HIPE{\_}download.shtml}).
The HIFI data have been corrected for fringes using the 
\textit{HIFIFitfringe routine}.
The single pointing and maps have been exported to the CLASS
package\footnote{available at http://www.iram.fr/IRAMFR/GILDAS/} for further
processing. The HIFI mixers are double side band (DBS) and therefore 
sensitive to the emission in two
frequency bands. Therefore the detected continuum levels are the combination
of the continuum emission in these two side bands
while the line emission and absorption is detected in one side band only. Correction for the relative gain of each mixer in the two side bands is included in the calibration pipeline. 
The HIFI calibration accuracy is better than 10\%.
As the foreground emission is extended relative to the
HIFI beam at 1.9~THz, we present all data in this paper in
units of  antenna temperature  denoted T$_A$. 
As an example of the resulting spectra and maps,  
the data toward W49N are shown  in Fig. \ref{fig:specw49n} and
Fig. \ref{fig:mapw49n}. The data for the 
other sources are presented in  Appendix \ref{app:spec}. 
For each source, we display the spectrum observed in load chop mode
toward the central position  with the correct SSB continuum level,
 as well as the mean spectrum toward
the OFF positions, obtained as the difference between the load chop
and DBS spectra. This shows the level of contamination by emission
signal in the OFF beam. A linear baseline has been removed from
the spectra. 

We derived the continuum intensities from the DBS spectra. We used the HIFI 
efficiencies given in \citet{roelfsema} to convert the continuum values from
Kelvin to Jy, namely 464, 469, and 506 Jy/K at 0.49, 0.81, and 1.9~THz,
respectively. The continuum values are listed in Table \ref{tab:continuum}. 
For the [\CII] data at 1.9~THz, we provide a comparison with the continuum intensities
 measured in the central pixel of the PACS footprint.

\begin{figure*}
\rotatebox{0}{
\resizebox{6cm}{!}{
\includegraphics{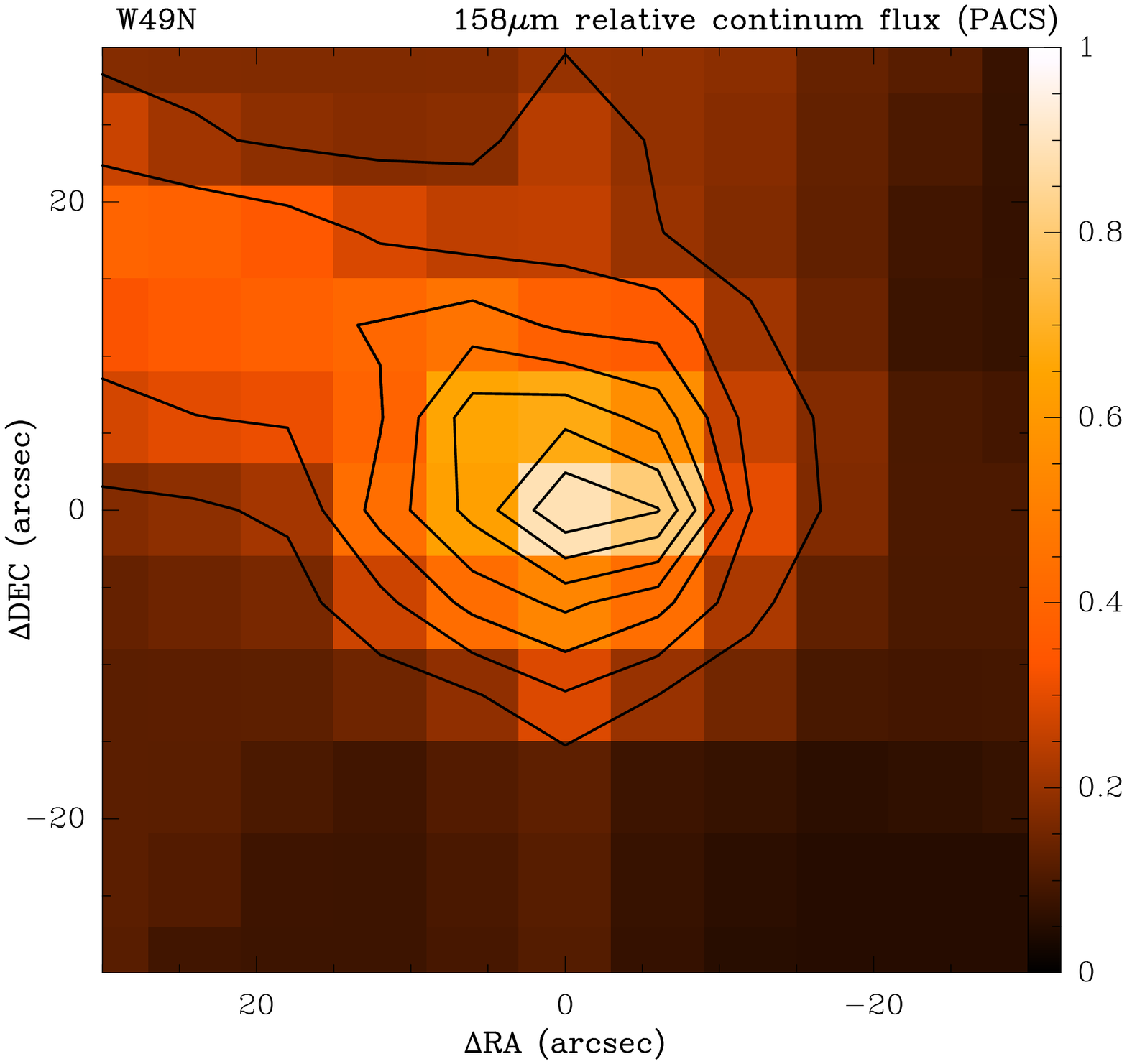}}}
\rotatebox{0}{
\resizebox{6cm}{!}{
\includegraphics{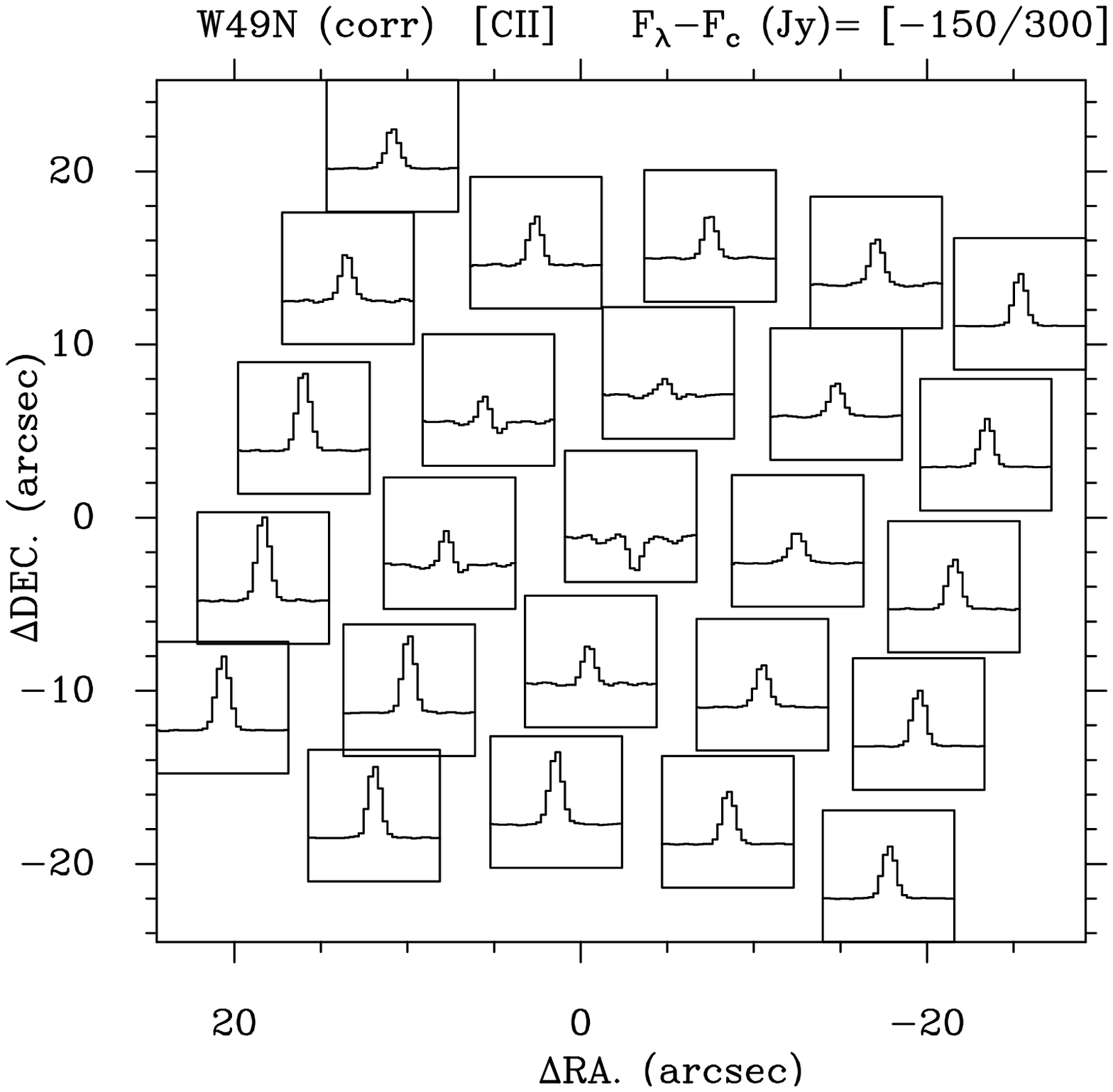}}}
\rotatebox{0}{
\resizebox{6cm}{!}{
\includegraphics{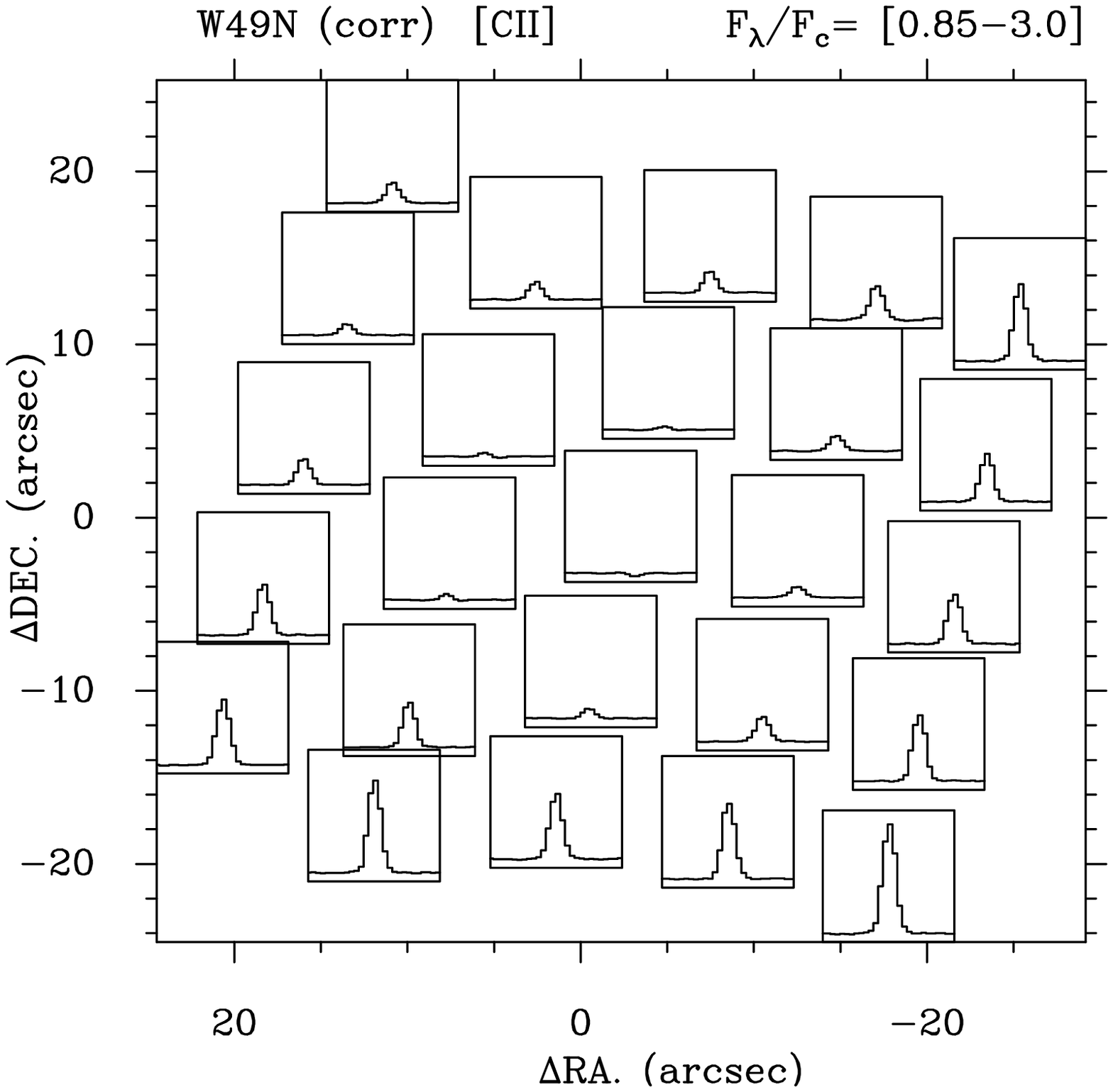}}}
\caption{\label{fig:pacs-w49n} PACS data towards W49N. For all maps, the
  offsets are given relative to the central position listed in Table
  \ref{tab:sources}. The data have been corrected from the weak emission in
  the OFF position. Left : Continuum
  emission at 158~$\mu$m. Contour levels are drawn at 0.2, 0.3, 0.4 ... 0.9 relative
to the maximum. Middle : [\CII] emission in the 25 PACS spaxels. The continuum
emission has been subtracted. The vertical scale runs from -150 to 300
Jy. Right : Map of the line to continuum ratio in the 25
PACS spaxels. The vertical scale runs from 0.85 to 3.0. Note the very low
contrast in the central spaxels. }
\end{figure*}

\subsection{PACS data processing}

The  PACS spectra presented in this work were obtained between 2010 and 2012 
using the Range Spectroscopy SED mode. This mode is designed to 
cover the full PACS wavelength range in Nyquist 
sampling to get the complete far-IR Spectral Energy Distribution (SED). 
The observation ObsIDs are tabulated in
Table A.1. The PACS spectrometer uses photoconductor detectors and
provides 25 spectra  over a 47$''$$\times$47$''$ field-of-view (FoV) arranged 
as 5$\times$5 spatial pixels  (``spaxels''), each with an angular size of $\sim$9.4$''$ on the sky \citep{Pog10}. 
The measured width of the PACS spectrometer point spread function (PSF) increases with   wavelength for $\lambda$$\gtrsim$100\,$\mu$m.
In particular, only $\sim$50\%\,($\sim$70\%) of a point source emission would fall in the central spaxel 
at $\sim$158\,$\mu$m (63\,$\mu$m). 
The resolving power  in the  R1 and B3A grating orders 
are $R$$\sim$1200  ($\sim$250\,km\,s$^{-1}$) at  158\,$\mu$m and $R$$\sim$3300
($\sim$90\,km\,s$^{-1}$) at  63\,$\mu$m. The total observing time  was $\sim$1.4\,h per source in the R1 order and $\sim$1.3\,h in the B3A order.

Observations were carried out in the  ``chop-nod'' mode with the largest
chopper throw of $\pm$6~arc-min. This mode offers a reasonable measurement of
the continuum level toward bright, compact star forming cores.
In the observed regions, however, the reference nod positions are 
often contaminated by extended [C{\sc ii}]158\,$\mu$m and [O{\sc i}]63\,$\mu$m
line emission. Accurate  calibration of the line intensities toward each
source is thus not trivial and requires subtraction of the reference position
emission. The continuum rms noise of the spectra toward each continuum peak
position  was typically $\sim$5-10\,Jy/spaxel. Except for W33A (the faintest
of the sample) the  continuum peak in each source is a few thousand Jy/spaxel,
close but still below the saturation levels of PACS detectors.

The PACS data were processed in HIPE 11 \citep{Ott10}. The $\sim$158,
$\sim$145 and $\sim$63\,$\mu$m continuum levels in each observed position were
extracted using the standard \textit{ChopNodRangeScan} pipeline for chopped
Range Spectroscopy observations. Table \ref{tab:continuum} shows a summary of the 158\,$\mu$m continuum levels measured with PACS  toward the continuum peak of each source. The absolute flux calibration uncertainty is estimated to be 
$\sim$30\,$\%$\footnote{See PACS Spectroscopy performance and calibration,” PACS/ICC document ID PICC-KL-TN-041, by B. Vandenbussche et al.}. 
 Toward the peak positions, the continuum fluxes have been extracted using the PACS point source correction  applied to the central spaxel fluxes.
Owing to the relatively large distance to these sources, this is a good first
approximation (specially for their bright and compact far-IR continuum
peak). Indeed, the resulting  fluxes are in very good agreement with those
obtained from HIFI at the map  center. Maps of the relative continuum emission
with respect to the peak position were constructed for each observed source
and are shown in Figure \ref{fig:pacs-w49n} and Figs \ref{fig:pacs-w28a} to
 \ref{fig:pacs-w3i}. The PACS maps were re-centered with respect to the HIFI map (0$''$, 0$''$) coordinates. 
In most sources, the (0$''$, 0$''$)  position agrees with the continuum peak.

\begin{figure}
\rotatebox{0}{
\resizebox{7cm}{!}{
\includegraphics{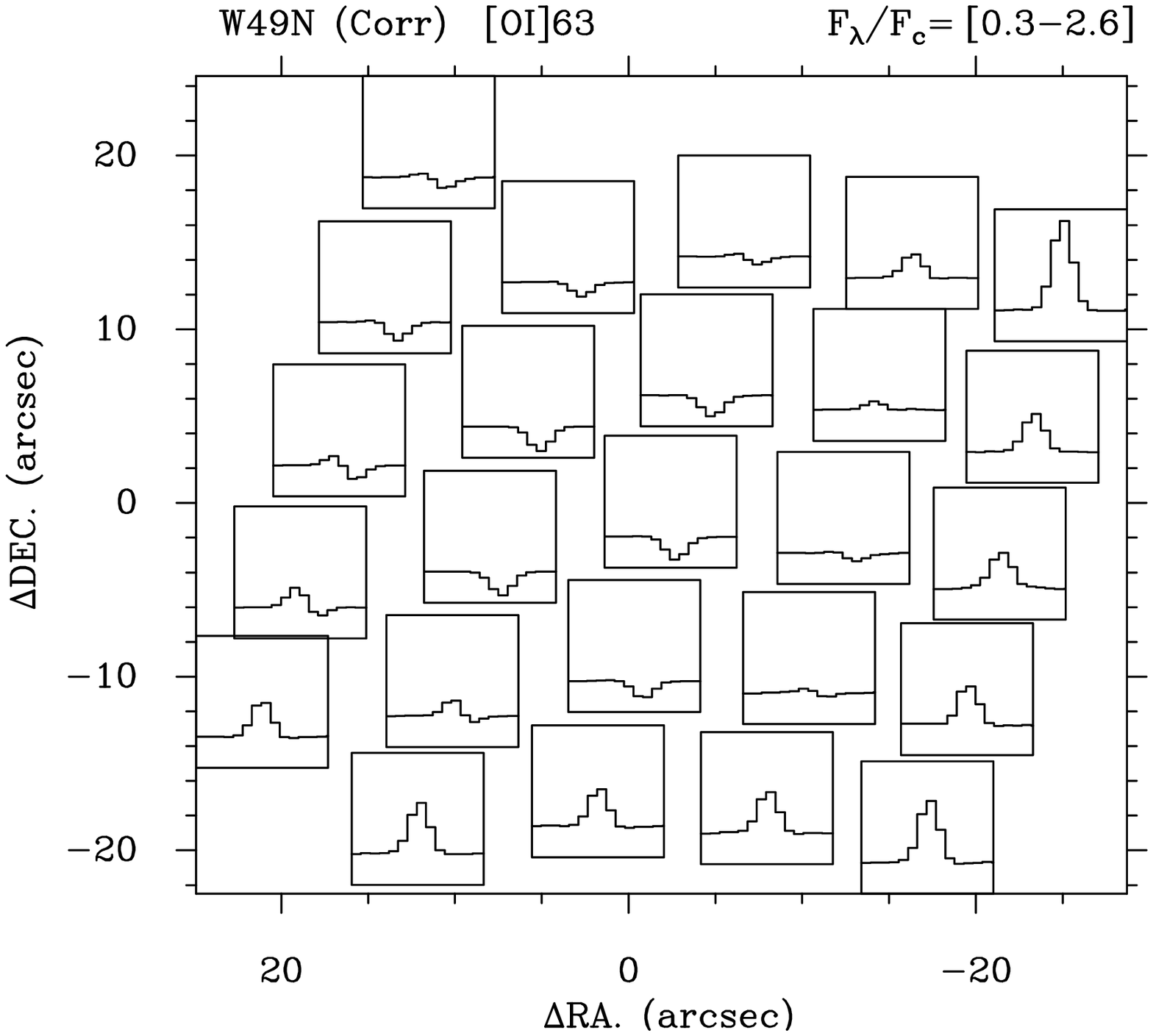}}}
\rotatebox{0}{
\resizebox{7cm}{!}{
\includegraphics{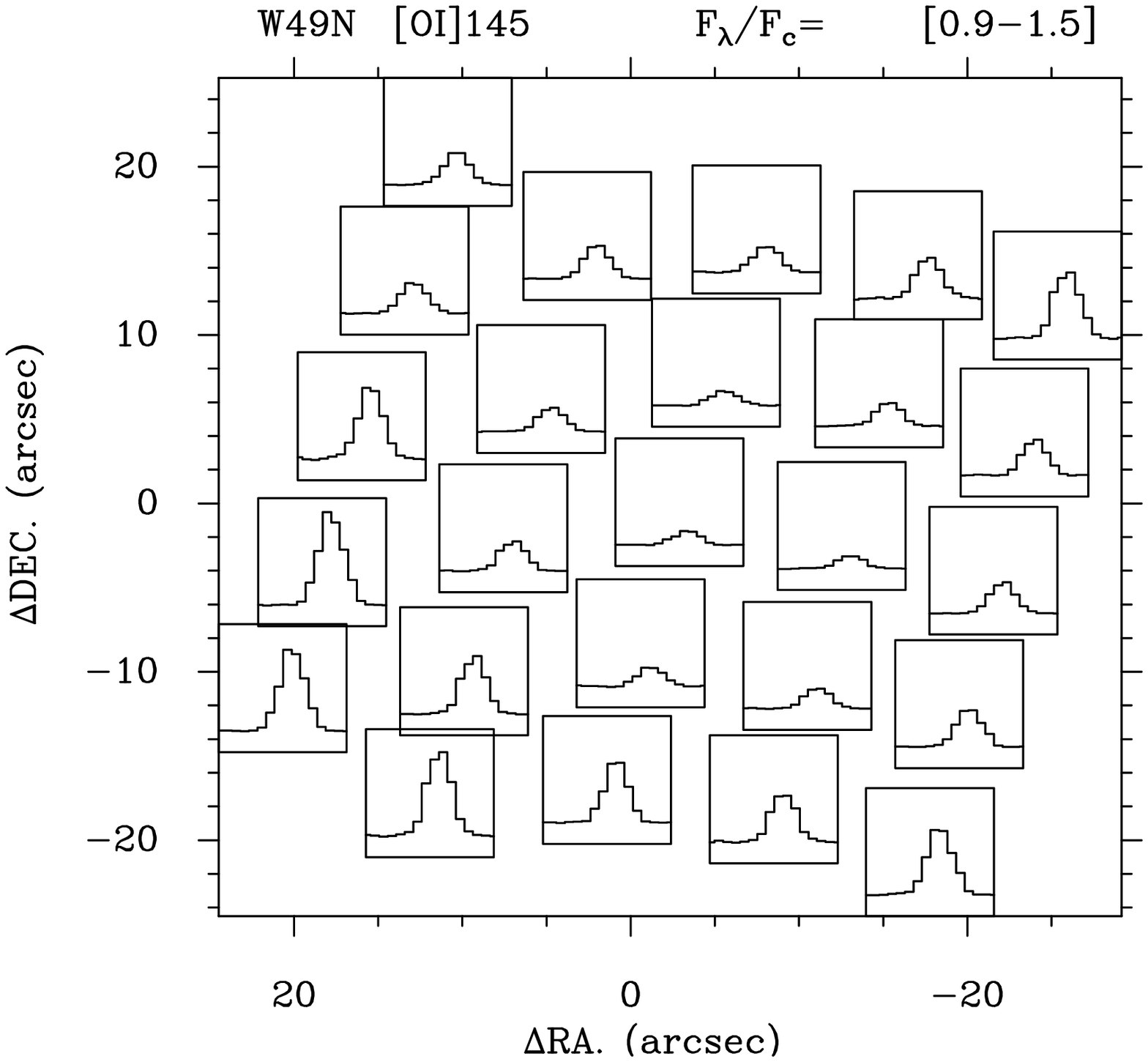}}}
\caption{\label{fig:w49-oi}
Line to continuum ratio spectral maps obtained with PACS for the oxygen fine
structure lines towards W49N. Top: The $^3P_1-^3P_2$ line at 63 $\mu$m,
Bottom:  the $^3P_0-^3P_1$ line at 145 $\mu$m. The vertical scale is indicated
in each map. }
\end{figure}

In order to correct for the [C{\sc ii}] 158\,$\mu$m and  [O{\sc i}] 63\,$\mu$m
line contamination in the reference nod positions we used the
\textit{ChopNodSplitOnOff} pipeline. For W49N, W51 and W31C, the [C{\sc
  ii}] 158\,$\mu$m line emission in the A and B nod positions  is almost the
same ($\sim$10-30$\%$ of the intrinsic ON source emission). The [O{\sc i}] 63\,$\mu$m line emission in the nod positions is typically $\sim$0-10\%\, of the ON source emission.
 For these sources we could correct for the extended emission  by taking into
 account the line emission in each reference spaxel. The final spectral map
 thus contains the continuum emission plus the corrected  [C\,{\sc
   ii}] 158\,$\mu$m and [O{\sc i}]63\,$\mu$m line emission/absorption. 
A polynomial baseline of order 3 was subtracted from each spectrum to obtain the continuum--subtracted data, and each spectrum was divided by the continuum to obtain the line/continuum spectra.
 Figure \ref{fig:pacs-w49n}, and Figures \ref{fig:pacs-w28a} to
\ref{fig:pacs-w3i} show the resulting  [\CII] maps  for each source. 
We also display maps of the relative continuum emission at 158$\mu$m (with
respect to the peak position). Figure \ref{fig:w49-oi} shows the line to
continuum maps  for the [\OI] 63 $\mu$m and 145 $\mu$m lines toward W49N.

\begin{table*}
\caption{\label{tab:continuum} Summary of continuum measurements}
\begin{tabular}{lccccccc}
\hline
\hline
Source & $T_{490}$ & $S_{490}$(HIFI)$^1$ &$T_{809}$ & $S_{809}$(HIFI)$^2$ &
$T_{1900}$ & $S_{1900}$(HIFI)$^3$ & $S_{1900}$(PACS)$^4$ \\ 
  &  (K)  & (10$^2$ Jy/beam) & (K) & (10$^2$ Jy/beam)     &  (K)  & (10$^2$
  Jy/beam) & (10$^2$ Jy/beam)\\
\hline
W28A & $0.25 \pm 0.02$ & $1.16 \pm 0.1$ &$1.18 \pm 0.1$ & $5.50 \pm 0.5$ & $6.6
\pm 0.5$ & $33 \pm 2.5$ & $30 \pm 4.5$\\
W31C & $0.39 \pm 0.03$  & $1.81 \pm 0.14$ & $1.96 \pm 0.05$ & $9.20 \pm 0.2$  &
$6.7 \pm 0.5$ & $34 \pm 2.5$&  $33.8 \pm 5.1$\\
W33A & $0.17 \pm 0.01$ & $0.79 \pm 0.05$ &... &... & $1.7 \pm 0.2$ & $8.6 \pm 
1.0$ & $7.1 \pm 1.1$ \\
G34.3+0.15 &$0.61 \pm 0.02$ & $2.83 \pm 0.1$ &  $2.27 \pm 0.1$ &$10.6 \pm 0.5$
&  $6.8 \pm 0.5$ & $34 \pm 2.5$& $36.2 \pm 5.4$\\ 
W49N & $0.7 \pm 0.02$ & $3.25 \pm 0.1$  & $2.45 \pm 0.1$ & $11.5 \pm 0.5$&
$8.6 \pm 0.6$ & $44 \pm 3.0$ & $39.4 \pm 5.9$ \\
W51 & $0.8 \pm 0.02$ & $3.71 \pm 0.1$ & $2.9 \pm 0.1$ &$13.6 \pm 0.5$ &$7.6
\pm 0.6$ & $38 \pm 2.5$& $39.2 \pm 5.9$ \\
DR21(OH) & $0.36 \pm 0.02$ & $1.67 \pm 0.1 $ & ...& ... &$3.8 \pm 0.5$ & $19
\pm 2.5$ & $15.7 \pm 2.4$\\
W3-IRS5 &$0.2 \pm 0.03$  &$0.93 \pm 0.1$  &... &... &$4.0 \pm 1.0$& $20.0 \pm
5$ & $23.7 \pm 3.6$\\
\hline
\end{tabular}

$^1$ using 464 Jy/K, $^2$ using 469 Jy/K,  $^3$ using 506 Jy/K, $^4$ using the point source calibration.
\end{table*}

\section{Results}
\label{sect:results}

\subsection{Spatial structure}

\begin{figure}
\rotatebox{0}{
\resizebox{8cm}{!}{
\includegraphics{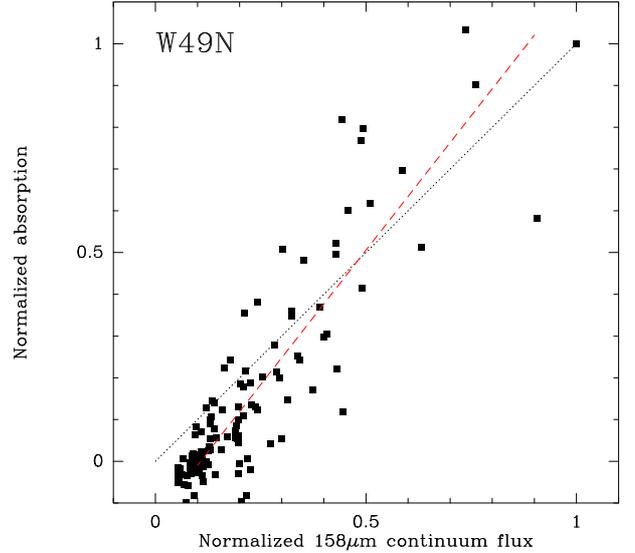}} }

\caption{\label{fig:w49n-absmap}
Comparison of the velocity integrated
absorption ($\int_{16.5}^{80}(T_A-T_C) dv$) measured in the HIFI map relative to the 
velocity integrated absorption at the map
center, with the continuum flux measured in the PACS map relative to the
map center. The dashed red line shows the linear regression line 
and the dotted black line a 1:1 relationship.   }
\end{figure}

We have made use of the  information in both the PACS and HIFI data
to investigate the spatial variations of the foreground absorption. 
The continuum emission was accurately measured at the map center with HIFI.
However, since the continuum information obtained in the HIFI map is not
fully reliable,  the HIFI map provides only the velocity integrated emission 
or  absorption (in \Kkms) at each position. We have therefore made use of the PACS spectral
map data to determine the spatial variation of the far infrared
continuum emission in each source.
As the continuum emission is expected to vary smoothly at 158$\mu$m, we
have interpolated the measured continuum fluxes in each PACS spaxel to
obtain a far infrared continuum map of each source. Because 
this method is not expected to provide an accurate absolute
calibration, we only make use of the spatial variation
of the fluxes relative to the map center.  Figure \ref{fig:pacs-w49n} (left
panel) presents the resulting continuum map for W49N. 
Figure \ref{fig:w49n-absmap} presents the comparison of the velocity integrated absorption relative to the 
velocity integrated absorption at the map center for each pixel of the
HIFI map. The relative integrated absorption and continuum emission
(each normalized by value at the center of the map) show an excellent linear correlation, with a correlation coefficient of 0.88 for W49N and a slope close to unity. 
This behavior is expected for an extended and uniform foreground absorber.
While small amplitude variations are not excluded given the limited signal
to noise ratio, the diffuse gas along the line of sight behaves to first order as a
uniform screen across the angular extent of the W49N background
source. Therefore the molecular column densities derived from the ground state
hydride  absorption lines  reported elsewhere with different angular resolutions are directly comparable to the present data.
The same behavior is found for the other sources in the sample.

\subsection{[\CII] line to continuum ratio}

All sources exhibit  complex line profiles in the velocity ranges
 associated with the star forming regions, with multiple velocity components,
the possibility of self-absorption and in some cases broad line wings
likely tracing  UV-irradiated outflows, e.g. for W49N or for W28A (Fig.~\ref{fig:specw28a},\citet{gusdorf}). Nevertheless
the velocity integrated [\CII] signal from the massive star forming 
regions is always positive.  W33A is the most extreme case since the [\CII] emission almost vanishes at the HIFI map center as illustrated in  Fig.~\ref{fig:mapw33a}. The presence of diffuse material in the foreground leads to absorption of the
 background emission. At the PACS spectral resolution, the 
resulting [\CII] signal   in the central spaxel can appear either in absorption
(towards W49N) or in emission (towards W51) depending on the relative
strengths of the emission and absorbing signals. 
This shows the need for  a sufficient spectral resolution to separate 
the different contributions to the [\CII] signals in sources with 
a complex velocity structure.

\begin{figure}
\rotatebox{0}{
\resizebox{7.5cm}{!}{
\includegraphics{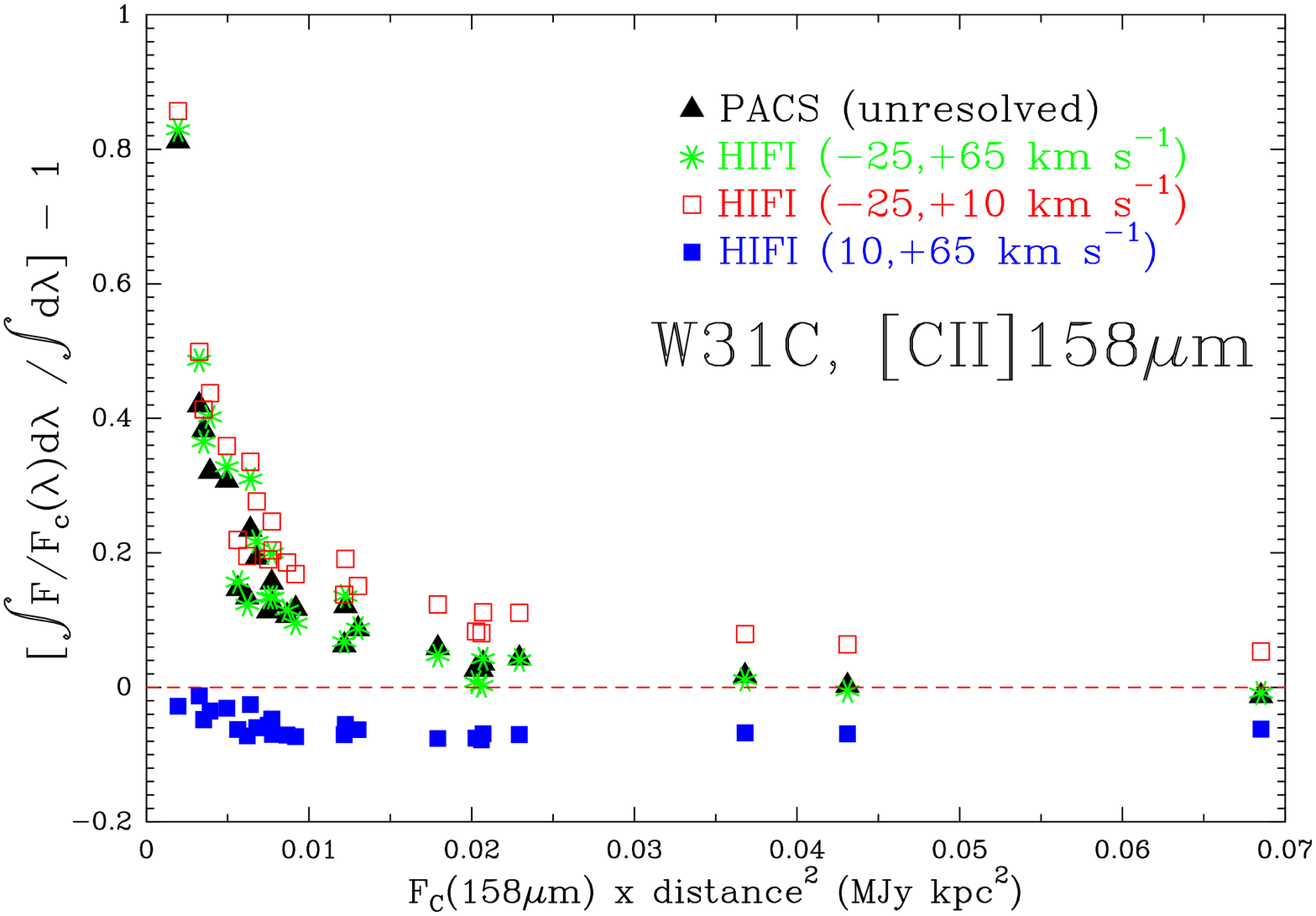} }}
\\
\rotatebox{0}{
\resizebox{7.5cm}{!}{
\includegraphics{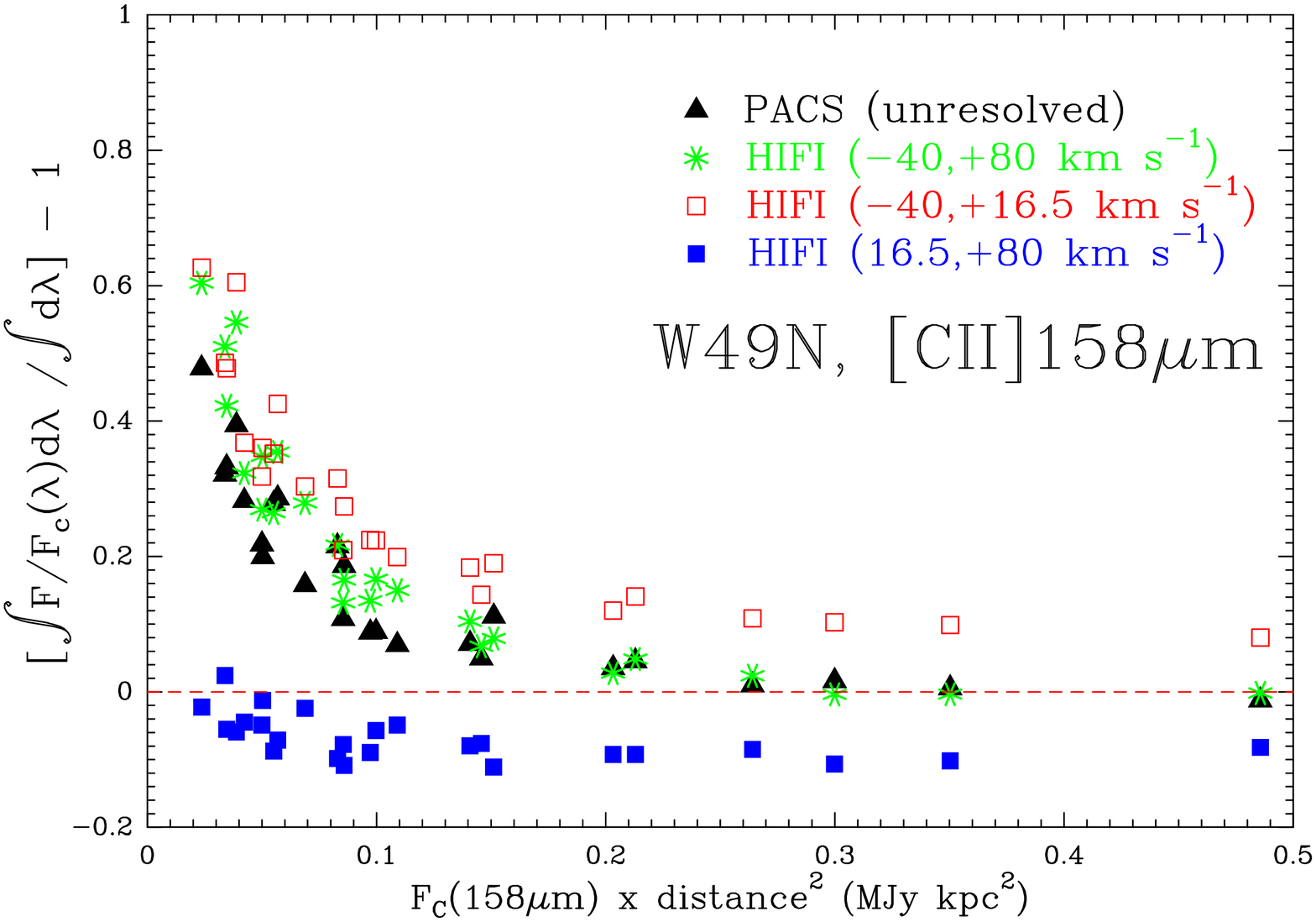} }}
\\
\rotatebox{0}{
\resizebox{7.5cm}{!}{
\includegraphics{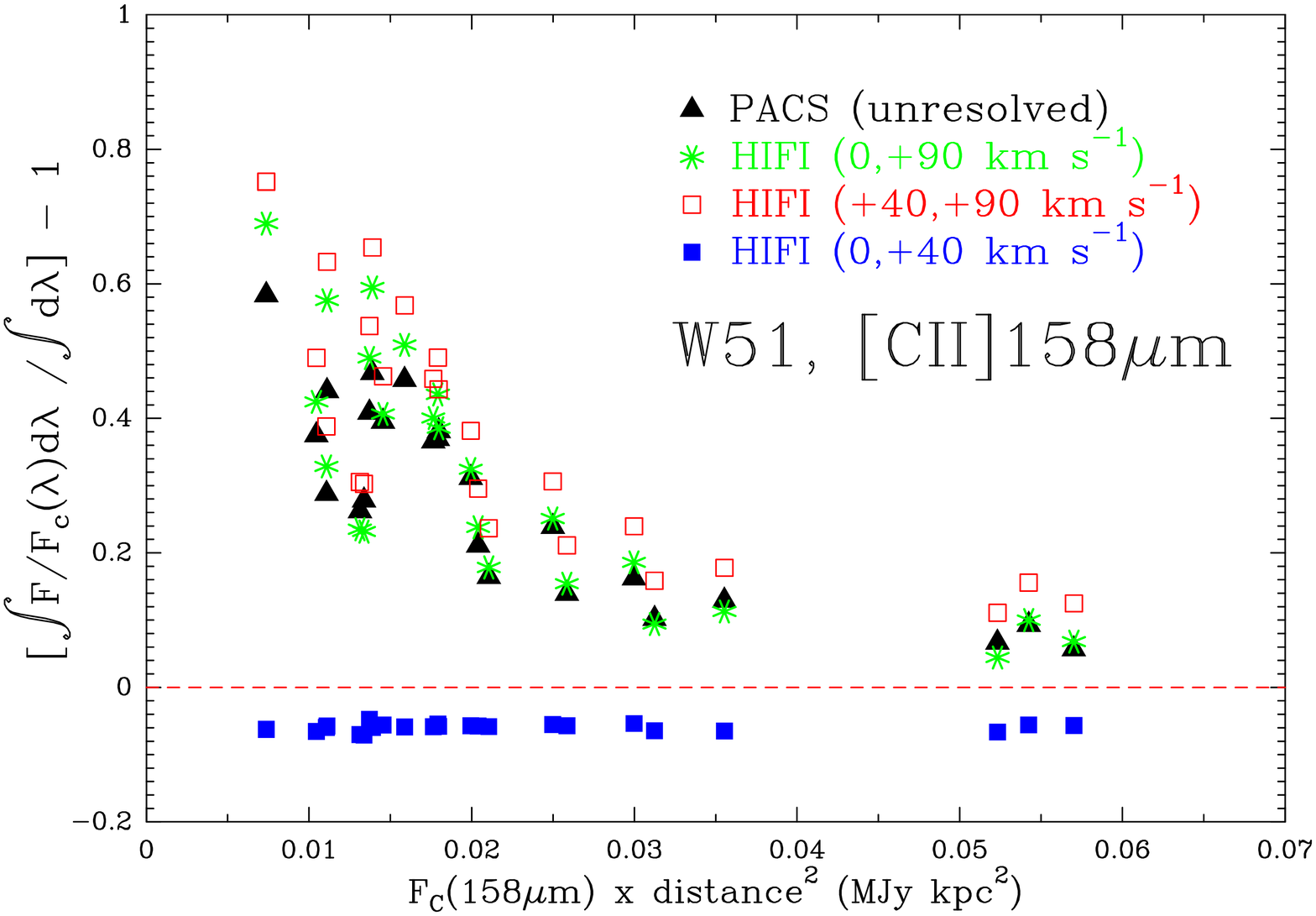} }}
\caption{\label{fig:relative_cii} Relative [\CII] emission as a function of
  the 158$\mu m$ continuum luminosity for W31C, W49N and W51. For each source,
  the black triangles show the spectrally unresolved PACS data, the
  blue squares the contribution of the foreground absorption measured by HIFI,
  the red squares the intrinsic source emission measured by HIFI, and the green
  stars the total emission and absorption measured by HIFI. The velocity
  intervals for the integration are given in each plot. In all plots the points on the left side correspond to the map edges while those on the right side correspond to the massive star forming regions associated with the dust continuum peak.}
\end{figure}

The [\CII] line is used as a tracer of the star formation rate in external
galaxies,
including distant systems at high redshifts \citep{stacey:10}. We have used the present data to
estimate the [\CII] emission relative to the far infrared continuum  in the three
massive star forming regions W31C, W49N and W51. As explained above, 
the PACS [\CII] and [\OI] data could be  corrected from the low level emission 
from the OFF position for these three sources only.  The data are shown in Figure
\ref{fig:relative_cii} in which each PACS spaxel is displayed as an independent  point.
For each source  the PACS data where the contributions from the
source and the foreground absorption are merged are shown as black squares, and
the same quantity measured by HIFI  as green stars. The HIFI data allow
 separation of the intrinsic source contribution (empty red squares) and 
the absorption
from the foreground gas (filled blue squares with negative values). In all plots
 the horizontal scale is the product of the spaxel flux by the distance to the source, with the low values on the left side corresponding to the map edges and the high values on the right side to the continuum peak associated with the massive star forming region. The 
vertical scale is the relative [\CII] emission $I_{rel}(\ion{C}{II})$, defined as
\begin{equation}
I_{rel}(\ion{C}{II}) = \frac{{\int_{157.5}^{158.0} \frac{F(\lambda)}{F_c(\lambda)}d\lambda}}{\int_{157.5}^{158.0}
  d\lambda} - 1~. \label{eq:1}
\end{equation}

 To compute $I_{rel}(\ion{C}{II})$ from the PACS observations we used a wavelength  window of
$\delta \lambda = 0.5$~$\mu$m starting at 157.5 $\mu$m and ending at 
158.0~$\mu$m . This wavelength interval is  broader than the
intrinsic line widths and comparable to the full frequency bandpass of the HIFI
band 7 receiver.

As both the line and continuum values are detected with the same instrument,
$I_{rel}(\ion{C}{II})$  is independent of calibration errors or uncertainties in the PSF.
The ratio of the [\CII] line luminosity $L([\ion{C}{II}])$, to the far infrared
luminosity $L_{FIR}$ can be
simply derived from $I_{rel}(\ion{C}{II})$ and depends weakly on the shape of the far
infrared spectral distribution of the source.  Using infrared emission
  spectra computed with the  Draine \& Li dust model \citep{draine:07} and
  choosing the parameters appropriate to
our sources, we determine 
\begin{equation}
\frac{L([\ion{C}{II}])}{L_{FIR}} \sim 3.1 \pm 0.35 \times 10^{-3} I_{rel}(\ion{C}{II})~.
\end{equation}
The data for W31C, W49N and W51 show an excellent agreement between the
unresolved data from PACS and the velocity resolved HIFI data. The dip in
[\CII] detected in the PACS maps towards W49N and W31C is fully explained by
the foreground absorption. These plots also confirm the constancy of the
absorption across the extent of the PACS map as the blue points fall
nearly on a horizontal line. The slight decrease of the absorption 
at lower continuum flux  is likely to reflect the competition between  the weak
emission filling in the absorption at lower continuum fluxes, as expected from
the physical conditions along these sight-lines (see below).

It is also interesting to notice that the intrinsic emission from the massive
star forming regions presents a strong spatial variation of their relative
 [\CII] emission, from $I_{rel}(\ion{C}{II}) \sim 0.5$ at the map edges down to
 $I_{rel}(\ion{C}{II}) \sim 0.06$ at the peak of the far infrared emission,
or from $L([\ion{C}{II}])/L_{FIR}$ $\sim 1.5 \times 10^{-3}$ down to 
 $L([\ion{C}{II}])/L_{FIR}$ $\sim 2 \times 10^{-4}$.
This  behavior is reminiscent of the deficit in [\CII] emission observed in compact and
luminous infrared galaxies such as Arp~220 or NGC~4418 
\citep{gracia-carpio} with deeply
embedded star forming regions and possible Active Galactic Nuclei. As for
massive star forming regions, the heating sources are deeply embedded in large
dust column densities. Therefore the emerging signal in the fine structure
lines is heavily dependent on the geometry of the heating radiation
relative to the location of the dense molecular gas, and the emergent intensity can be severely attenuated. 
The presence of absorption by a foreground layer, as in the  PRISMAS sources,
will also contribute to dimming the emergent [\CII] signal, especially at low
spectral resolution.

\subsubsection{Diffuse ISM absorption toward NGC~4418}
To illustrate the effect of foreground absorption, we have used the recent \ion{H}{I} absorption observations towards the  NGC~4418 galaxy nucleus \citep{costagliola} to evaluate the [\CII] line
opacity associated with the foreground absorption. The HI spectrum presents
two components with total column densities of $5.8 \times 10^{21}$~\pscm \ and
$1.5 \times 10^{22}$~\pscm. Assuming that the gas producing the HI absorption
is diffuse, it is expected to produce an absorption signal in [\CII] as well.
We can estimate the magnitude of this absorption from the HI column densities
using a carbon to 
hydrogen abundance ratio of $x_C = 1.4\times$10$^{-4}$. These numbers
lead to expected integrated line opacities of $\sim 5.8 \kms$ and $\sim 15
\kms$ for the [\CII] line associated with each velocity component.
To further estimate the impact of this absorption, we use 
the measured line widths of 130 and 160 \kms \ to derive 
the relative emission $I_{rel}(\ion{C}{II})$ as defined in Eq.\ref{eq:1}, and
obtain -0.04 and -0.11. Note that the figures are negative because 
they correspond to absorption. 
We therefore obtain comparable figures for the
absorption caused by the diffuse gas along the line of sight to massive
star forming regions, and for the diffuse gas in the foreground of NGC~4418
nucleus.  This simple example shows that the
interpretation of low [\CII] emission, or even of the non detection of [\CII], 
in luminous
infrared galaxies is difficult. The presence of absorption lines, either from 
HI in the centimeter domain, or molecular lines in the far infrared domain should be
used as a warning for the possibility of a [\CII] absorption, which would
affect the emission from the dense star forming regions. In that case the use
of classical PDR modeling to interpret the detected signal should be done
with caution. A detailed knowledge of the source geometry and its environment
is needed to obtain meaningful results.

\subsection{Atomic Oxygen fine structure lines}

While the discussion is focused on [\CII], it should be noted that
the presence of foreground absorption is expected to affect the [\OI] line at 63
$\mu$m even more severely than  [\CII], as this line is expected to
have a larger opacity for the same total hydrogen column density.
Indeed, using a carbon to hydrogen ratio of $x_C = 1.4 \times 10^{-4}$, 
an oxygen to hydrogen ratio of $x_O = 3.1 \times 10^{-4}$ \citep{sofia:04}, the
spectroscopic parameters listed in Table \ref{tab:lines}, and assuming
low excitation as appropriate for diffuse gas, we obtain
$\tau([$\ion{O}{I}$_{63}])$ = 1.6~$\tau([\ion{C}{II}])$.

 Spectrally resolved line profiles of the [\OI] 63~$\mu$m line are clearly needed  to accurately evaluate
the impact of foreground absorption, or self-absorption in the sources
themselves. 

\begin{figure}
\rotatebox{0}{
\resizebox{7.5cm}{!}{
\includegraphics{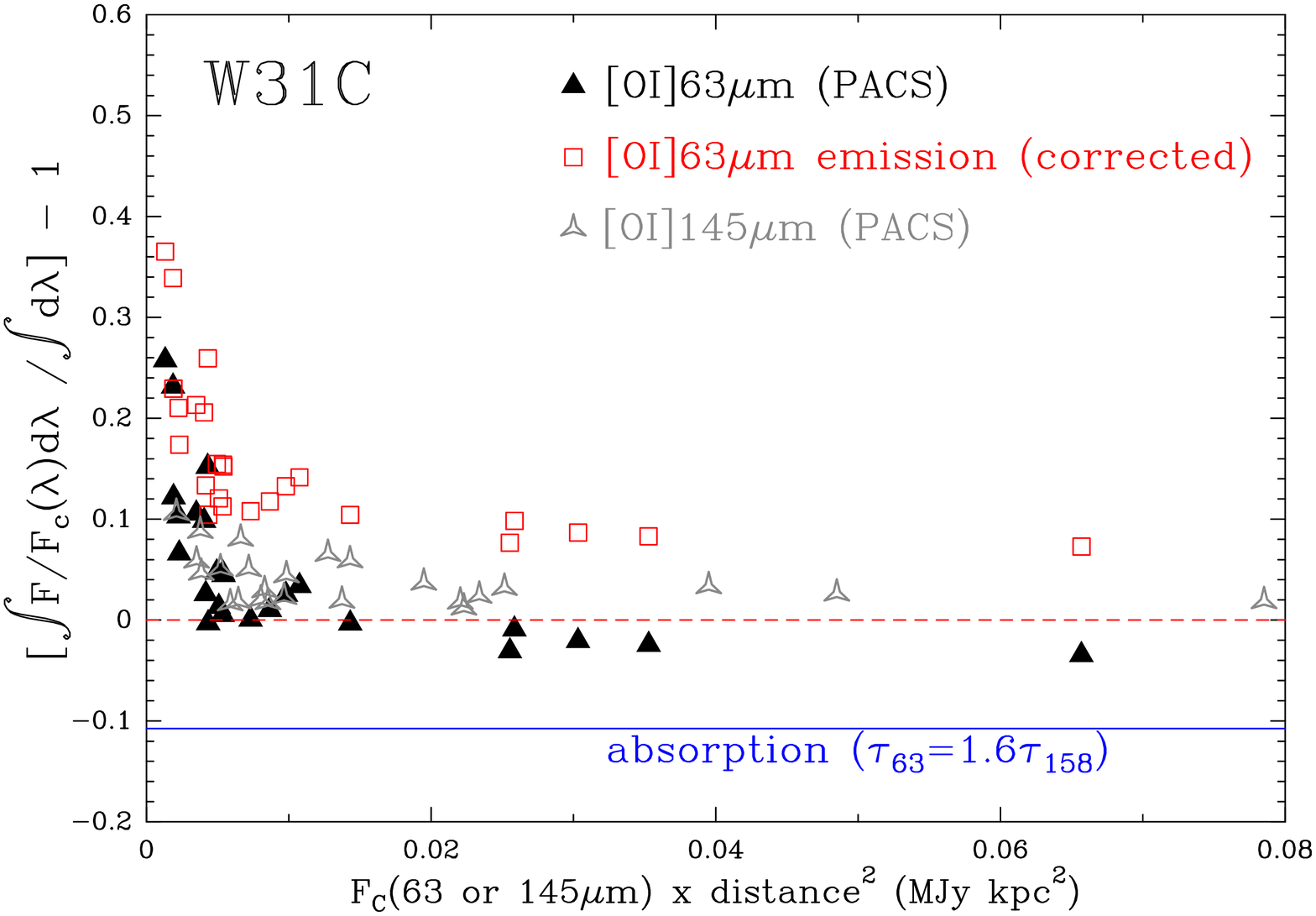} }}
\\
\rotatebox{0}{
\resizebox{7.5cm}{!}{
\includegraphics{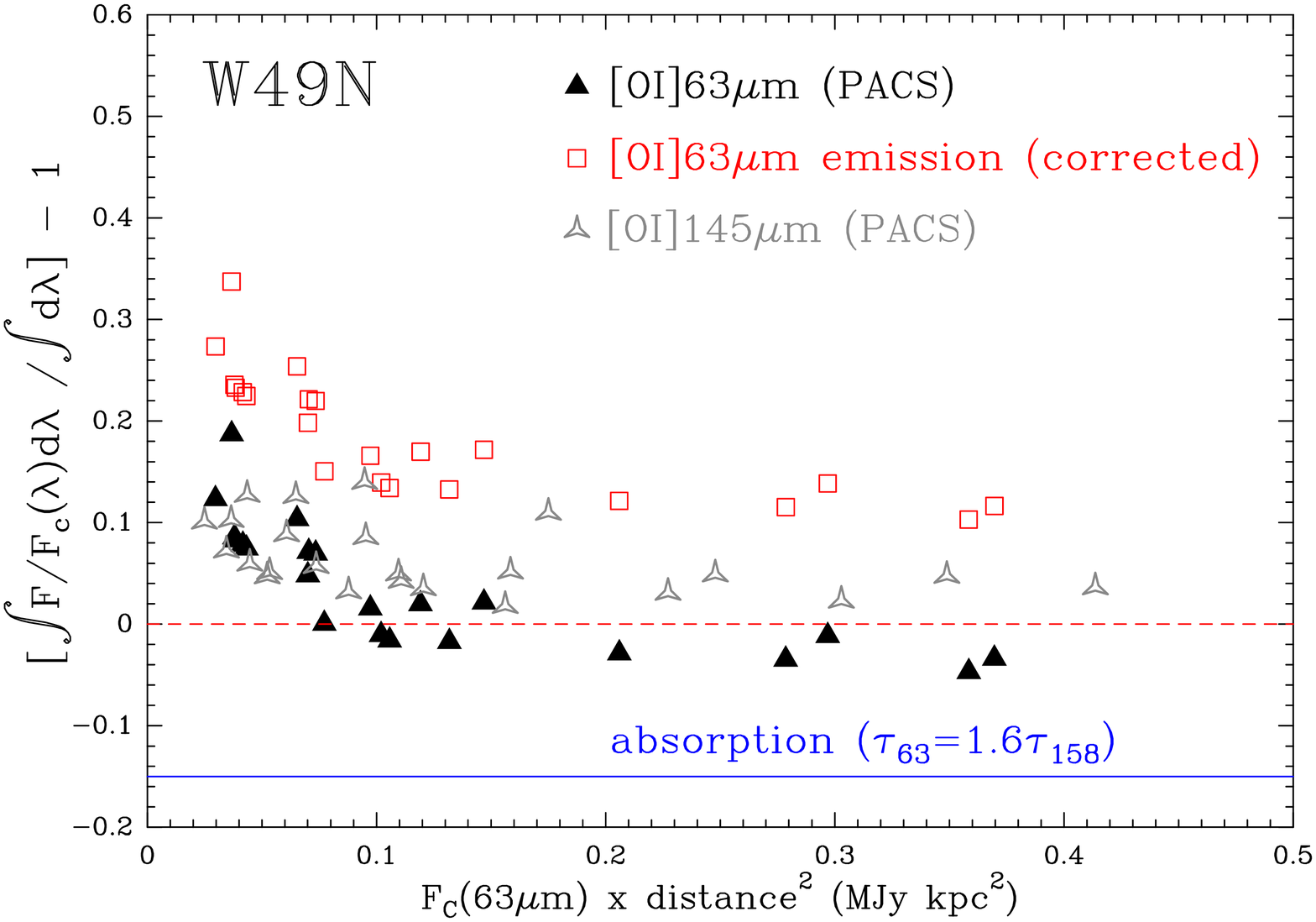} }}
\\
\rotatebox{0}{
\resizebox{7.5cm}{!}{
\includegraphics{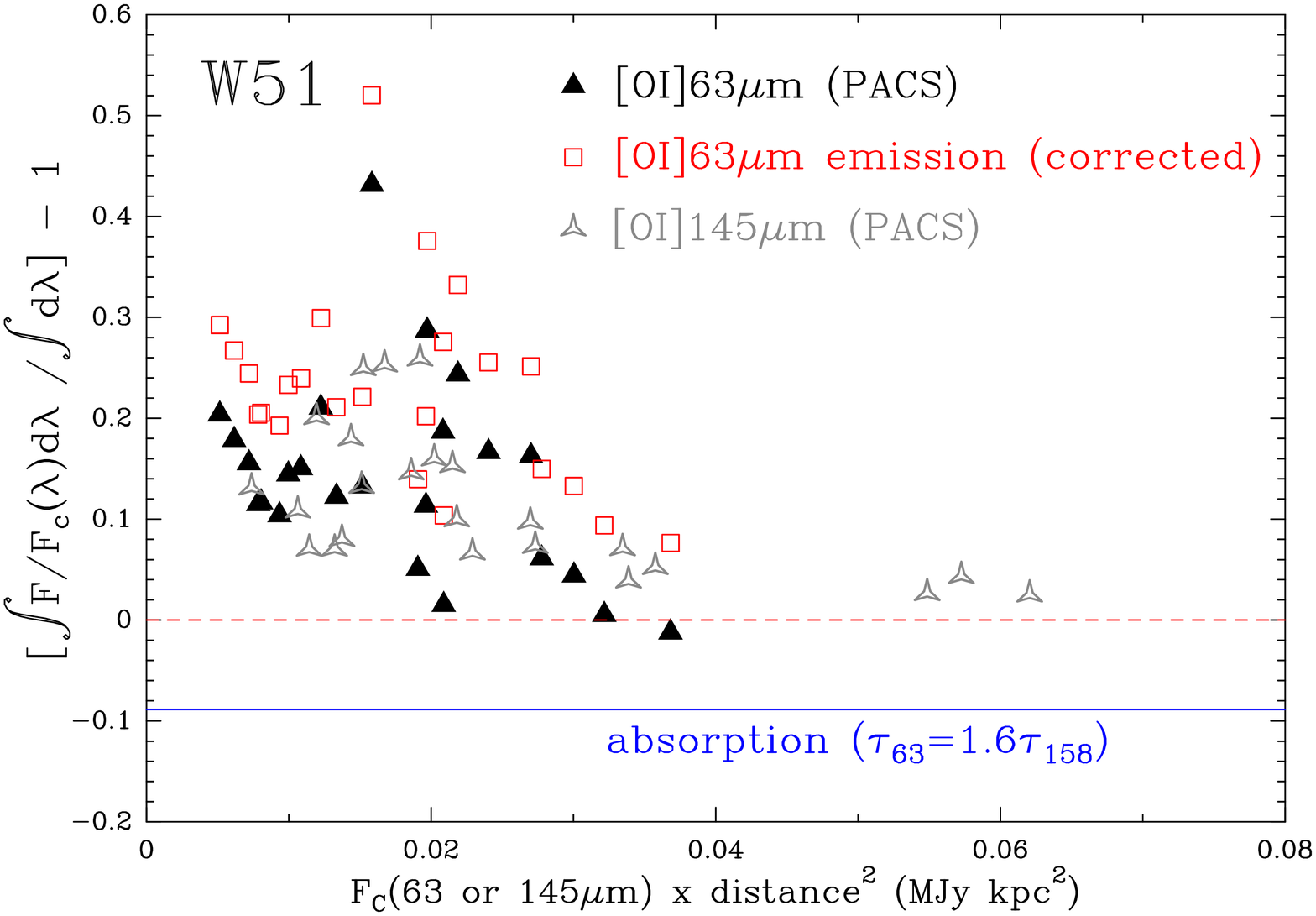} }}
\caption{\label{fig:relative_oi} As for Figure \ref{fig:relative_cii}, but for
  the two [\OI] fine structure transitions. The blue line shows the estimated
  level of absorption (negative relative emission) caused by the diffuse
gas along the line of sight. It is obtained 
from the mean of the [\CII] absorption scaled by a factor 1.6, the ratio of
the [\OI] 63~$\mu$m and [\CII] line opacities for the adopted carbon and 
oxygen abundances. }
\end{figure}

Using the PACS data shown in Fig \ref{fig:w49-oi}, we have computed the
relative emission of the [\OI] fine
structure lines at 63~$\mu$m and 145~$\mu$m, as a function of the continuum
luminosity at the same wavelength. 
The relative emission for both oxygen lines
is defined as in Eq.\ref{eq:1}. 
\footnote { To compute $I_{rel}$ from PACS observations we used a wavelength  window of
$\delta \lambda = 0.5$~$\mu$m for [\CII], 0.45~$\mu$m for [\OI$_{145}$] and
0.2~$\mu$m for [\OI$_{63}$], with the following lower and upper limits:
157.50 -- 158.00~$\mu$m for [\CII], 63.28 -- 63.08~$\mu$m for [\OI$_{63}$], and 145.30 -- 145.75~$\mu$m for [\OI$_{145}$].  These wavelength intervals are  broader than the intrinsic line widths.}
In a similar manner as for [\CII], 
 the relative [\OI] line emission can be related to the ratio of the
 the line luminosity to the far infrared luminosity ratios, but with
different scalings. We obtain 
\begin{equation}
\frac{L([\ion{O}{I}_{63}])}{L_{FIR}} \sim  10^{-3} I_{rel}(\ion{O}{I}_{63})
\end{equation}
 and 
\begin{equation}
\frac{L([\ion{O}{I}_{145}])}{L_{FIR}}\sim 3.1 \times 10^{-3}I_{rel}(\ion{O}{I}_{145}).
\end{equation}
 Figure \ref{fig:relative_oi} presents the relative emission of the 
[\OI] lines for the W31C, W49N and W51 sources.  
In this figure the black triangles show the data for the ground state line at 63~$\mu$m and the gray symbols those for the excited line at 145~$\mu$m.
As expected from the [\CII] behavior, and the larger opacity of the 
[\OI] 63$\mu$m line, the oxygen maps show an
even stronger effect of the absorption by the foreground gas than do the
[\CII] maps. To evaluate the effect of the foreground gas and restore the
intrinsic emission from the background source, we have estimated the 
absorption caused by the foreground gas using the mean
value of the [\CII] absorption, scaled by 1.6, the ratio of the line opacities
for the adopted carbon and oxygen abundances. It is shown as a blue line in
Fig \ref{fig:relative_oi}. The corrected points are shown as empty red squares.
While the original PACS [\OI] data at 63~$\mu$m   present either 
a pure absorption (negative $I_{rel}(\ion{O}{I}_{63})$), or a weak [\OI] emission
comparable to that of the  [\OI] 145 $\mu$m line, the corrected points are 
always positive,  with $I_{rel}(\ion{O}{I}_{63})$  stronger than $I_{rel}(\ion{O}{I}_{145})$.
For the massive sources,  $I_{rel}(\ion{O}{I}_{63})$ ranges from $\sim 0.4$ at the
edge of the observed regions with moderate far infrared emission 
down to $\sim 0.1$ at the peak of the far infrared emission. 
 $L(\ion{O}{I}_{63})/L_{FIR}$ thus decreases from  $\sim 4 \times 10^{-3}$ down to 
$\sim 10^{-4}$.

This shows that in a situation where a bright source is observed through 
a foreground of
diffuse gas, the use of the [\OI] 63~$\mu$m line alone for quantitative models is
not recommended either. Line ratios comparing the two fine structure lines of
oxygen at 63 and 145~$\mu$m, or [\OI] with [\CII] will need to be corrected
for absorption before being    of quantitative use. Luminous infrared galaxies showing a
[\CII] deficit are also expected to be deficient in the 63~$\mu$m [\OI] line
if the deficit is caused by a foreground layer of diffuse gas.

\subsection{C$^+$ Column densities}

The spectroscopic parameters of the detected lines are summarized in Table
\ref{tab:lines}. 
The C$^+$ column density of the diffuse gas can be obtained from the velocity integrated  absorption through the relationship 
\begin{equation}
N({\rm C^+}) = \frac{8 \pi \nu^3 Q(T_{ex}) }{g_uA_{ul}c^3}\frac{1}{1-exp(-\frac{h\nu}{k_BT_{ex}})} \int \tau([\ion{C}{II}]) dv~,
\end{equation}
where $Q(T_{ex})$ is the partition function at the excitation temperature
$T_{ex}$ ($Q(T_{ex}) = g_l + g_ue^{-h\nu/k_BT_{ex}}$), $A_{ul}$ is the Einstein coefficient listed in Table \ref{tab:lines}, and $\nu$ is the line frequency. 
As discussed by \citet{goldsmith:12}, the excitation
temperature of [\CII] is significantly lower than the kinetic temperature
in the diffuse gas because the densities are too low to produce a collisional excitation rate comparable to the spontaneous decay rate.  For $T_{ex}$ lower than $\sim 25$~\K, the above expression can be written
\begin{equation}
\label{lowtex}
N({\rm C^+}) \ (\pscm) =  1.4 \times 10^{17} \times  \int \tau ([\ion{C}{II}]) dv \ \ (\kms)~.  
\end{equation}
The above limit in the excitation temperature
corresponds to a fraction of C$^+$ ions in the ground state larger
than 95\%. At higher excitation temperatures, equation \ref{lowtex} underestimates the C$^+$ column density.

In the diffuse interstellar gas, C$^+$ is the gas phase carbon reservoir. The
total gas column density can therefore be derived, assuming that the
gas phase carbon abundance relative to hydrogen  is constant. 
Using the value derived by \citet{sofia:04}, $1.4 \times 10^{-4}$,
we obtain :
\begin{equation}
N({\rm H}) \ (\pscm) \sim 10^{21} \int \tau ([\ion{C}{II}]) dv \ \ (\kms)~.  
\end{equation}
This formula shows the excellent sensitivity of the absorption measurements
that allow the detection of low column density regions, $\sim 10^{20}$\pscm \
corresponding to 0.05 magnitudes of visible extinction.

\subsection{Atomic carbon column densities}
\label{subsec:ci}
For neutral carbon, most spectra show pure emission, the sole
exception being the 40 \kms \ velocity feature in the 
$^3P_2 - ^3P_1$ spectrum towards W49N. 
 In LTE the atomic carbon column densities can be derived from the
  expression appropriate for emission, 
\begin{equation}
N({\rm C}) (\pscm) = 1.4 \times 10^{16 }\int T_{1-0}dv \ \ (\Kkms)~,
\end{equation}
where $T_{1-0}$ is the main beam temperature of the ground state fine
structure line at 492~GHz. This formula assumes that the emission is
optically thin and the level populations can be described by 
a single excitation temperature, $T_{ex}$. It is accurate within
10\% for $16 \K <  T_{ex} < 120 \K$. The column density is underestimated
 for higher or lower values of $T_{ex}$, especially at the lower end where the
 populations of the excited levels become very small. This may be the case for
instance in low density diffuse neutral gas. Therefore we performed statistical
equilibrium calculations to derive the atomic carbon column densities, using
the column density derived from the LTE formula as a lower limit.

The velocity feature at $\sim 40 \kms$ towards W49N deserves a specific
analysis because it shows up in absorption in the $^3P_2 - ^3P_1$ line at
809~GHz, but is not detected neither in emission nor in absorption in the $^3P_1
- ^3P_0$ line at 492~GHz.  
This behavior corresponds to the very special case where the
excitation temperature of the ground state line is equal to the
apparent continuum value of the background source. At 809~GHz, the 
rise of the thermal dust continuum emission
results in a higher continuum temperature than  the line 
excitation temperature, hence the appearance of an absorption line.

The radiative transfer equation for a plane parallel isothermal cloud located
in front of a background continuum source producing a continuum antenna temperature  $T_c$ can be written
\begin{equation}
T = T_ce^{-\tau} + J(T_{ex})(1-e^{-\tau})~,
\label{eq:cont}
\end{equation}
where $J(T_{ex}) = \frac{h\nu}{k_B}\frac{1}{e^{h\nu/k_BT_{ex}} - 1}$. In this
formula the first term represents the absorption of the background radiation
and the second one the emission of the plane parallel foreground cloud. 
These two terms can partially cancel each other, with the emission filling in
the absorption. In the situation that $J(T_{ex}) = T_c$, equation \ref{eq:cont}
yields $T = T_c$ and the foreground cloud becomes invisible.
For W49N, the continuum value at 492~GHz, expressed in main beam temperature
units, is $0.9  \pm 0.03 \K$. This  translates
to an excitation temperature of $T_{ex} = 7 \pm 0.2 \K$. Such a low
excitation temperature is consistent with the presence of  cold 
molecular gas, in agreement with the detection of deep absorption in
CN, NH$_3$ and H$^{13}$CO$^+$ \citep{godard:10,persson}. Indeed, as
discussed by \citet{tatematsu}, the 492~GHz [\CI] line is almost
thermalized at the edges of dense molecular cores. The carbon column density
for this particular feature is derived from the $^3P_2 - ^3P_1$ line using
$T_{ex} = 7 \K$, $N({\rm C}) = (2.5 \pm 0.2) \times 10^{17}$ \pscm. This large value
is consistent with the strong H$^{13}$CO$^+$ absorption \citep{lucas:00} and
confirms that this feature is associated to translucent material 
(A$_V \sim 3-5$~mag)  rather than diffuse gas (A$_V \sim 1$~mag).

\subsection{Physical conditions from the [\CII] data}
\label{physcon}
The combination of the emission and absorption allows
us to derive the excitation conditions of ionized carbon since
it provides both the line opacity and peak temperature.
Following \citet{goldsmith:12} we have derived the excitation temperature
for the velocity components detected both in absorption and in
the OFF emission spectra.
The well--known equations in the case that the foreground gas is uniform in terms of excitation temperature and optical depth are:
\begin{equation}
T_A(ON) = T_{c}e^{-\tau} + J(T_{ex})(1-e^{-\tau})
\end{equation}
\begin{equation}
T_A(OFF) = J(T_{ex})(1-e^{-\tau})
\end{equation}
\begin{equation}
\Delta T_A = T_A(ON) - T_A(OFF) = T_{c}e^{-\tau}
\end{equation}
where $T_c$ and $J(T_{ex}$) are defined above. ON - OFF spectrum provides the line opacity.   This can then be used to analyze the OFF emission spectrum, which yields  $T_{ex}$ since the opacity is known. Given the large  separation
of the C$^+$ fine structure levels, 92~K, the modest density and column
density of the diffuse medium produces only  weak emission, typically at the
0.5~K level. The OFF position spectra displayed in the bottom panel of 
Fig. \ref{fig:specw49n} and Fig
\ref{fig:specw28a} to \ref{fig:specw3}
 indeed show emission at this level in the same  velocity range as the
prominent absorption detected in the ON source spectra.

\begin{figure}
\rotatebox{0}{
\resizebox{9.cm}{!}{
\includegraphics{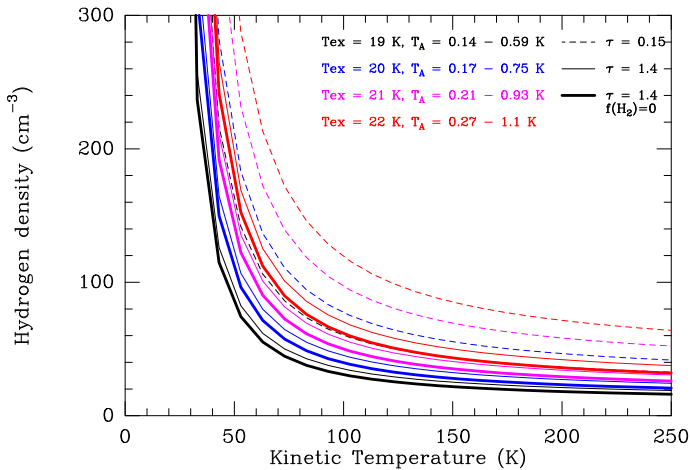} }}
\caption{\label{fig:ciiex}Contours of constant excitation temperature of the [\CII] transition in the diffuse neutral medium with values typical of those deduced here (indicated at the top of the figure).  The dashed lines show the results for an optically thin line ($\tau = 0.15$), while the solid lines are for a mildly optically thick line ($\tau = 1.4$).  For curves with normal weight lines,  n(HI)=n(H$_2$), while the results for a pure atomic medium are given in bold lines; we see that the form of hydrogen makes little difference in terms of [\CII] excitation.  }
\end{figure}

The derived optical depths and excitation temperatures  are given in Table \ref{tab:tex}. Despite the range
in [\CII] opacities, the excitation  temperatures cluster around the
median value of $T_{ex} = 20$~\K \ and a mean value of $T_{ex} = 20.4 \pm 1.2$~\K. The weak feature towards
W3-IRS5 has been omitted from this average as the S/N ratio is
poor and the accuracy of the opacity determination is lowered by the presence
of base line undulations. 

The derivation of the excitation temperature
relies on several hypotheses. In particular it assumes that the
same component is responsible for the emission and absorption. It therefore
ignores possible contribution of the warm ISM phases, 
the Warm Neutral Medium (WNM) and the Warm Ionized Medium (WIM) to the
[\CII] signals. Although weaker than the CNM features, emission from the
WIM has been detected by {\sl Herschel} \citep{velu} at a level comparable to
the features discussed here. The WNM is not expected to produce emission
features given the low densities and electron fraction, but could contribute
to the broad absorption features, and therefore lead to an overestimate
of the line opacity. In this case, the excitation temperatures could be
slightly underestimated. 

The expected opacity from a WNM feature can be estimated from the average
physical conditions of this phase as derived by \citet{heiles:03}.  The mean density $\sim 0.3$~\pccm, and the average 
HI column density per kpc of  $N(\rm{WNM}) = 8.7 \times  10^{20}$~\pscm.
With a C$^+$ fractional abundance of $1.4 \times 10^{-4}$, the corresponding C$^+$ column
density is $N({\rm C^+_{WNM}}) \sim 1.2 \times 10^{17}$~\pscm kpc$^{-1}$, corresponding to an integrated
line opacity of $\int \tau([\ion{C}{II}])dv \sim 0.9 $ \kms kpc$^{-1}$.
 The line width is constrained by the 
typical velocity gradient along a sight-line through the Galactic Plane, about 
$10 \kms$ per \kpc \ for the considered Galactic longitudes, and by the turbulent and thermal broadening,  to be $ \geq 10 \kms$,
hence the maximum opacity of each feature is   $\tau_{max} \leq 0.09$. 

A similar calculation holds for the WIM, with an expected 
higher C$^+$ abundance given the partial destruction of dust grains
in hot gas.
 This simple calculation therefore shows that the warm phases 
make only a minor contribution to the absorption, except for long path-lengths
in velocity--crowded regions. They may contribute to the weak and broad 
absorption features detected  between  the deep and narrow features.
This minor contribution with $\tau \leq 0.1$ will not affect the
determination of the excitation temperature except for the most optically thin
features.

Figure \ref{fig:ciiex} presents loci of constant
excitation temperature of the [\CII] transition in the diffuse ISM taking into
account collisions with H, H$_2$ and electrons. In these calculations, the
electron fraction $x_e$ is set to $3 \times 10^{-4}$ \citep{draine}, about twice 
the adopted gas-phase elemental
carbon abundance  and two hypothesis 
have been made for 
the relative fractions of atomic and molecular hydrogen : pure atomic hydrogen
or equal amounts of H and H$_2$. 
For a typical kinetic temperature of 100~\K, the mean excitation temperature
derived above (20.4~K) corresponds to gas densities between 
40 and 80 \pccm depending on the
[\CII] line opacity and the fraction of hydrogen in molecular form. These
conditions clearly correspond to those of the diffuse ISM and confirm that
the bulk of the detected [\CII] emission and absorption is consistent with 
diffuse gas.

\begin{table*}
\begin{center}
\caption{\label{tab:tex}[\CII] Excitation temperatures deduced from the
  emission in the OFF  spectra$^a$}
\begin{tabular}{lcccccc}
\hline
\hline
Source & $V_{LSR}$ & $T_{peak}$  & Width & Area  & $\tau$ & $T_{ex}$ \\
       & (\kms) & (\K) & (\kms) & (\Kkms) &  & (\K) \\
\hline
W28A & [15;25] &  $0.53 \pm 0.2$ & $2 \pm 0.6$ & $1.1 \pm 0.4$ & $0.7 \pm 0.10$ & 
$20.8 \pm 1.4$ \\
W31C & [15;30] & $0.56 \pm 0.1$ & $11 \pm 4$ &  $6.7 \pm 3$  & $1.3 \pm 0.06$ & $19.5 \pm
1.7 $\\
     & [30;50] & $0.63 \pm 0.1$ & $14 \pm 4$ & $9.6 \pm 3$  & $1.4 \pm 0.06$ 
& $19.8 \pm 1.0 $\\
G34.3+0.15 & [10;20] & $0.43 \pm 0.1$ & $1.2 \pm 0.5$ & $0.54 \pm 0.2$  
& $0.65 \pm 0.06$ & $20.0 \pm 0.9 $\\
	& [20;40] & $0.56 \pm 0.1$ & $6.6 \pm 1.2$ & $3.9 \pm 0.7$ & 
$0.83 \pm 0.06$ &   $20.4 \pm 0.5 $\\
 W49N    & [10;50] & $0.53 \pm 0.2$ & $3.0 \pm 0.9$ &$1.6 \pm 0.5$ & $0.4 \pm
 0.1$ & $22.6 \pm 0.5$ \\
 & [50;70] & $0.63 \pm 0.1$ & $17 \pm 3.8$ &$11.3 \pm 1.7$ & $1.3 \pm 0.06$ 
 & $19.8 \pm 1.6 $\\
W51 & [1:10] & $0.33 \pm 0.15$ & $12.2 \pm 2.4$ &  $4.3 \pm 0.8$ & $0.58 \pm 0.06$
& $19.3 \pm 0.2$\\
    & [16:35] & $0.35 \pm 0.15$ & $1.9 \pm 0.6$ & $0.71 \pm 0.3$ & $0.34 \pm 0.06$ & 
$21.4 \pm 2.0$ \\
DR21(OH) & [5;25] & $0.70 \pm 0.3$ & $20.3 \pm 2.4$ & $14.6 \pm 1.5$ & $1.5 \pm
0.09$ & $19.9\pm 1.0$\\
W3-IRS5 & [-25;-15] & $0.71 \pm 0.36$ & $4.7 \pm 1.9$ & $3.6 \pm 1.7$ & $0.56
\pm 0.09$ & $23.1 \pm 1.6$ \\
        & [-15;15]& $1.1 \pm 0.4$ & $18.2 \pm 3$ & $21.2 \pm 3$ & $0.11 \pm
        0.09$ & $41 \pm 3.0$  \\
\hline
\end{tabular}

\end{center}
$^a$  The optical depths are determined from the ON and OFF difference spectra with assumption of uniform foreground gas; see text.
\end{table*}

We have used the HI Galactic Plane survey
\citep{mcclure,kalberla:05,kalberla:10} and new JVLA data \citep{winkel} 
to obtain the gas kinetic temperature and have 
derived the volume densities ($n = n(\ion{H}{I})+n({\rm H_2})$) 
from the C$^+$ excitation temperature using the relationship
\begin{equation}
n =\frac{\beta A_{ul}}{\gamma_{ul}} \frac{e^{T_0/T_{kin}}} {e^{T_0/T_{ex}} - e^{T_0/T_{kin}}} ~.
\end{equation}
In this formula, $T_0 = \frac{h\nu}{k_B}$, 
$\beta$ is the escape probability, which can be estimated
from the line opacity $\tau$ ($\beta = (1-e^{-\tau})/\tau$), 
and $\gamma_{ul}$ is the collisional de-excitation rate coefficient so that
$C_{ul}=n\gamma_{ul}$. 
We neglect the effect of any background radiation field on the excitation as
the CMB and other contributions in the diffuse ISM are likely unimportant at
the frequency of the [\CII] line. The [\CII] excitation is computed using
the collisional cross sections for atomic and molecular hydrogen, helium, 
and for electrons. We have assumed that all carbon is ionized with
a gas-phase carbon abundance of $x_C = 1.4 \times 10^{-4}$
\citep{sofia:04}. This leads to an electron abundance  $x_e = 3 \times
  10^{-4}$ taking into account all sources of electrons \citep{draine} .
We used the collisional cross sections summarized  by
\citet{draine}. 
The results are displayed in Table \ref{tab:phases}. They apply to
the absorption signals, for which the atomic and molecular column densities
have been derived \citep{godard:12}.  Atomic and
molecular hydrogen have only slightly different [\CII] collision rate coefficients \citep{wiesenfeld:14}, so the molecular H$_2$ fraction is not very important for determining the excitation of the [\CII] fine structure line. 

The median density derived  from the [\CII] data is 
 $n \sim 60$~cm$^{-3}$ and the mean value, ignoring the W3-IR5 sight-line
is  $n = 56 \pm 15$~cm$^{-3}$.
The thermal pressure ranges from  $p/k_B = 3.8 \times 10^3$ \Kpccm \ to 
$2.2 \times 10^4$ \Kpccm \ with a median value of $5.9 \times 10^3$ \Kpccm. 
In addition to the gas density, Table \ref{tab:phases} also lists the
 ionized carbon column density N$_{tot}({\rm C^+})$ and the kinetic temperature
 derived from the \ion{H}{I} data.

{\small {\bf
\begin{table*}
\begin{center}
\caption{\label{tab:phases} Physical conditions derived from the [\CII] spectra}
\begin{tabular*}{17cm}{lcccccccccc}
\hline
\hline
Source & $V_{LSR}$ &  Width & $\int  \tau\delta v$  & $N_{tot}(\rm{C^+})$$^a$
 & $T_{kin}$$^b$ & $f({\rm H_2})$$^c$& $n$$^d$& Length$^e$  & $L$ & Filling \\
	&	 &         &                       & (10$^{18}$  
 &             &    &    & $s$    &  & factor$^f$ \\
	&(\kms)   & (\kms)   & (\kms)             & \pscm) 
&  (\K) & & (cm$^{-3}$)&  (pc)& (\kpc) & (\%)\\
\hline
W28A & [15;25]  & $4.9 \pm 0.7$ & $3.7 \pm 0.5$ & $0.58 \pm 0.1$ 
 & 100 & 0.51  & $67 \pm 10$ & 24 & 1.1 & 2\\   
W31C & [10,23] & $10.0 \pm 2.4$ & $13.5 \pm 3.0$ & $1.9 \pm 0.3$ 
 & 105 & 0.68 & $39 \pm 5$  & 85 &  1.5 & 6 \\
   & [23,34]  & $7.9 \pm 1.8$ & $12.9 \pm 5$ & $1.8 \pm 0.4$ & 
 100 & 0.74 & $38 \pm 5$  & 116&  0.9 & 13 \\
   &[34,61]  &  $11.8 \pm 2.5$ & $17.7 \pm 3.7$ & $2.5 \pm 0.5$ & 
    90 & 0.45 & $45 \pm 5$  & 107 & 1.3 & 8 \\
G34.3+0.15 & [8,20] &  $7.2 \pm 1.5$ & $5.0 \pm 0.8$ & $0.70 \pm 0.1$ &
 95 & 0.52 & $63 \pm 10$  &   18 &  0.85 & 2 \\
	   & [20,35]  & $7.5 \pm 1.0$  & $6.6 \pm 0.7$ &  $0.92 \pm 0.1$ &
           85 & 0.34 & $68 \pm 10$   & 30& 1.0 & 3 \\ 
W49N & [25,50]  & $12.8 \pm 0.8$ & $20.0 \pm 0.9$ &  $2.8\pm 0.5$ &   
 90 & 0.51 & $80 \pm 10$   & 43 & 1.8 & 2\\
     & [50,80]  & $19.0 \pm 1.0$ & $27.0 \pm 1.5$ & $3.8 \pm 0.3$ 
 & 100 & 0.59 & $38 \pm 5$ &  134 & 2.6 & 5 \\
W51	& [1,10] & $4.5 \pm 1$ & $2.8 \pm 0.3$ & $0.39 \pm 0.05$ 
 & 90 & 0.55 & $58 \pm 5$   & 13 & 0.7 & 2\\
	& [10,16] & $5.0 \pm 1.0$  & $1.4 \pm 0.2$ &  $0.2 \pm
                0.05$   & 150 & 0.19 & $56 \pm 10$   & 6 & 0.5 & 1 \\	
	& [16,35] & $10.6 \pm 1$ & $3.8 \pm 0.5$ & $0.53 \pm 0.07$ &
     130    &  0.06  & $70 \pm 5$  & 11 & 1.5 & 1 \\
DR21OH & [0,20] & $6.1 \pm 1.0$ & $12.9 \pm 1.5$ & $1.8 \pm 0.3$ & 
 95 & 0.61 & $36 \pm 5$   & 81 & 0.9 & 9 \\
W3-IRS5 & [-25;-15]  &  $1.3 \pm 1.$ & $0.78 \pm 0.5 $ & $0.11 \pm 0.06$
 & 40 & 0.52 & $560 \pm 50$ & 0.7 & 0.9 & 0.07\\
\hline
\end{tabular*}

$^a$ $N_{tot}({\rm C^+})$ is the C$^+$ column density in the diffuse gas along the line of sight to the background sources, computed for the given LSR velocity interval.
$^b$  HI data from the Galactic Plane survey
http://www.astro.uni-bonn.de/hisurvey/profile/index.php , and from
\citet{winkel}. 
$^c$ $f({\rm H_2}) = \frac{2n({\rm H_2})}{n(\ion{H}{I}) +2n({\rm H_2})} $ is the fraction of hydrogen in molecular form.
$^d$ $n = n({\ion{H}{I})} + n({\rm H_2})$ is the gas density.
$^e$ $s = \frac{N_{tot}({\rm H})}{n}(1-f({\rm H_2})/2)$ is the total size of the CNM absorption features in the considered velocity interval.
$^f$ $ff = s /L$ is the filling factor of the CNM absorption features in the considered velocity interval.
\end{center}
\end{table*}
}}

\subsection{Physical conditions from the [\CI] data}
To analyze the \ion{C}{i} observations, we used the offline version of RADEX,
a radiative transfer code based on the escape probability approximation
\citep{vdtak:07}. For each velocity component,
 we compared the observed line intensities to  predictions obtained
from a grid of RADEX models covering a pre-defined set of input parameters. 
The observed quantities include, for each velocity component, the integrated
intensity of the \ion{C}{i} ($^3P_1 - ^3P_0$) line, as well as that of the
 ($^3P_2 - ^3P_1$)
transition, whenever observed. In the most favorable cases, we thus could use
both the ($^3P_1-^3P_0$) integrated intensity value and the ($^3P_1-^3P_0$) to 
($^3P_2 - ^3P_1$) integrated
intensity ratio. In cases where the $^3P_2-^3P_1$ line was not detected, we  used the $^3P_1-^3P_0$ integrated intensity and a lower limit to the integrated intensity ratio ($^3P_1 - ^3P_0$) to ($^3P_2 - ^3P_1$). In cases where the 
($^3P_2 - ^3P_1$)) line was not
observed, we used only the ($^3P_1 - ^3P_0$) integrated intensity. We also determined
the line width, $\Delta \varv$, and the peak  temperature for each
velocity component, from Gaussian fits  over the observed
spectral range.

The radiative transfer calculations were performed using collisional
rates with ortho and para H$_2$, H, H$^+$, He and electrons \citep{Schroeder91,Launay77,Roueff90,Staemmler91,Johnson87}.  For the model grid, the equivalent
temperature of the background radiation field was set to the microwave
background field, $T_{\rm CMB} = 2.73$~K. The grid covers a range of gas densities
$n$  from   10 to 1000~cm$^{-3}$, a range 
of kinetic temperatures $T_{kin}$ from  40 to 200~\K \ with  eight values
  for  $n$ and six for $T_{kin}$, corresponding to the
parameters of the diffuse gas. We also sampled the effect of the
composition of the gas by computing models for three values of the fraction
of hydrogen in molecular form $f({\rm H_2}) =  {2n}({\rm H_2}) / (2n({\rm H_2}
+ n(\ion{H}{I})) = 0.1, 0.4, 0.9$, and of the total column density of carbon  using  eight values of the  C column density: $0.5, 0.75,1.0, 3.0, 6.0, 9.0,
  20., 50. \times  10^{16}$~cm$^{-2}$.  As for the [\CII] excitation
  calculations, we assumed that C$^+$
is the main reservoir of carbon in these clouds, hence the
electron density is derived from the total density using the same
 carbon abundance relative to hydrogen as above : $n({\rm e}) =
1.4\times10^{-4} n({\rm \ion{H}{I}})+ 2.8\times10^{-4} n({\rm
H_2})$). The helium density is set to 10\% of the hydrogen density
 and the H$^+$ density is set to its
minimum value $2 \times 10^{-3}$ \pccm.  We have checked that increasing
  the electron density to $n({\rm e}) = 3 \times 10^{-4} (n({\rm \ion{H}{I}})+ 2 n({\rm H_2}))$ \pccm \ and the proton density to $n({\rm H^+}) = n({\rm e}) - n({\rm C^+})$\pccm \ does not change the model predictions for neutral carbon fine structure lines. 
For each velocity component, the observed quantities were
compared to the appropriate model results. 
The fits are reported in Table \ref{tab:ci}  and illustrated in Fig. \ref{fig:histos}.They make use of the kinetic
temperature derived from the HI data.
Since collisional excitation rates with atomic hydrogen are larger than 
those with molecular hydrogen, the derived physical parameters show
a clear dependence on the fraction of gas in molecular form. We have used the
available information on the fraction of hydrogen in molecular
form to select the best fit models. For about half of the 
velocity components, the fits are not well constrained and a broad
interval of gas pressures is consistent with the data, while the range
of best fit gas pressures is relatively narrow for a second half of the models.
 The pressure distribution is bimodal with about two thirds 
of the velocity components with pressures
values ranging between $10^3$ and $1.5 \times 10^4$  \K\pccm, and the remaining  
   components with significantly larger pressures.  As the excitation analysis of the
carbon lines does not return a single pressure value but rather an interval of
acceptable values, we have derived the range of pressures using the median
values of the minimum and maximum pressure values for each velocity component.
 We derive a pressure range of $2200 - 7800$  \K\pccm, which is consistent
 with the median pressure value derived from the analysis of the [\CII]
 data.

\begin{table*}
\begin{center}
\caption{\label{tab:ci}Physical conditions from the [\ion{C}{i}] data}

\begin{tabular*}{17.5cm}{lccccccc}
\hline
\hline
Source	& V$_{LSR}$\tablefootmark{(1)}	& Width\tablefootmark{(1)}
& $\int T d\varv$ (1--0)\tablefootmark{(2)}	& $\int T d\varv$ (2--1)\tablefootmark{(2)}	& N(C)
& $n$	 	& $p/k_B$ \\
		& (\kms)	&	(\kms)	& (\Kkms)	& (\Kkms)	& (10$^{16}$\pscm)	& (\pccm)		& (10$^3$ \K\pccm )	\\ 
\hline
W28A		& 20		& 4.0		& 4.40$\pm$1.20	& $< 3.30$\tablefootmark{(3)}	
& $15\pm 5$	& 20 - 200 &  0.7 - 11 \\
		& 24		& 1.5		& 0.60$\pm$0.30	&
                $<0.78$\tablefootmark{(3)}	 & 1.8$\pm$1.3	&   40 - 1000
                & 1.0 - 100\\ 
W31C		& 16.5	& 5.0	& 0.87$\pm$0.09	& $<0.42$\tablefootmark{(4)}	& 2$\pm$1	
	& 40 - 400	&  2.2 - 34		\\
                & 17.5	& 3.5		& 0.84$\pm$0.08	& 0.16$\pm$0.02 &
                $3\pm 0.5$	& 40 - 70		& 2.2 - 3.4		\\
                & 23		& 3.5		& 1.33$\pm$0.13	&
                0.41$\pm$0.04 & $2\pm0.5$ & 200 - 400	&  11 - 20	\\
                & 24.5	& 1.5		& 0.53$\pm$0.05	& 0.24$\pm$0.02 &
                $0.75\pm 0.5$	& 400 - 1000			& 22 - 29		\\
                & 27.5	& 2.5		& 2.70$\pm$0.27	& 0.63$\pm$0.06  &
                $6\pm2$ & 70 - 200		& 4.4 - 7.2		\\
	        & 30		& 2.5	& 2.30$\pm$0.23	& 0.59$\pm$0.06    &
                $4.5\pm1$	& 70 - 130	&  5.5 - 9.4 \\
               & 34		& 5.0		& 2.00$\pm$0.20	& 0.43$\pm$0.04
& $10\pm2$				& 40 - 70 &  2.2 - 3.5 \\
                & 37.5	& 4.0		& 1.59$\pm$0.16	&
                $<0.48$\tablefootmark{(4)}	& 10$\pm$2
                & 20 - 40		& 0.9 - 1.5		\\
            	& 41		& 2.5	& 0.83$\pm$0.08	& $<0.12$\tablefootmark{(4)}	& 3$\pm$1
                &       40 - 70  	&         1.7 - 2.2		\\
                & 45		& 1.5	& 0.24$\pm$0.02	&
                $<0.13$\tablefootmark{(4)}	& 0.75$\pm$0.25	& 40 -
                200&  1.7 - 8.4		\\		
W33A		& 24		& 2.5		& 1.35$\pm$0.28	& --	& $3\pm1$ & 
40 - 200		& 2.7 - 7.8		\\			
G34.3		& 27		& 1.5		& 0.56$\pm$0.06	&  0.51$\pm$0.05	& 0.75$\pm$0.25
& 1000	& 90 - 100		\\
               & 47.5	& 1.5		& 0.44$\pm$0.04	& $<1.67$\tablefootmark{(3)}	&               0.75$\pm$0.25	
& 130 - 200	& 9 - 13 		\\
W49N		& 51	& 1.5	& 0.07$\pm$0.02	&
$<0.07$\tablefootmark{(4)}	& 0.5$\pm$0.2	& 40 - 70	& 1.2 - 2.0		\\	
		& 54.5	& 3.5	& 0.35$\pm$0.04	& 0.06$\pm$0.01		&
                $1\pm0.5$		& 70 - 200		& 2.9 - 4.4		\\
		& 59		& 2.5		& 0.51$\pm$0.05	&
                0.80$\pm$0.03			& 2$\pm$1	& 40 - 200
                & 1.5 - 7.5	\\
		& 63		& 2.5		& 1.40$\pm$0.14	&
                0.53$\pm$0.05		& 2$\pm$1        & 200 - 1000
                & 12 - 26 	\\
                & 40\tablefootmark{(5)} &         &               &                & $30 \pm
                5$ & 100 - 600 & 3.0 - 4.5 \\ 
DR21(OH)	& 9		& 2.5		& 3.40$\pm$0.34	& --
& $15\pm5$	& 20 - 130		& 0.65 - 4.5	\\
		& 13.5	& 3.5		& 0.57$\pm$0.06	& --	& 2.0$\pm$1
                & 40 - 400	& 1.3 - 11		\\
W3-IRS5		& -21	& 1.0		& 0.11$\pm$0.01	& --	& $0.5\pm0.3$
&  40 - 70			& 1.3 - 2.0		\\	
\hline
\end{tabular*}
\tablefoottext{1}{extracted from the (1--0) spectrum.}
\tablefoottext{2}{The error bars are dominated by the calibration uncertainty which is set to 10\%.}
\tablefoottext{3}{upper limit due to blending and/or line confusion.}
\tablefoottext{4}{upper limit due to non detection.}
\tablefoottext{5}{velocity component detected in absorption in the $^3P_2 -
  ^3P_1$ line (see section \ref{subsec:ci}).}
\end{center}
\end{table*}

\section{Discussion}
\label{sect:discussion}

\subsection{Gas pressure}

\begin{figure}
\rotatebox{0}{
\resizebox{8cm}{!}{
\includegraphics{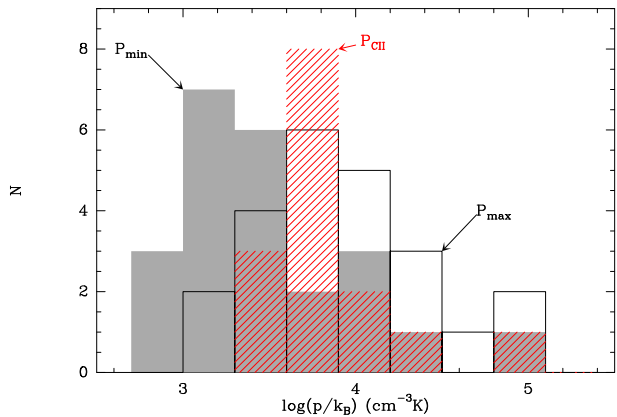} }}

\caption{\label{fig:histos} Histograms of the pressure values derived from the
  analysis of the [\ion{C}{II}] and  [\ion{C}{I}] data. The gray histogram
  represents the minimum values of the gas pressure derived from the
  [\ion{C}{I}] data, the black lines show the maximum values also derived from
  [\ion{C}{I}] and the dashed red histogram shows the pressure values derived
  from [\CII]. 
}
\end{figure}

Figure  \ref{fig:histos} presents histograms of the pressure derived from
the [\ion{C}{II}] and  [\ion{C}{I}] data. We show histograms of the minimum and maximum
pressure compatible with the [\ion{C}{I}] data given the wide range of
acceptable solutions. 

It is encouraging to see that 
the independent analysis of the C and C$^+$ excitation lead to
 similar physical conditions for the diffuse interstellar gas
in the Galactic plane.  The median pressure derived from the velocity components
detected in emission is $5.9 \times 10^3$ K\pccm \ for [\CII] and 
$2.2 - 7.8 \times 10^3$ K\pccm \ for the  [\CI] velocity components.  
There is a slight tendency for $n$ to decrease with increasing $\tau$, as
more opaque lines reach higher excitation temperatures for a given
density. This shows that the emission data may be slightly biased towards the
high pressure regions. In addition, the regions with larger fractions of
molecular gas tend to have slightly higher pressures than do the more atomic
components.  The pressures derived from the neutral carbon emission lines can be biased towards  relatively dense regions because the ionization equilibrium is driven to   higher C/C$^+$ ratio at higher densities. 
Because of the more efficient collisional excitation of atoms and ions, 
higher pressure regions are also 
more likely to produce line emission than low pressure regions. 
The true median pressure may therefore be slightly lower than the median value, and closer
to the lower bound of our pressure determination, $\sim 4.0\times10^3$ K\pccm. 

The median density derived from the [\CII] data is 60 \pccm. 
These figures are very comparable to the 
physical conditions of the diffuse ISM in the solar neighborhood, determined
from UV absorption  measurements \citep{jenkins1,goldsmith:13}. The
average pressure is also consistent with the 
theoretical predictions for a
Galacto-centric radius of 5 \kpc \  \citep{wolfire:03}.

Interestingly these density and pressure conditions are also those of the
low density gas responsible for the strong CH$^+$ absorption features \citep{edith:10,godard:12}. Models of the chemistry triggered by turbulent 
dissipation reproduce the observed CH$^+$ column densities 
in the diffuse medium if the dissipation bursts occur in the CNM  \citep{godard:09,  godard:14}.  The association with the CNM is further supported by the
close similarity of the velocity profiles of the CH$^+$ and [\CII] absorption lines in the diffuse ISM \citep{edith:10a}.

We conclude that the main phase responsible for the foreground absorption
is the so-called cold neutral medium. The diffuse medium in
this phase has a broad range of content of molecular gas, from
nearly pure atomic gas up significant amounts of molecular gas
\citep{godard:12,winkel}. The CNM is also responsible for the diffuse [CII]
emission of the Galactic plane.

\subsection{Filling factor}

The knowledge of the median density and total gas column allows us to
derive the average (1-dimensional) filling factor of this diffuse component along the sampled
sight-lines. The length $s$ and filling factor $ff$ 
of the CNM absorption features are listed in Table \ref{tab:phases}. 
These figures have been derived using the total gas
column N$_{tot}(\rm{H})$, the mean fraction of gas in molecular form $f({\rm H_2}) = 2n({\rm H_2})/[n(\ion{H}{I})+2n({\rm H_2})]$, and the
distance along the line of sight in a given velocity interval $L$ as

\begin{equation}
s = \frac{N_{tot}(\rm{H})}{n(\ion{H}{I})+2n(\rm{H}_2)} = \frac{N_{tot}(\rm{H})}{n}(1-f(\rm{H}_2)/2) ~,
\end{equation}

and
\begin{equation}
ff = \frac{s}{L} ~.
\end{equation}

In these formulas, $s$ represents the total length along the line of sight of
each velocity feature and $L$ is computed from the  Galactic rotation curve $\Theta(R)$
assuming a flat rotation curve with $\Theta_0 = \Theta(R_0)=220 \kms$ and a distance to the
Galactic Center, $R_0$ of 8.5 \kpc.
In the first Galactic quadrant and for $b=0^\circ$, the distance along the line of sight can be written as
\begin{equation}
L = R_0cos(l) \ \ (-/+) \ \sqrt{R^2-(R_0sin(l)^2)} ~,
\end{equation}
 where the $-/+$ signs refer
to the near and far distance branches and $R$ is the distance to the Galactic
center. $R$ is determined from $V_{LSR}$ \ assuming circular motions,
\begin{equation}
V_{LSR} = R_0sin(l)(\Theta(R)/R-\Theta_0/R_0) ~.\end{equation}

For the studied lines of sight, the mean distance gradient produced
by the rotation curve is about 100~pc per \kms. Since  the velocity features
we studied are broader than 1~\kms, they are likely not to be  associated with
a single and coherent  entity but most likely result from the superposition of
several  individual ``CNM cloudlets''. The derived density and temperatures
correspond to the average properties of these ``CNM cloudlets''. For all
velocity intervals, the combined length along the line of sight of the ``CNM
cloudlets'', $s$, is significantly
smaller than the path-length  along the sight--lines to the background
sources $L$. 
The filling factors for the selected velocity intervals range  from 0.07 to 
13 \% with higher filling factors in the largest opacity sight-lines, and a median value of   2.4\%.  These filling factors have been evaluated along paths of $\sim 1 - 2 $ kpc
 length along the 
sight--lines to the background sources.  The highest values may
result from a selection bias as we chose sight-lines known for the presence of
foreground absorption and adapted the velocity intervals to the line profiles.
Note that the approach we followed to derive the
filling factor leads to an upper limit as we assign all the absorbing gas
to the CNM.  
Overall, these figures are 
consistent with the expected low volume filling factor of 
the CNM \citep{heiles:03,kalberla}. These data therefore confirm that although
most of the mass resides in the cold phases, these phases occupy a small
fraction of the volume of the Galaxy. 

The line profiles provide additional information
on the spatial distribution of the CNM : for all sources, the [\CII] line 
profiles show well defined velocity components, which correspond to localized
regions in the Galactic plane. Therefore the diffuse medium is not  
uniformly distributed in space but spatially concentrated, most likely in the
spiral arms as found in emission line surveys (e.g. \citet{pineda} for [\CII]).

\subsection{Gas phase carbon abundance and C$^+$ in the ISM phases}
The C and C$^ +$ column densities listed in Tables \ref{tab:phases} and \ref{tab:ci}
show that C$^+$ dominates over atomic  carbon in the diffuse gas, although a
direct comparison  between the velocity components
present in emission and absorption  is not straightforward. 
Nevertheless it is possible to compare the total C$^+$ and C column
densities along each sight-line. We find N(C)/N(C$^+$) ranging from  1 to 28
\%, with a median value of 6 \%.

\begin{figure*}
\centering
\rotatebox{0}{
\resizebox{8cm}{!}{
\includegraphics{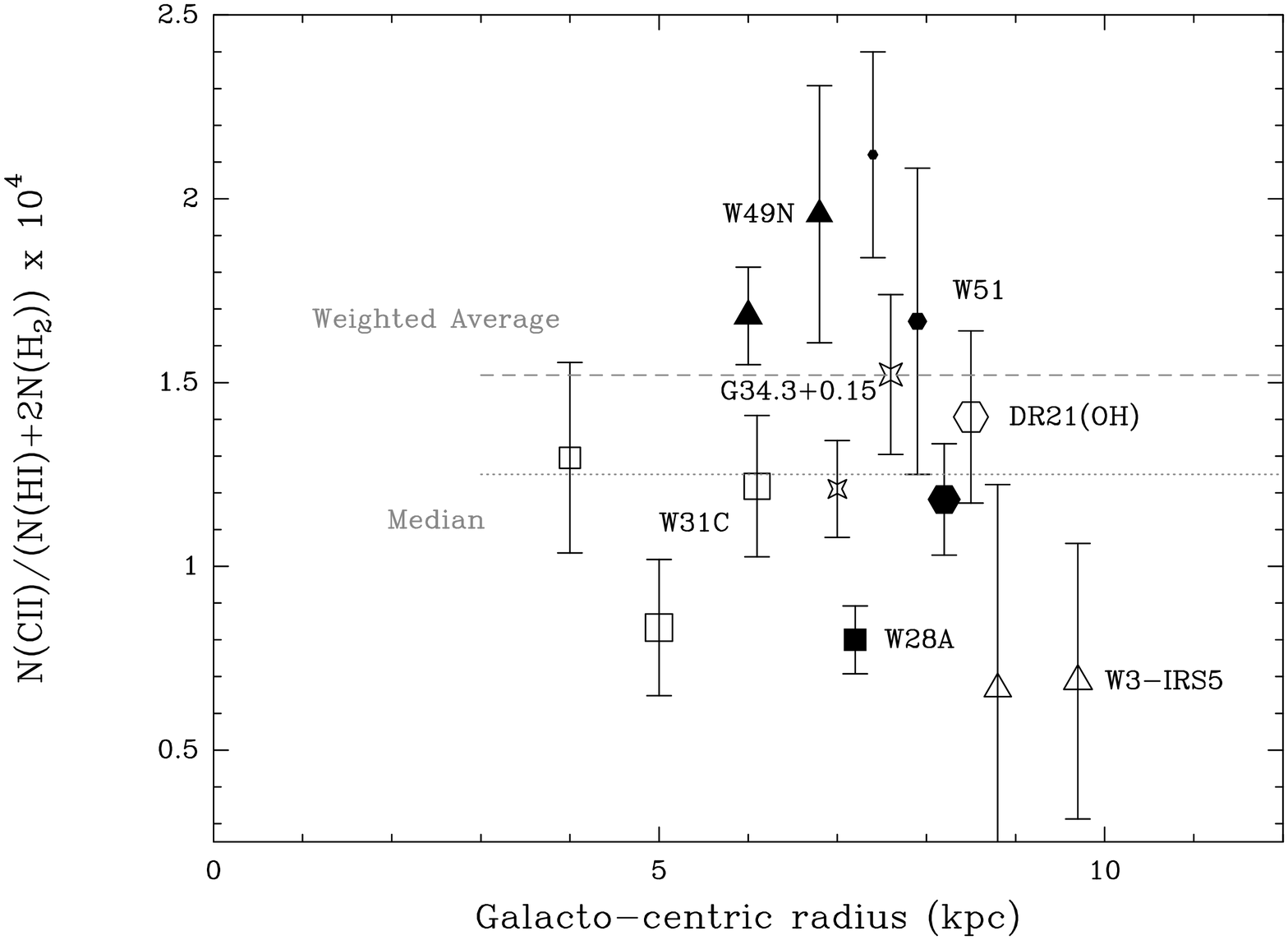} }}
\rotatebox{0}{
\resizebox{8.cm}{!}{
\includegraphics{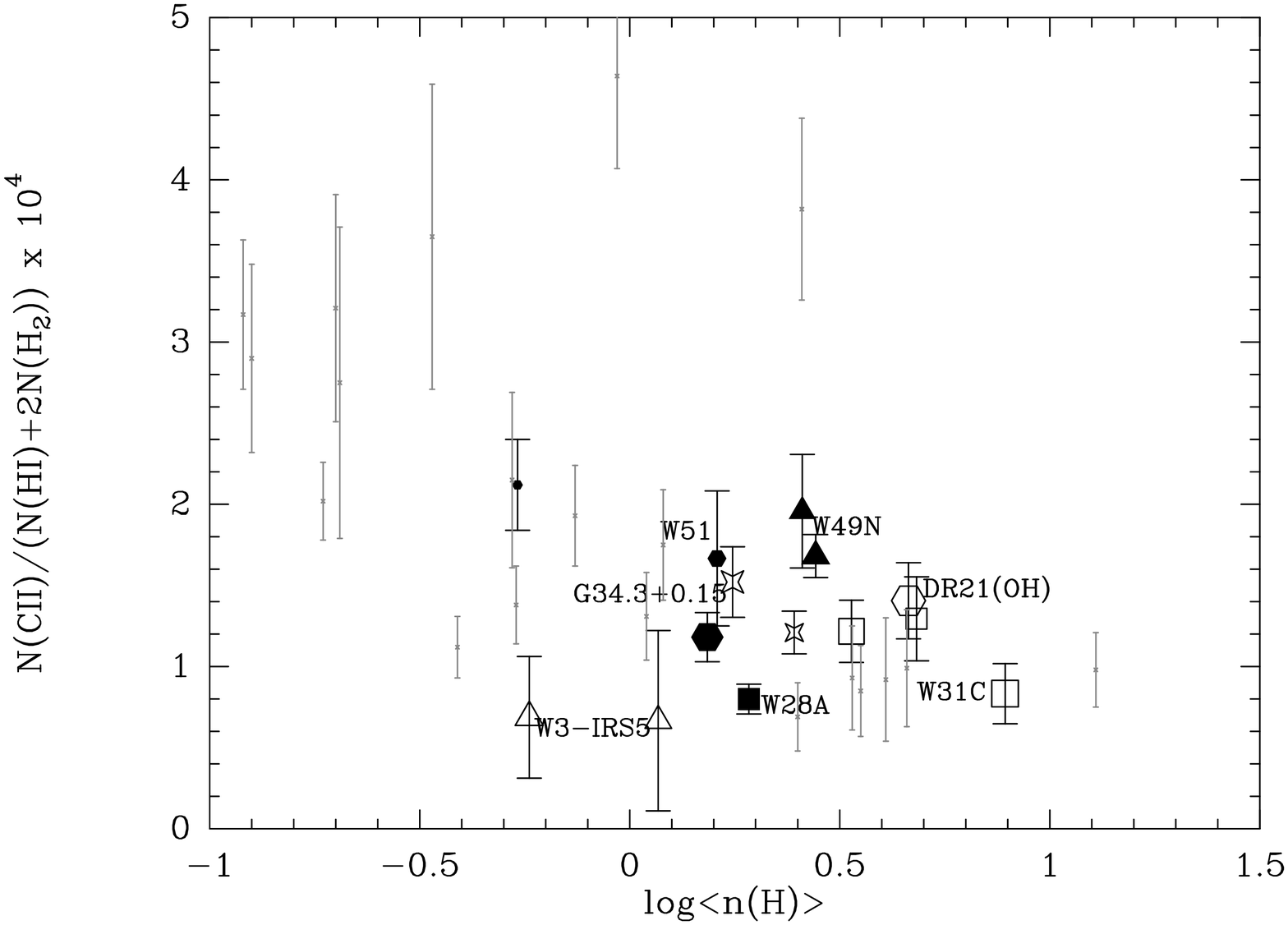} }}

\caption{\label{fig:ciicomp}Left : Ratio of the total C$^ +$ column density to that
  of the total
gas column measured by adding HI (from 21cm data) and H$_2$
  (derived from CH and HF) as a function of Galacto-centric radius. 
The symbol size
   is proportional to the fraction of gas in molecular form ranging from
  $\sim 0.1$ to $\sim 0.8$ for the velocity intervals considered here, with
  each sight-line plotted with a different symbol.
 The weighted average ($1.5 \times 10^{-4}$) is
  indicated with a dashed line and the
  median value ($1.25 \times 10^{-4}$) with a dotted line. Right panel :
  comparison of the {\sl Herschel} data (black symbols with source name) with the FUV absorption data from
  \citet{parvathi:12} (gray dots with error bars). The gas-phase carbon abundance is plotted as a function
  of the logarithm of average gas density $log_{10}(<n({\rm H})>) = log_{10}(N_{tot}({\rm H})/L)$. 
 }
\end{figure*}

Because the gas-phase carbon abundance has been shown to spatially vary as a consequence of
dust grain processing \citep{parvathi:12} we have investigated the gas-phase
carbon abundance by comparing the derived C$^ +$ column densities with 
 independent estimates of the hydrogen column
densities derived from HI for the cold atomic gas and CH and HF for the
diffuse molecular gas. Figure \ref{fig:ciicomp} shows the ratio of the
C$^+$ and total hydrogen column densities as a function of Galacto-centric radius.
In this figure, the size of the symbol scales with the fraction of gas
in molecular form $f(\rm{H}_2)$. We do not find any strong effect with
$f(\rm{H}_2)$. The carbon abundance may be lower towards W3-IRS5 sampling the outer
galaxy but the achieved S/N ratio is too low to draw a definitive
conclusion. The median value is C/H $= 1.25 \times 10^{-4}$ and the
weighted average C/H $ = 1.5 \times 10^{-4}$. The right panel of 
Figure \ref{fig:ciicomp} illustrates how the {\sl Herschel} values are fully
consistent with the measurements derived from FUV absorption lines towards
massive stars by \citet{parvathi:12}, especially comparing  the same range of
average hydrogen densities, $<n({\rm H})> = N_{tot}({\rm H})/L$, along the lines of sight, namely $log_{10}(<n({\rm H})>) = [-0.3,0.9] $.

There are several sources of uncertainties in this comparison, which
contribute to the observed scatter :

\begin{itemize}
\item At large molecular hydrogen column densities, C and CO will make a significant
contribution to the carbon budget. We do not expect this effect to be very
strong because we do not see any significant effect
with $f(\rm{H}_2)$ and the total atomic carbon column density does not rise above
15\% of the total C$^+$ column density except towards W28A. 
\item The warm interstellar phases may be contributing to the [CII] absorption. 
Ionized carbon is expected to be 
present in the Warm Neutral Medium (WNM) and Warm Ionized Medium (WIM). Given
the low densities, the excitation of the fine structure line will be
moderate, leading to an absorption signal towards a strong continuum source. 
Additional information from other fine structure lines, 
[NII] at 205 $\mu$m and [\OI] at 63$\mu$m, is required to identify the
relative contributions of these two warm phases. {\sl Herschel} HIFI data of
the [NII] at 205 $\mu$m line have been analyzed by \citet{persson:14}, and
indeed show absorption towards W31C and W49N confirming the presence of 
 the warm ionized medium along the same sight-lines. The WIM produces an
excess [CII] absorption at the 7 - 10\% level along both sight-lines. 
\item The estimations of the molecular hydrogen column densities are based on
CH and HF measurements assuming a constant abundance relative to H$_2$. Optical and
FUV measurements towards bright stars  have shown that the relationship between CH
and H$_2$ is linear and exhibits a significant scatter of 0.2dex \citep{sheffer:08}, larger than the measurement uncertainties, and which probably reflects small variations in the physical conditions
along the probed sight-lines. As shown by \citet{levrier:12}, 
averaging over long path-lengths tends to reduce the scatter, but local
variations of the  CH abundance relative to H$_2$ cannot be excluded. The same can be true for HF as well. 
\end{itemize}

Overall, the excellent agreement between the gas phase carbon abundance
deduced from FUV and FIR data confirms the validity of the derivation of
atomic and molecular gas column densities and supports
 the association of the [\CII] absorption with CNM gas,
with minor contributions from the WIM.

\subsection{C$^+$ Cooling}
For an optically thin line, the cooling rate is proportional to the upper state population, 
and the cooling rate per C$^+$ ion is given by $\Lambda({\rm C^+}) = h\nu A_{ul} f_u$, 
where $f_u$ is the fraction of the carbon ions in the upper level of the fine structure transition.
The determination of the excitation temperature, $T_{ex}$, of the [\CII] line
allows to compute the C$^+$ cooling of the diffuse medium.
The low values from Table \ref{tab:tex} indicate that only a few percent of the population is in the upper level.
With the excitation temperature known, we can write the energy loss per H nucleon as
\begin{equation}
\Lambda({\rm C^+}) = \frac{ h\nu A_{ul} g_u e^{-\frac{h\nu}{k_BT_{ex}}}}{Q(T_{ex})} \frac{N({\rm C^+})}{N_{tot}({\rm H})}~.
\end{equation}

Taking the mean [\CII] excitation temperature (Section \ref{physcon}),  $T_{ex} = 20.4 \K$,  and assuming that all
carbon is in ionized form so that the ratio of the  C$^+$ to H column
densities is equal to $x_C$ gives
\begin{equation}
\Lambda({\rm C^+}) = 9 \times 10^{-26} \frac{x_C}{1.4 \times 10^{-4}} \ erg s^{-1}H^{-1}~. 
\end{equation}

 Using a carbon abundance of $1.5 \times 10^{-4}$ as derived from the
  present data, we obtain a cooling
  rate per H atom
about three times larger than the mean value derived by \citet{gry:92} for local
lines of sight and five times larger than the cooling rate for low velocity
clouds derived by \citet{lehner:04} from a 43 object sample. The difference
between the FUV and {\sl Herschel} measurements can be understood as arising
from the difference in the environments. The FUV data mostly target gas at
high Galactic latitudes in the solar neighborhood while the far infrared {\sl
  Herschel} data mainly probe the Galactic plane, where the star formation
activity and the heating rate are expected to be larger.
The {\sl Herschel}-derived cooling rate is in very good agreement with the
models for $n \sim  60$ \pccm and Galacto-centric radii of 5 and 8.5~\kpc\
presented by \citet{wolfire:03}, which include an increase of the heating rate
in the inner Galaxy.

\section{Conclusion}
\label{sect:conclusion}

In this paper we have analyzed high-resolution spectra of the fine structure
lines of ionized and neutral carbon, complemented by lower spectral
resolution line maps of C$^+$ and O, towards massive star forming regions in
the Galactic plane. 
\begin{itemize}
\item The high spectral resolution data exhibit complex line profiles from
the background sources combined with absorption produced by the diffuse
interstellar medium along the line of sight. We show that
 at low spectral resolution this combination leads to very low [\CII] and [\OI]
 signals, which can be wrongly interpreted. We suggest that part of the [\CII]
 deficit in local luminous infrared galaxies is due to the presence of
 low density material in the foreground of the compact region producing the
far-infrared emission. This diffuse gas component is expected to absorb part
 of the [\CII] signal  from the compact star forming region close to the
 nucleus. The effect is expected to be even larger for the 63  $\mu$m fine structure line  of oxygen, which has a higher opacity for the same gas column density.
 
\item The {\sl Herschel} data have been combined with ground based HI data to
derive the physical conditions in the diffuse gas. The median pressure is
 $p/k_B = 5900$~\Kpccm \ and the median density $n \sim  60 $~\pccm, for a kinetic
temperature of $\sim 100$~\K.  These properties confirm that the
absorbing gas can be associated with the so-called cold neutral medium (CNM). 
The fraction of hydrogen in molecular form in this gas is highly variable. 
 
\item The excitation conditions derived from the [\CI] emission data are
  consistent with those deduced from [\CII]. The atomic carbon column
  densities remain significantly smaller than the ionized carbon column
  densities, confirming that C$^+$ is the main carbon reservoir along the
lines of sight in most cases.

\item The line profiles show that the CNM is not uniformly distributed but
  follows the Galactic Spiral arms. The comparison of the gas densities and
  column densities has allowed us to determine the 
filling factor of the CNM along the lines of sight in the Galactic Plane.
 The derived median value of the filling factor is 2.4\%.
This value may be slightly biased upward by the source selection criteria.

 \item On average, the gas-phase abundance of carbon derived from the
comparison of the  [\CII] absorption with the sum of the atomic and molecular hydrogen
  column densities, using CH and HF as tracers of H$_2$, 
  is in good agreement with previous measurements using FUV data, leading to
  C/H $\sim 1.5 \times 10^{-4}$. Some sight-lines may present a
  small excess [\CII] absorption  which indicates that two other ISM
phases, the warm neutral medium and the warm ionized medium, may contribute
to the [\CII] absorption. 

\item The [\CII] cooling per hydrogen nucleon in the Galactic plane is 
$\Lambda({\rm C^+}) \sim 9.5 \times 10^{-26}$~ergs$^{-1}$H$^{-1}$, in good agreement
with predictions from \citet{wolfire:03}.

\end{itemize}

\begin{acknowledgements}

The Herschel spacecraft was designed, built, tested, and
launched under a contract to ESA managed by the Herschel/
Planck Project team by an industrial consortium
under the overall responsibility of the prime contractor
Thales Alenia Space (Cannes), and including Astrium
(Friedrichshafen) responsible for the payload module
and for system testing at spacecraft level, Thales Alenia
Space (Turin) responsible for the service module, and
Astrium (Toulouse) responsible for the telescope, with in
excess of a hundred subcontractors.
HIFI has been designed and built by a consortium of institutes
and university departments from across Europe,
Canada and the United States under the leadership of
SRON Netherlands Institute for Space Research, Groningen,
The Netherlands and with major contributions
from Germany, France and the US. Consortium members
are: Canada: CSA, U.Waterloo; France: CESR,
LAB, LERMA, IRAM; Germany: KOSMA, MPIfR,
MPS; Ireland, NUI Maynooth; Italy: ASI, IFSI-INAF,
Osservatorio Astrofisico di Arcetri-INAF; Netherlands:
SRON, TUD; Poland: CAMK, CBK; Spain: Observatorio
Astronmico Nacional (IGN), Centro de Astrobiologa
(CSIC-INTA); Sweden: Chalmers University of Technology
- MC2, RSS \& GARD; Onsala Space Observatory;
Swedish National Space Board, Stockholm University
- Stockholm Observatory; Switzerland: ETH Zurich,
FHNW; USA: Caltech, JPL, NHSC.
HIPE is a joint development by the Herschel Science
Ground Segment Consortium, consisting of ESA, the
NASA Herschel Science Center, and the HIFI, PACS and
SPIRE consortia.

MG, MR, AG and EF acknowledge support from the Centre National de
Recherche Spatiale (CNES). This work was partly funded by grant
ANR-09-BLAN-0231-01 from the French Agence Nationale de la Recherche as part
of the SCHISM project. J.R.G thanks Spanish MINECO for funding support
under grants CDS2009-00038, AYA2009-07304 and AYA2012-32032.
M.R. is supported by the 3DICE project, funded by an ERC Starting grant (grand
agreement number 336474). NRAO is operated by Associated Universities, 
Inc. under contract with  the National Science Foundation. This work was carried out in part at the Jet Propulsion Laboratory, which is operated by the 
California Institute of Technology for NASA.
 We thank the referee, E. Jenkins, for his comprehensive report which
helped us to significantly improve this paper.
\end{acknowledgements}

\bibliographystyle{aa}
\bibliography{cii_7oct}

\appendix
\section{Presentation of the Herschel Observations}
\label{app:obs}

The individual {\sl Herschel} observations analyzed in this paper
are listed in Table A.1.

\begin{table*}
\label{tab:obs}
\caption{Summary of observations}
\begin{tabular}{llll}
\hline
Source & Line & Obs mode & ObsID \\
\hline
\hline
W28A & [\CII] & DBS & 1342243682; 1342243683 \\
     & [\CII] & LC-OTF &  1342243684 \\
     & [\CI] $^3P_1 - ^3P_0$ & DBS & 1342216303; 1342216304; 1342216305\\
     &    [\CI] $^3P_1 - ^3P_0$       & LC & 1342218938\\
& [\CI] $^3P_2 - ^3P_1$ & DBS & 1342242856; 1342242857\\
&  [\CI] $^3P_2 - ^3P_1$     & LC & 1342242858\\
&     [\CII] & SED-R1 & 1342217940\\
&      [\OI] 63 & SED-B3A & 1342217941\\
\hline
W31C & [\CII]& DBS & 1342244100 ;1342244101\\
     & [\CII] & LC-OTF & 1342244102\\
     & [\CI] $^3P_1 - ^3P_0$ & DBS & 1342206657; 1342206658; 1342206659\\
     &  [\CI] $^3P_1 - ^3P_0$                    & LC & 1342251659 \\
     & [\CI] $^3P_2 - ^3P_1$ & DBS & 1342244106; 1342244107; 1342244108\\
     & [\CI] $^3P_2 - ^3P_1$                    & LC & 1342244109\\
&     [\CII] & SED-R1 & 1342217945\\
& [\OI] 63 & SED-B3A &  1342228531\\
\hline
W33A &[\CII] & DBS & 1342229846; 1342229847 \\
     & [\CII] & LC-OTF & 1342229848\\
     & [\CI] $^3P_1 - ^3P_0$ & DBS & 1342208038; 1342208039; 1342208040\\
     &  [\CI] $^3P_1 - ^3P_0$                    & LC & 1342218943\\
&     [\CII] & SED-R1 & 1342239714\\
& [\OI] 63 & SED-B3A & 1342239713\\
\hline
G34.3+0.15 & [\CII] & DBS & 1342244578; 1342244579 \\
           & [\CII] & LC-OTF &  1342244580\\
            & [\CI] $^3P_1 - ^3P_0$     & DBS & 1342219189; 1342219190; 1342219191\\
            &   [\CI] $^3P_1 - ^3P_0$                         & LC & 1342231776\\
            & [\CI]  $^3P_2 - ^3P_1$ & DBS & 1342244110; 1342244111; 1342244112\\
            &  [\CI]  $^3P_2 - ^3P_1$                & LC & 1342244113\\
&     [\CII] & SED-R1 & 1342209734\\
&     [\OI] 63 & SED-B3A & 1342209733\\
\hline
W49N & [\CII] & DBS & 1342231449; 1342231450\\
     & [\CII] & LC-OTF & 1342231451\\
& [\CI] $^3P_1 - ^3P_0$ & DBS & 1342206661; 1342206662; 1342206663\\
&   [\CI] $^3P_1 - ^3P_0$                   & LC & 1342244802\\
 & [\CI]  $^3P_2 - ^3P_1$ & DBS & 1342230253; 1342230254; 1342230255\\
 &    [\CI]  $^3P_2 - ^3P_1$                   & LC & 1342230256\\
&     [\CII] & SED-R1 & 1342207774\\
&     [\OI] 63 & SED-B3A & 1342207775\\
\hline
W51 &[\CII] &  DBS & 1342231446;1342231447 \\
    & [\CII] & LC-OTF &  1342231448\\
& [\CI] $^3P_1 - ^3P_0$     & DBS & 1342219179; 1342219180; 1342219181 \\
&  [\CI] $^3P_1 - ^3P_0$                         & LC & 1342232813\\
 & [\CI]  $^3P_2 - ^3P_1$ & DBS & 1342230243; 1342230244; 1342230245\\
 &  [\CI]  $^3P_2 - ^3P_1$                     & LC & 1342230246\\
&     [\CII] & SED-R1 & 1342193698\\
&     [\OI] 63 & SED-B3A & 134219697\\
\hline
DR21(OH) & [\CII] & DBS & 1342220468; 1342220469\\
         & [\CII] & LC-OTF & 1342220470\\
& [\CI] $^3P_1 - ^3P_0$ & DBS & 1342198320; 1342198321; 1342198322 \\
& [\CI] $^3P_1 - ^3P_0$                     & LC & 1342219171\\
&     [\CII] & SED-R1 & 1342209401\\
&     [\OI] 63 & SED-B3A & 1342209400\\
\hline
W3-IRS5 & [\CII]  & DBS & 1342248401\\
        & [\CII] & LC-OTF & 1342248400\\
        & [\CI] $^3P_1 - ^3P_0$ & LC & 1342247845\\
&     [\CII] & SED-R1 & 1342229092\\
&     [\CII] & SED-B3A & 1342229093\\
\hline
\end{tabular}

DBS stands for Double Beam Switch, LC for Load chop and LC-OTF for On The Fly
map in Load Chop mode. SED-R1 refers to the SED mode for the red channel of
the PACS spectrometer, SED-B3A refers to the blue channel of the PACS spectrometer.
\end{table*}

\section{Pointed data and spectral maps towards the individual sources}
\label{app:spec}
\subsection{W28A}

\begin{figure}[h!]
\resizebox{8cm}{!}{
\includegraphics{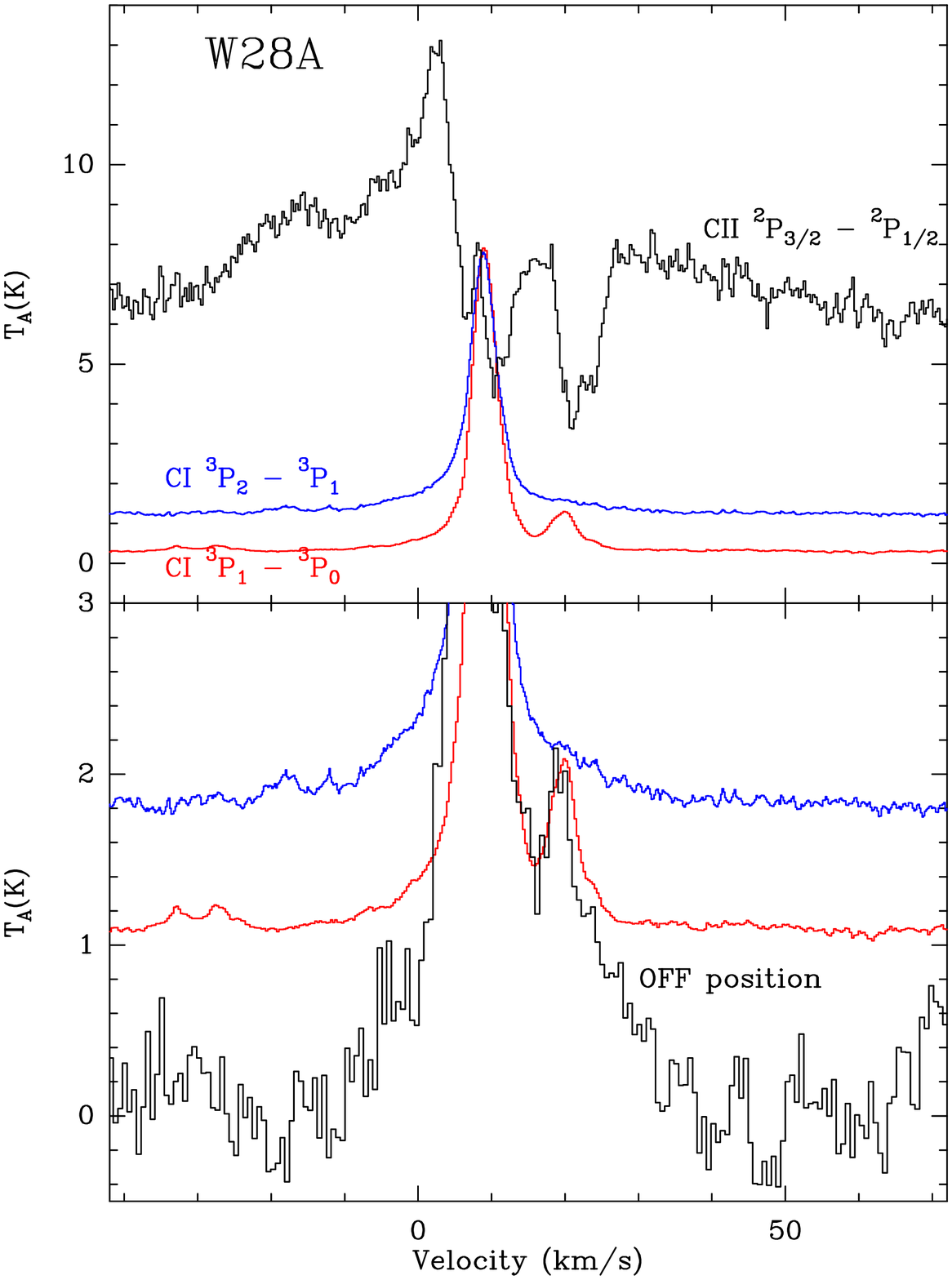}}
\caption{\label{fig:specw28a} Top : Herschel/HIFI spectra towards W28A.
The red line shows the [\CI]$^3P_1 - ^3P_0$ line at 492~GHz, the blue line shows
the [\CI]$^3P_2 - ^3P_1$ line at 809~GHz and the black line the
[\CII]$^2P_{3/2}-^2P_{1/2}$ line at 1.9~THz. The horizontal axis is the LSR
velocity in \kms, and the vertical axis the antenna temperature in
Kelvins. The continuum level for [\CII] corresponds to the SSB continuum level. 
Bottom : zoom on the [\CI] lines (red, blue as above), and average [\CII] 
spectrum of the OFF positions (black). The continuum levels have been shifted
for clarity in the bottom panel.  }

\end{figure}

\begin{figure}[h!]
\resizebox{8.5cm}{!}{
\includegraphics{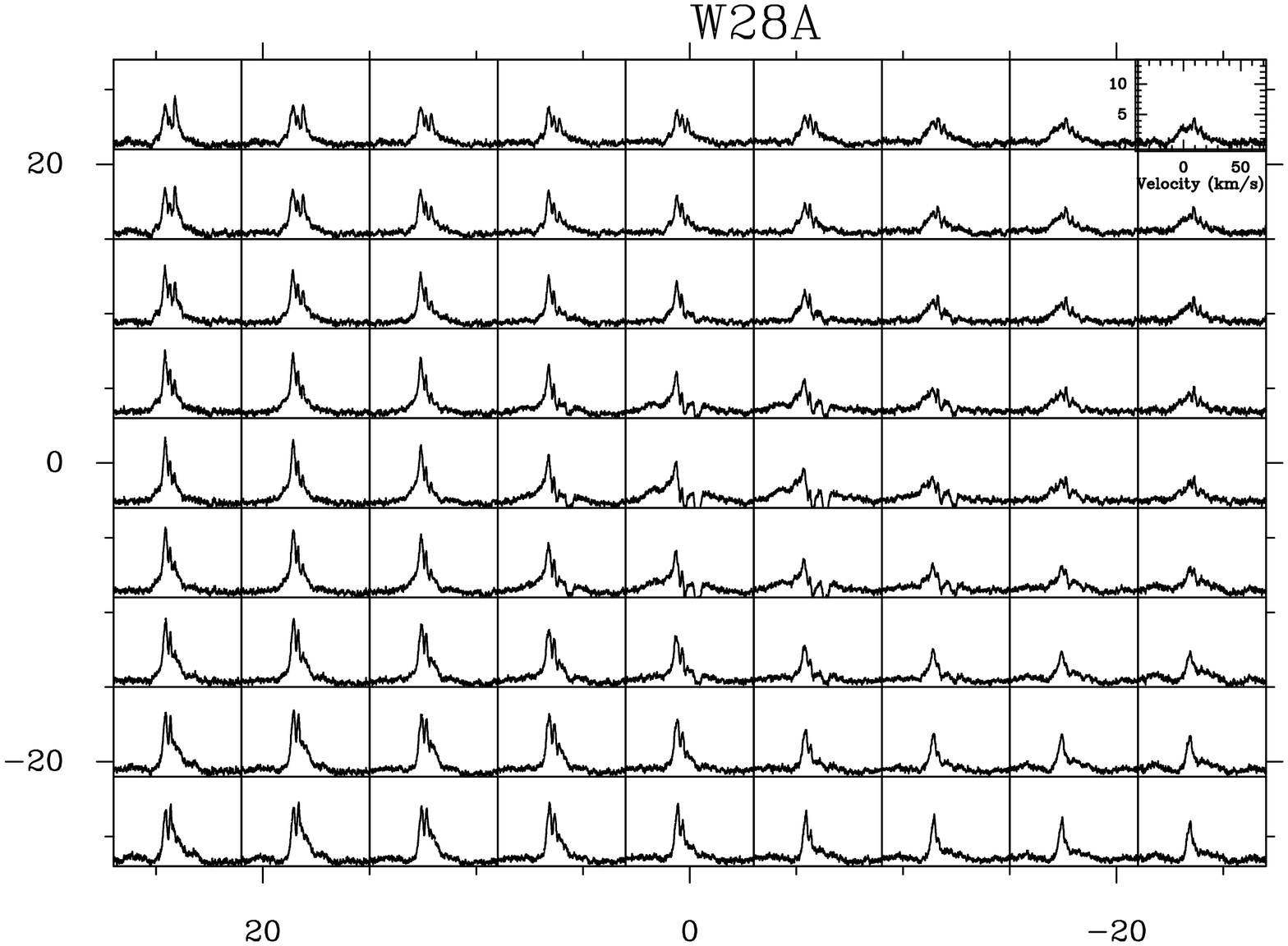}}
\caption{\label{fig:mapw28a} Montage of [\CII] spectra towards W28A. A 
  baseline has been subtracted from all spectra. The horizontal axis is the LSR
  velocity in  \kms, which runs from -42~\kms \ to 72~\kms, and  the  vertical axis is the antenna temperature in
  Kelvins, which runs from -1~\K \ to 14~\K. 
 The x--axis shows the right ascension offset in arc-sec and the
  y--axis the declination offset in arc-sec, relative to the source position given
in Table \ref{tab:sources}}.
\end{figure}

\begin{figure}[h!]
\resizebox{7.cm}{!}{
\includegraphics{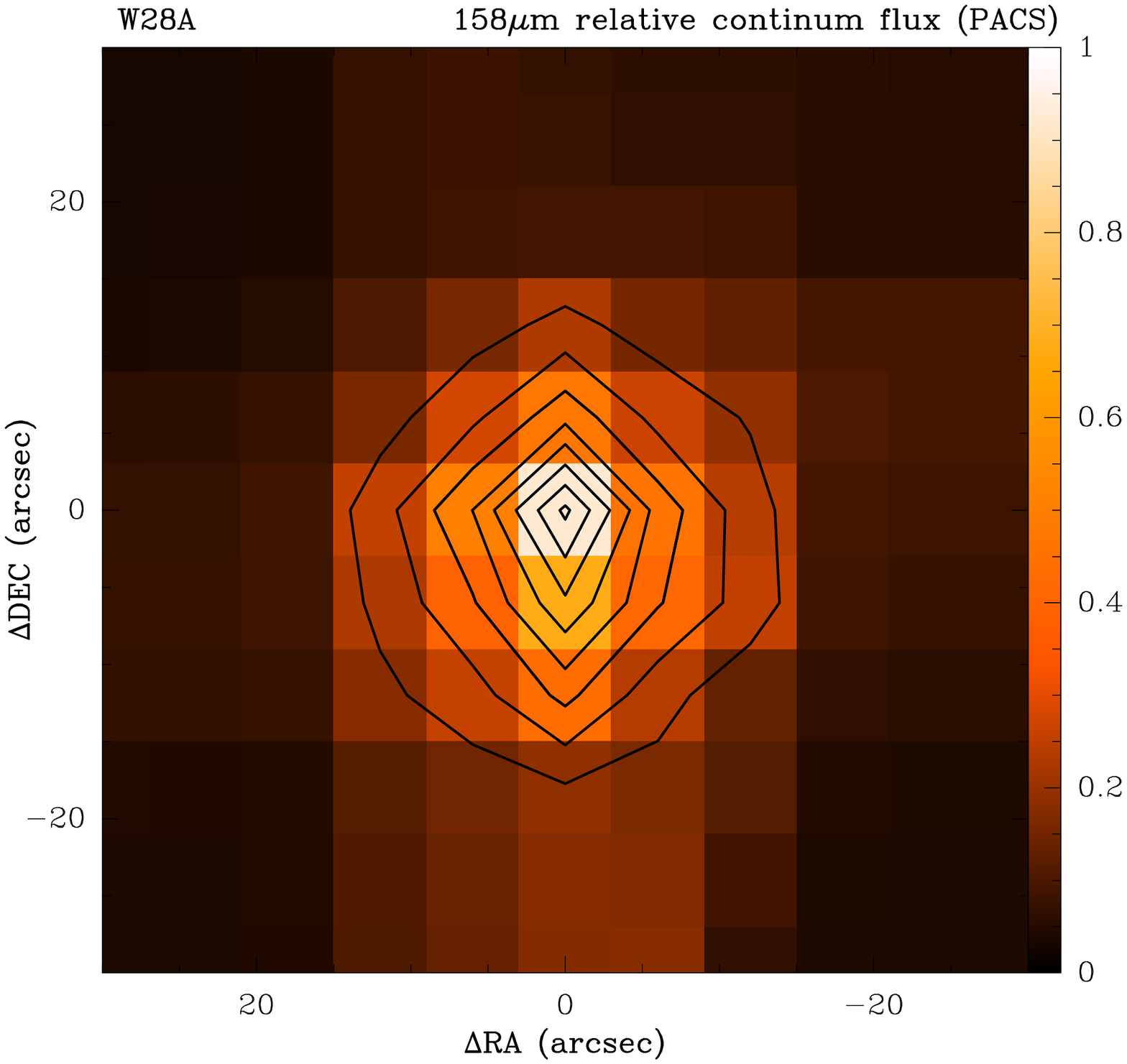}}
\\
\resizebox{7cm}{!}{
\includegraphics{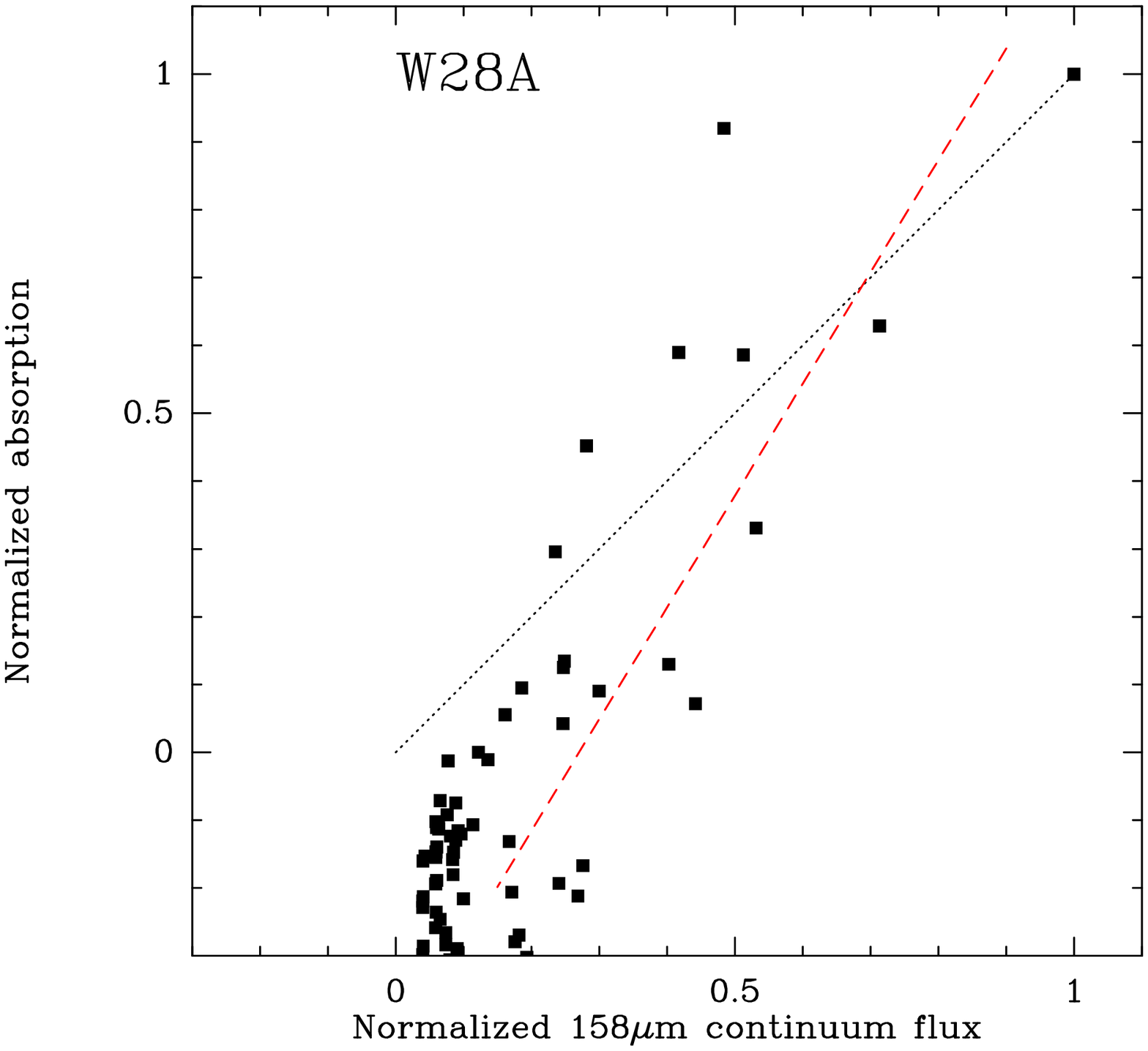}}
\\
\resizebox{7.cm}{!}{
\includegraphics{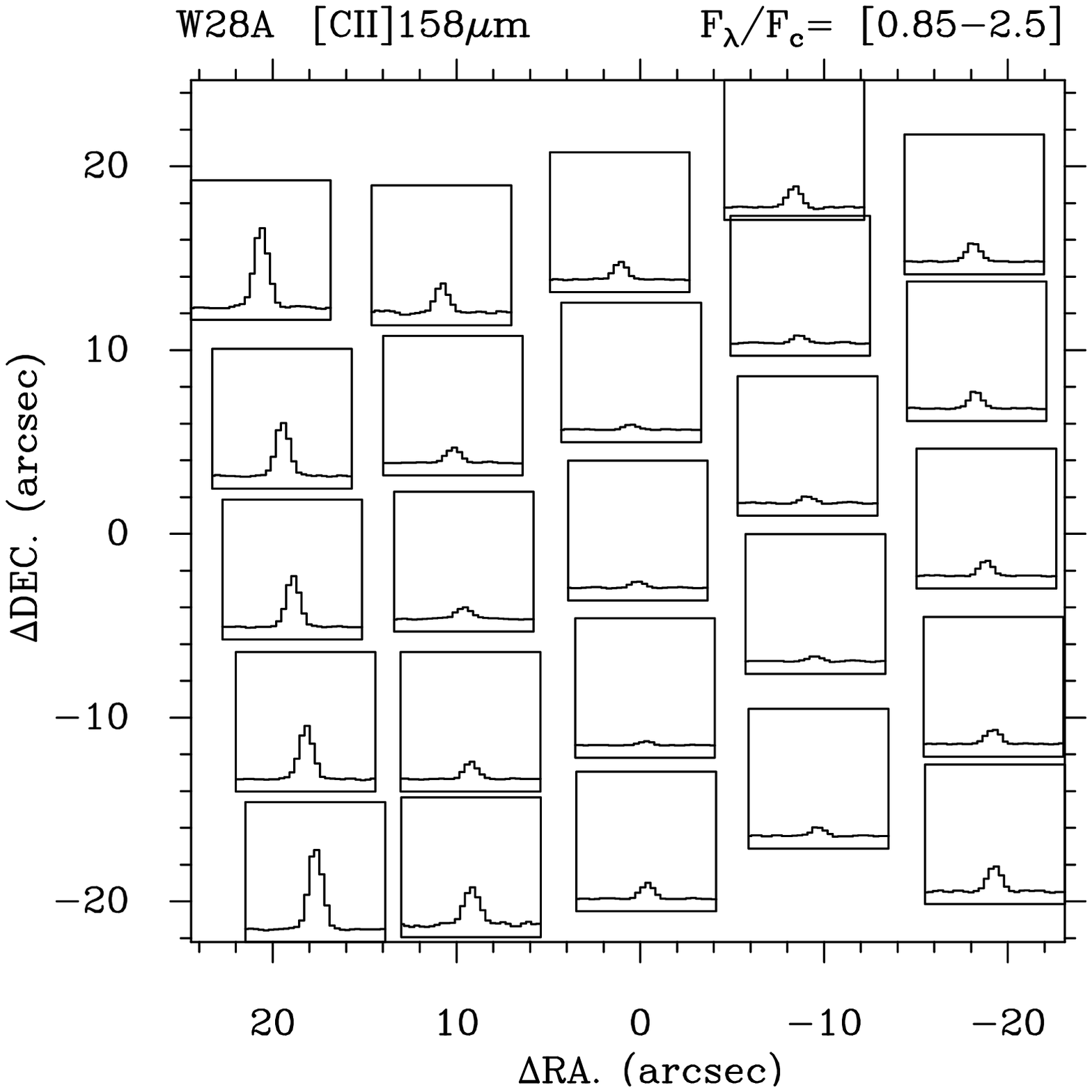}}
\caption{\label{fig:pacs-w28a} PACS data towards W28A at 158~$\mu$m. 
For all maps, the  offsets are given relative to the central position listed in Table  \ref{tab:sources}. Top: Continuum
  emission at 158~$\mu$m. Contour levels are 0.2, 0.3, 0.4 ... 0.9 relative
to the maximum.  Middle: Comparison of the integrated
absorption measured in the HIFI map relative to the absorption at the map
center, with the continuum flux measured in the PACS map relative to the
map center. The dashed red line shows the linear regression line 
and the dotted black line a 1:1 relationship. Bottom: Map of the line 
to continuum emission/absorption in the 25
PACS spaxels. The vertical scale runs from 0.85 to 2.5. The data have not been
corrected for the possible contamination in the OFF position.
}
\end{figure}

W28A (G5.89-0.4) is a massive star formation region at a distance of 1.3 \kpc
\ \citep{motogi}. The HIFI spectra are shown in Figure \ref{fig:specw28a} and
\ref{fig:mapw28a}. W28A hosts a powerful outflow with prominent velocity
wings extending from -40 to 60 \kms \ \cite{gusdorf}. 
The foreground gas is associated with the absorption feature at $\sim 20$~\kms
in the [\CII] spectrum, and with the emission line at the same velocity in the [\CI]
$^3P_1 - ^3P_0$ spectrum. This gas is probably local. 
 The PACS spectral maps data are presented in Figure \ref{fig:pacs-w28a}.

\subsection{W31C}

\begin{figure}[h!]
\resizebox{8cm}{!}{
\includegraphics{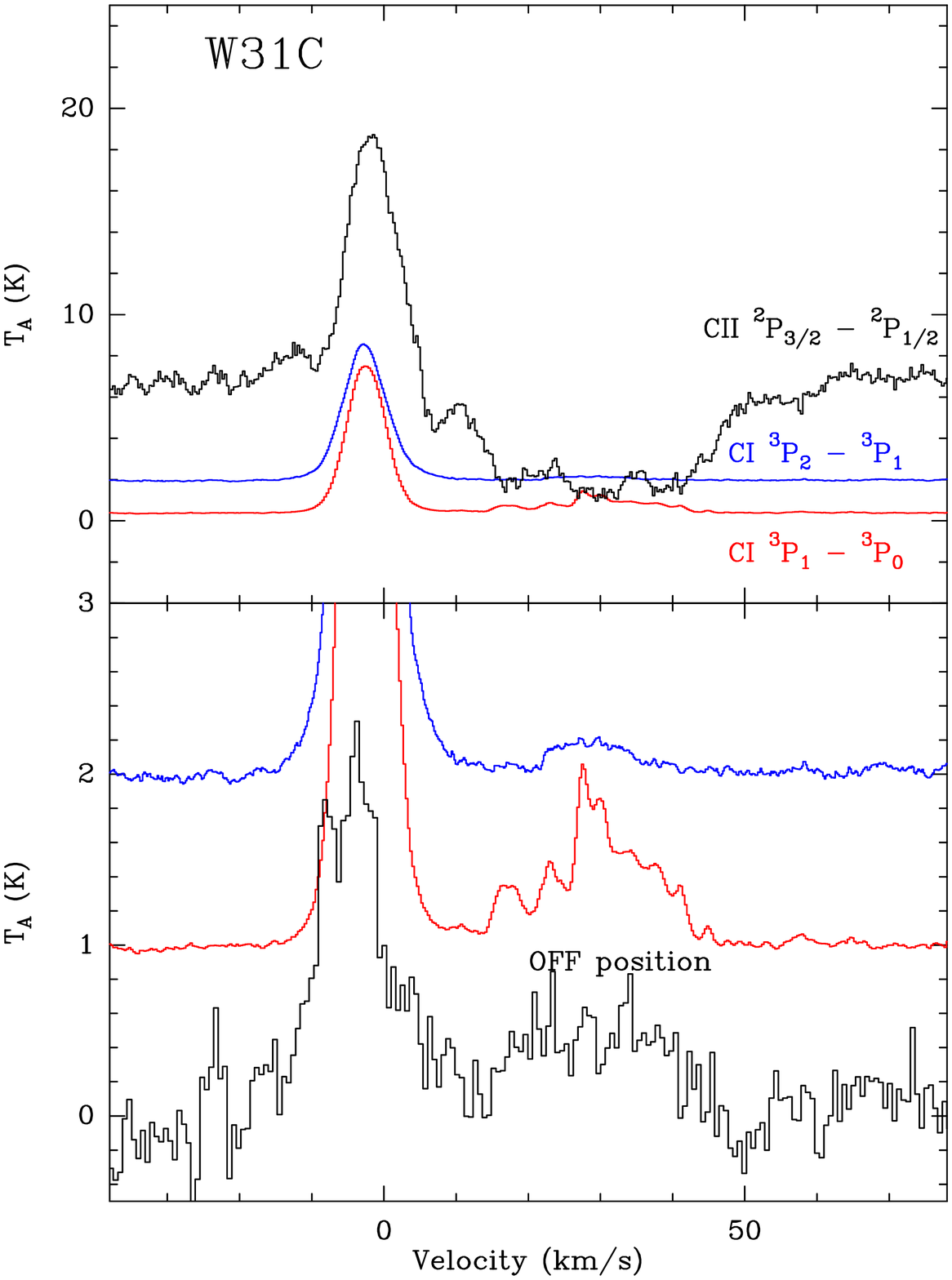}}
\caption{\label{fig:specw31c} Top : Herschel/HIFI spectra towards W31C.
The red line shows the [\CI]$^3P_1 - ^3P_0$ line at 492~GHz, the blue line shows
the [\CI]$^3P_2 - ^3P_1$ line at 809~GHz, and the black line the
[\CII]$^2P_{3/2}-^2P_{1/2}$ line at 1.9~THz. The horizontal axis is the LSR
velocity in \kms \ and the vertical axis the antenna temperature in
Kelvins. The continuum level for [\CII] corresponds to the SSB continuum levels. 
Bottom : zoom on the [\CI] lines (red and blue as above), and average [\CII] 
spectrum of the OFF positions (black). The continuum levels have been shifted
for clarity in the bottom panel.  }
\end{figure}

\begin{figure}
\resizebox{8.5cm}{!}{
\includegraphics{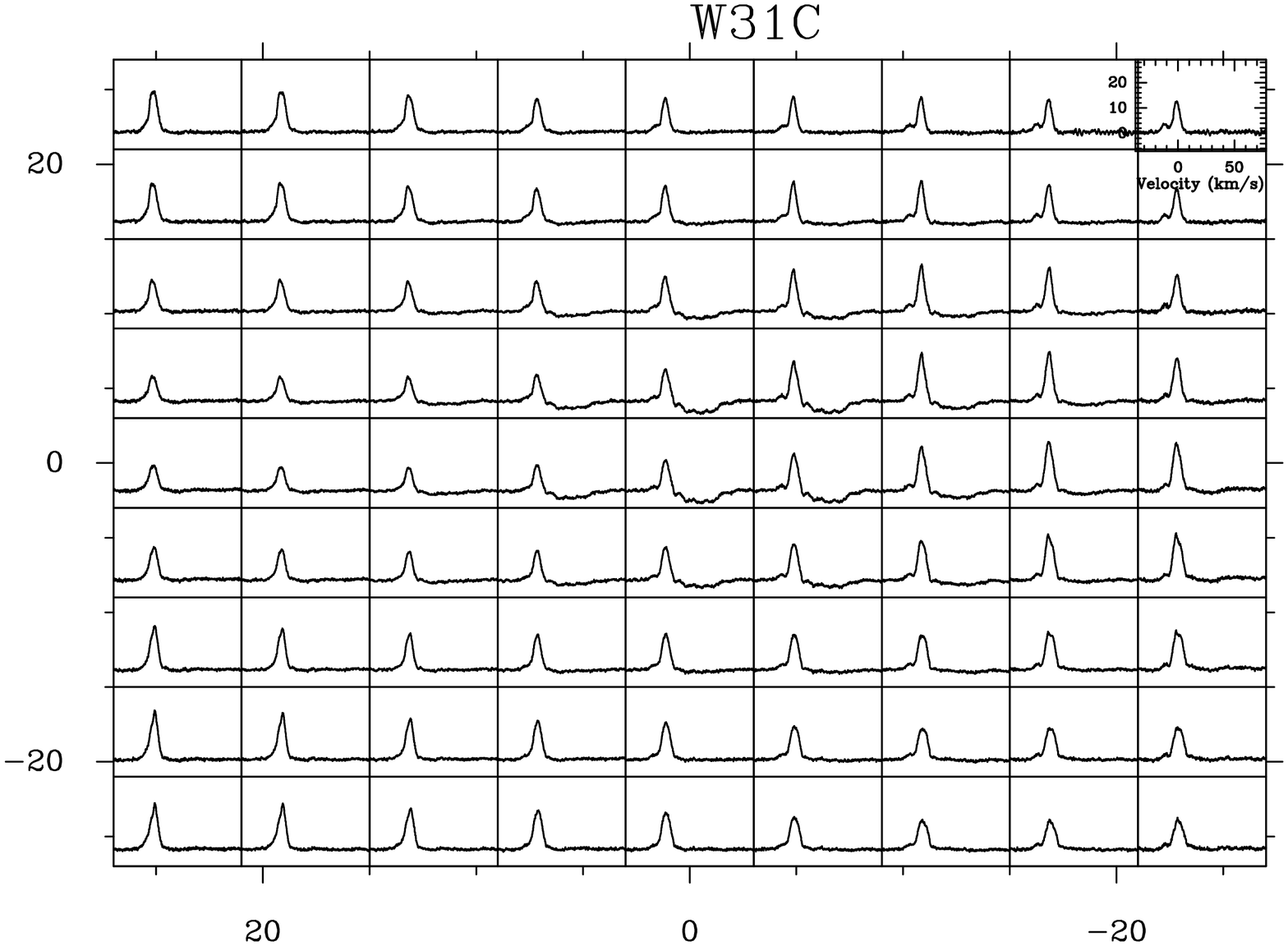}}
\caption{\label{fig:mapw31c} Montage of [\CII] spectra towards W31C. A 
  baseline has been subtracted from all spectra.  The horizontal axis is the LSR
  velocity in  \kms, which runs from -38~\kms \ to 78~\kms,  and   the  vertical axis is the antenna temperature in
  Kelvins, which runs from -7~\K \ to 29~\K.  
 The x--axis shows the right ascension offset in arc-sec and the
  y--axis the declination offset in arc-sec, relative to the source position given
in Table \ref{tab:sources}.
}
\end{figure}

\begin{figure*}[h!]
\resizebox{6cm}{!}{
\includegraphics{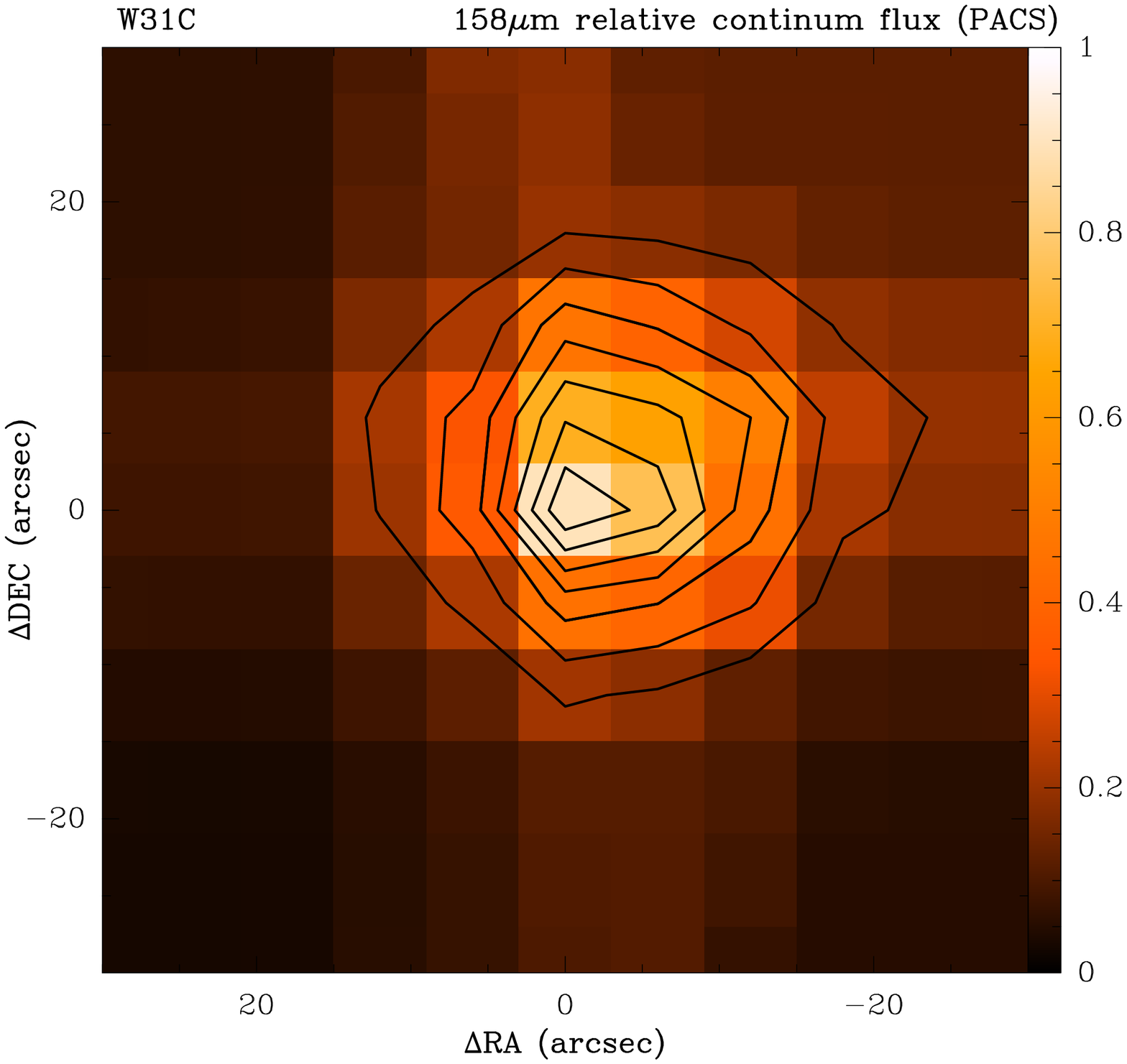}} 
\resizebox{6cm}{!}{
\includegraphics{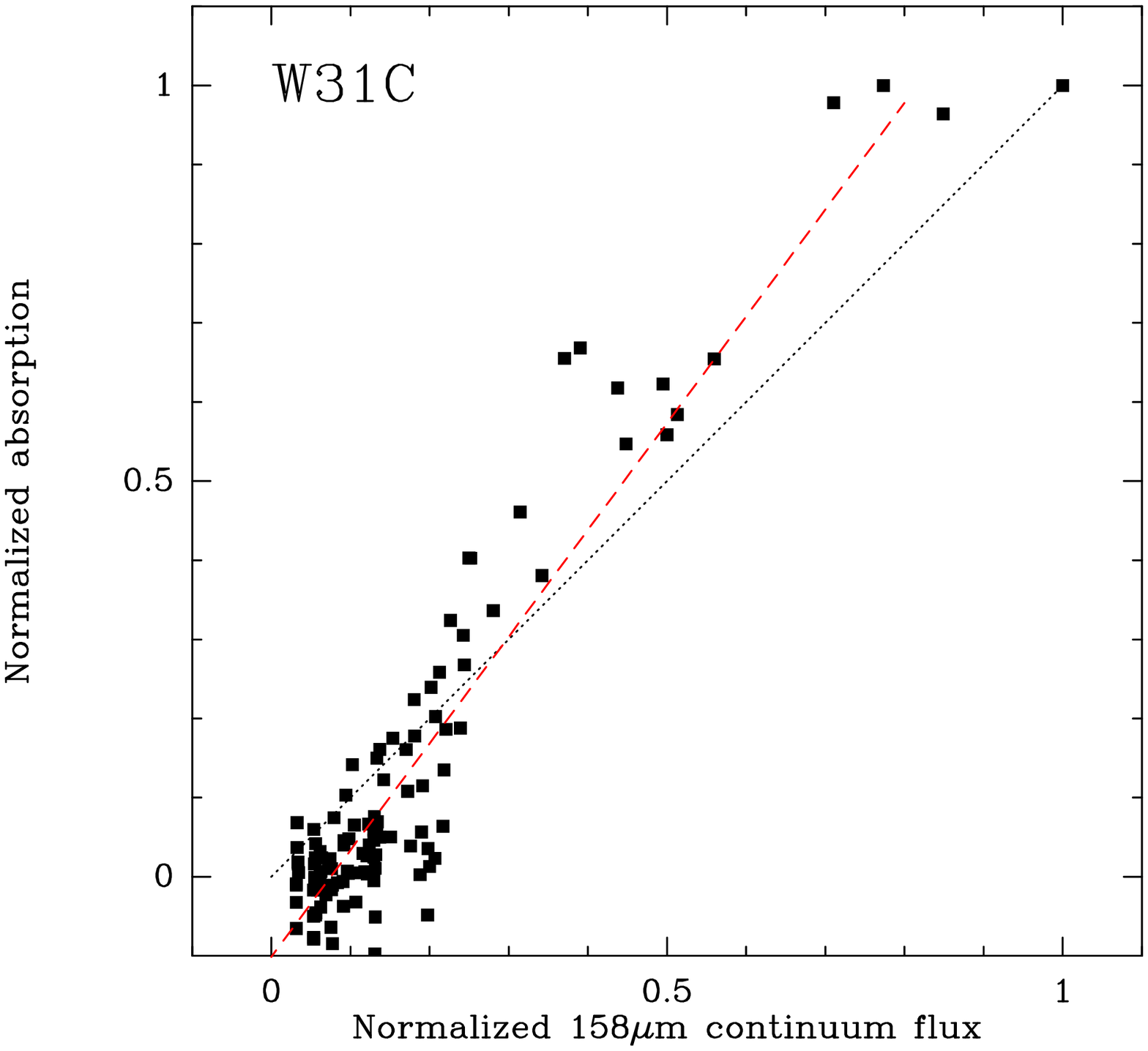}}

\resizebox{6cm}{!}{
\includegraphics{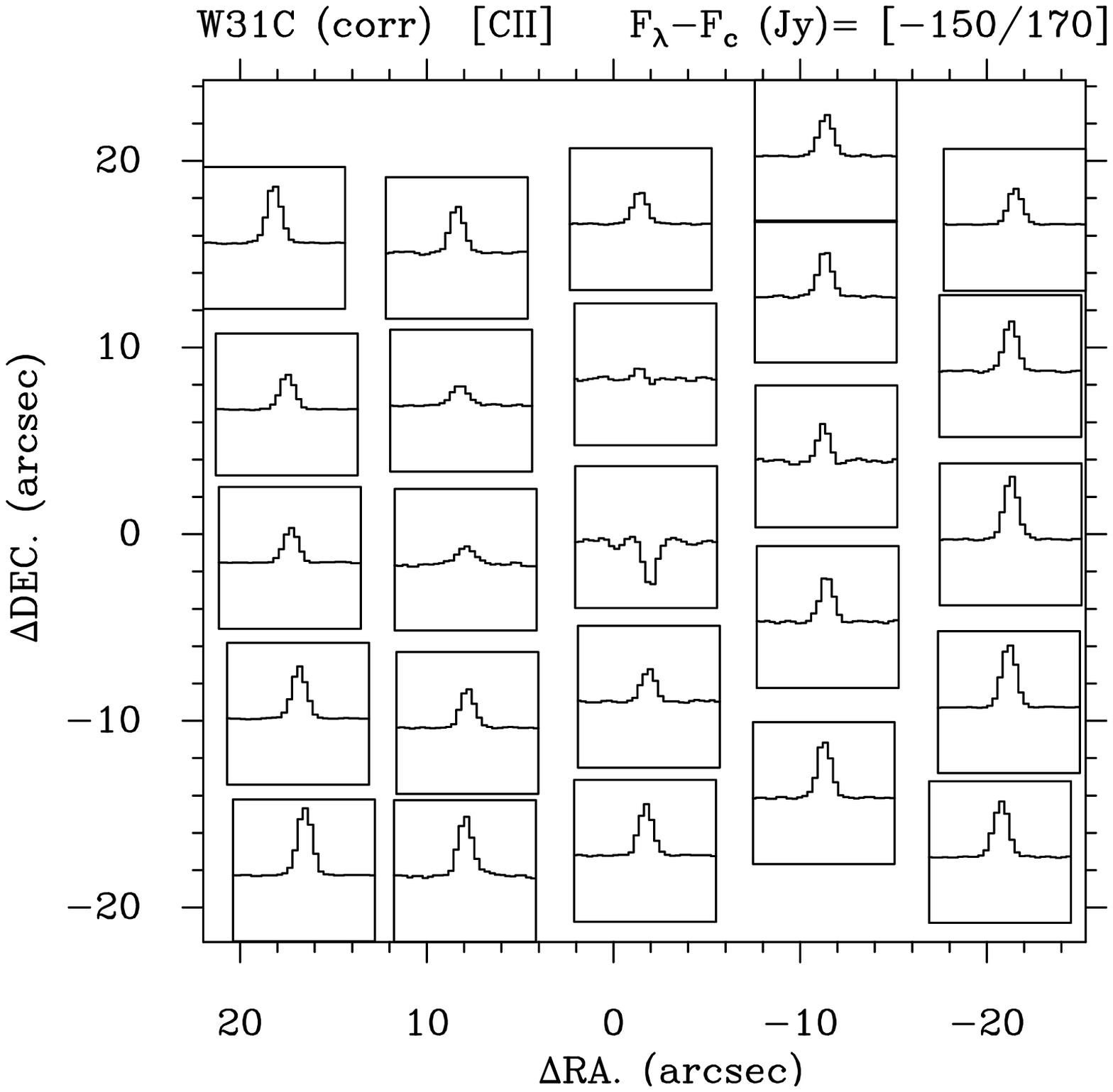}}
\resizebox{6cm}{!}{
\includegraphics{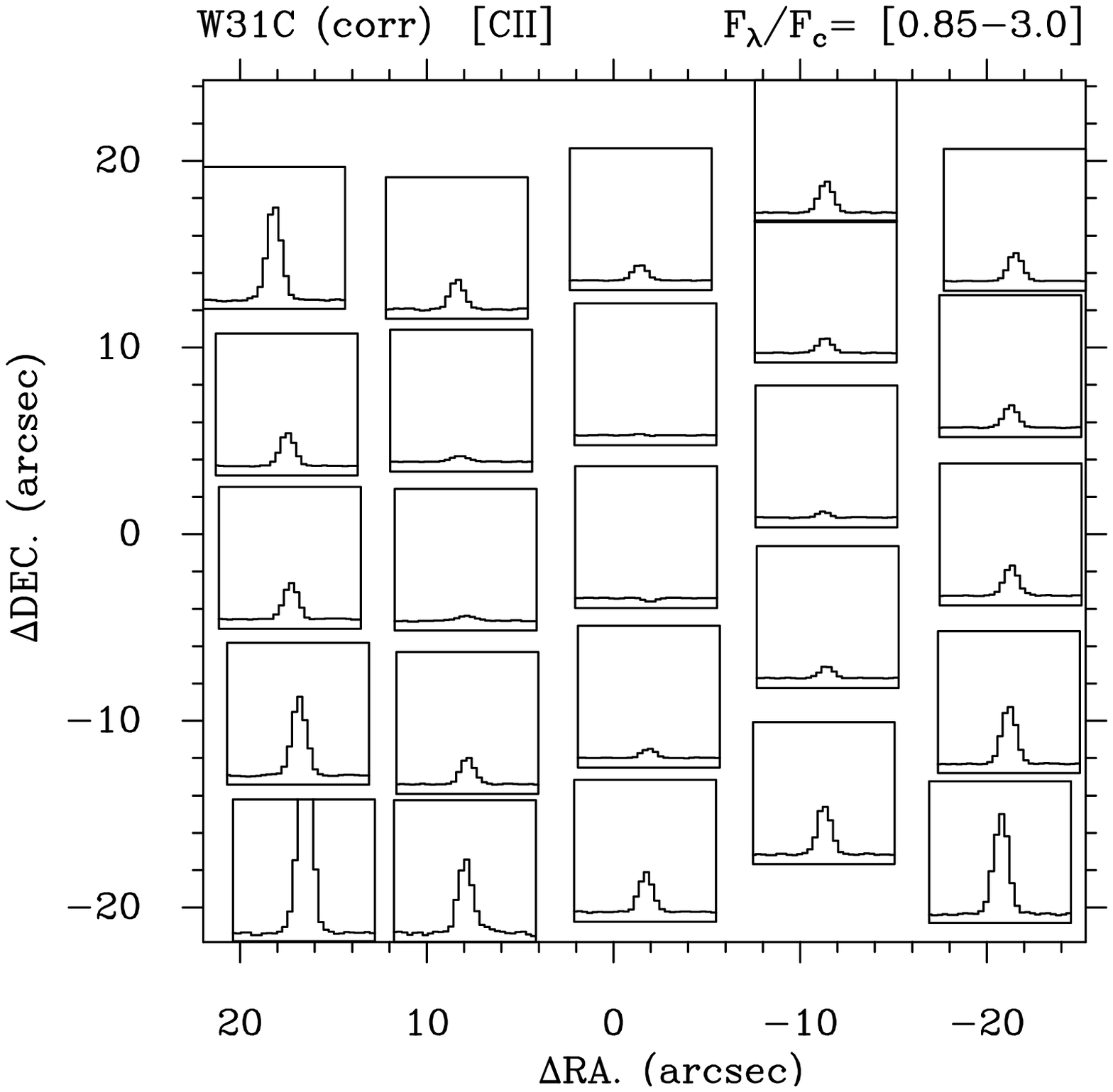}}
\resizebox{6cm}{!}{
\includegraphics{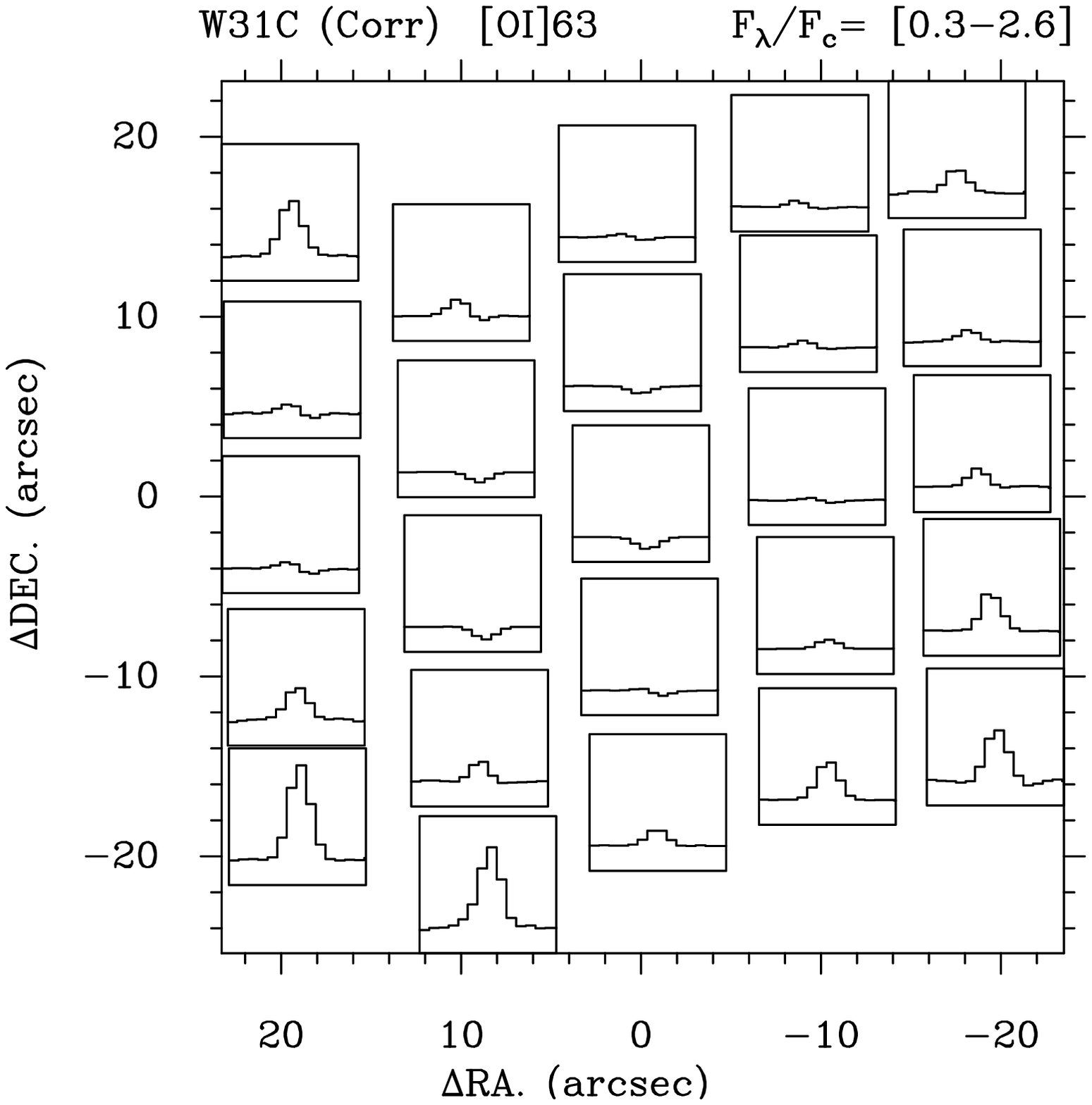}
}
\caption{\label{fig:pacs-w31c} PACS data towards W31C. For all maps, the
  offsets are given relative to the central position listed in Table
  \ref{tab:sources}. Top left: Continuum
  emission at 158$\mu$m. Contour levels are drawn at 0.2, 0.3, 0.4 ... 0.9 relative
to the maximum.  Top right: Comparison of the integrated
absorption measured in the HIFI map relative to the absorption at the map
center, with the continuum flux measured in the PACS map relative to the
map center. The dashed red line shows the linear regression line 
and the dotted black line a 1:1 relationship.
Bottom left: [\CII] emission in the 25 PACS spaxels. The continuum
emission has been subtracted. The vertical scale runs from -150 to 170
Jy. Bottom middle: Map of the [\CII] line to continuum emission/absorption in the 25
PACS spaxels. The vertical scale runs from 0.85 to 3.0. Note the very low
contrast in the central spaxels. Bottom right: Map of the [\OI] 63 $\mu$m 
line to continuum emission/absorption in the 25
PACS spaxels. The vertical scale runs from 0.3 to 2.6.}
\end{figure*}

W31C(G10.6-0.4) is a massive star forming region located in the Norma arm, at
a  distance of $5 $~\kpc \ \citep{sanna}. The
foreground material samples the Sagittarius and Scutum-Centaurus arms, in
addition to local material. The source velocity is about -2 \kms while the
foreground material is detected between $\sim 10$ and $\sim 50$ \kms.
The HIFI spectra are shown in Figures \ref{fig:specw31c} and
\ref{fig:mapw31c}. The PACS spectral maps near 158$\mu$m are presented 
in Figure \ref{fig:pacs-w31c}.

\subsection{W33A}

\begin{figure}[h!]
\resizebox{8cm}{!}{
\includegraphics{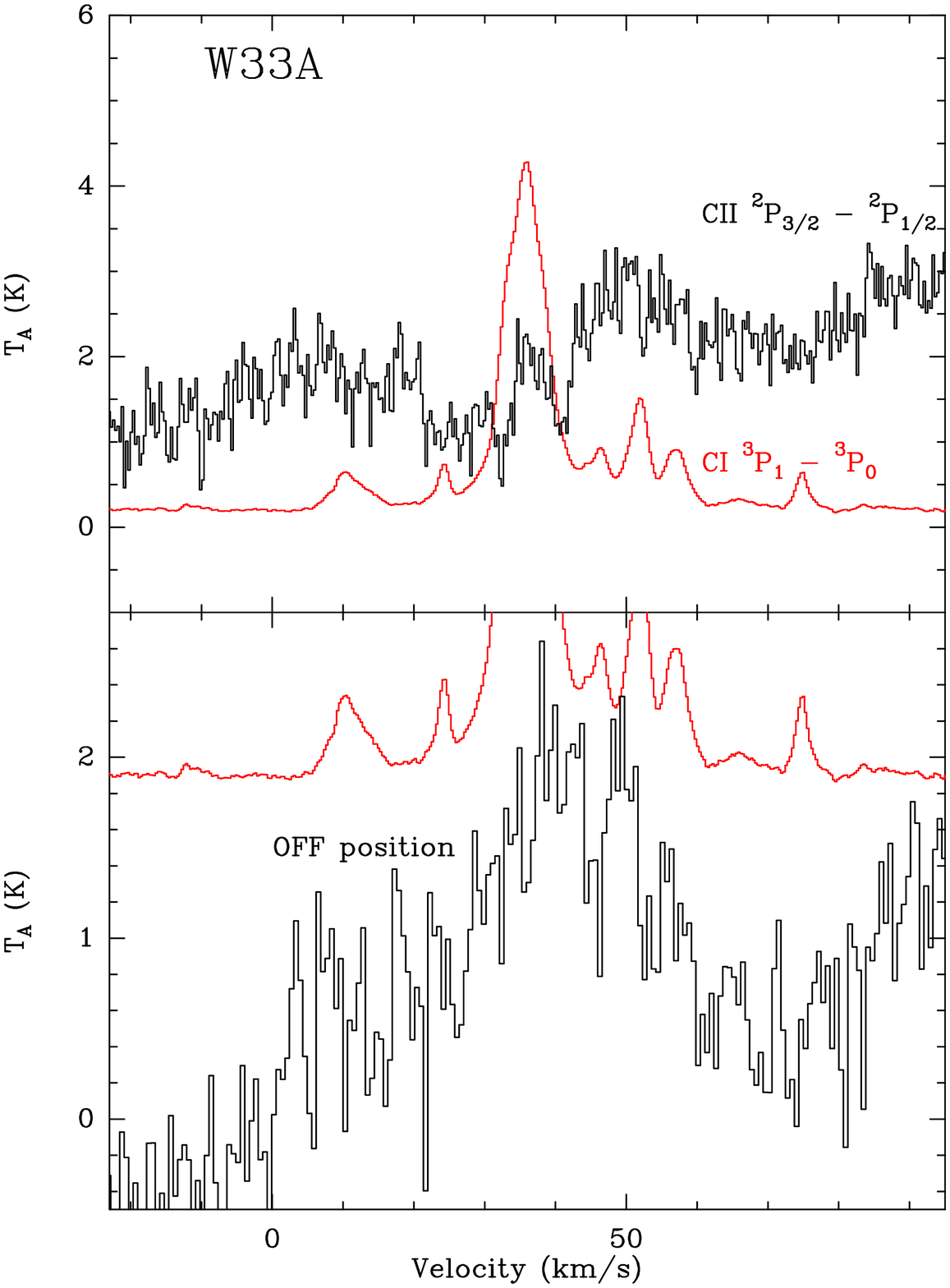}}
\caption{\label{fig:specw33a} Top : Herschel/HIFI spectra towards W33A.
The red line shows the [\CI]$^3P_1 - ^3P_0$ line at 492~GHz
and the black line the [\CII]$^2P_{3/2}-^2P_{1/2}$ line at 1.9~THz. 
 The horizontal axis is the LSR
velocity in \kms, and the vertical axis the antenna temperature in
Kelvins. The continuum level for [\CII] corresponds to the SSB continuum 
levels. Bottom : zoom on the [\CI] lines (red, as above), and average [\CII] 
spectrum of the OFF positions (black). The continuum levels have been shifted
for clarity in the bottom panel.  }
\end{figure}

\begin{figure}[h!]
\resizebox{8.5cm}{!}{
\includegraphics{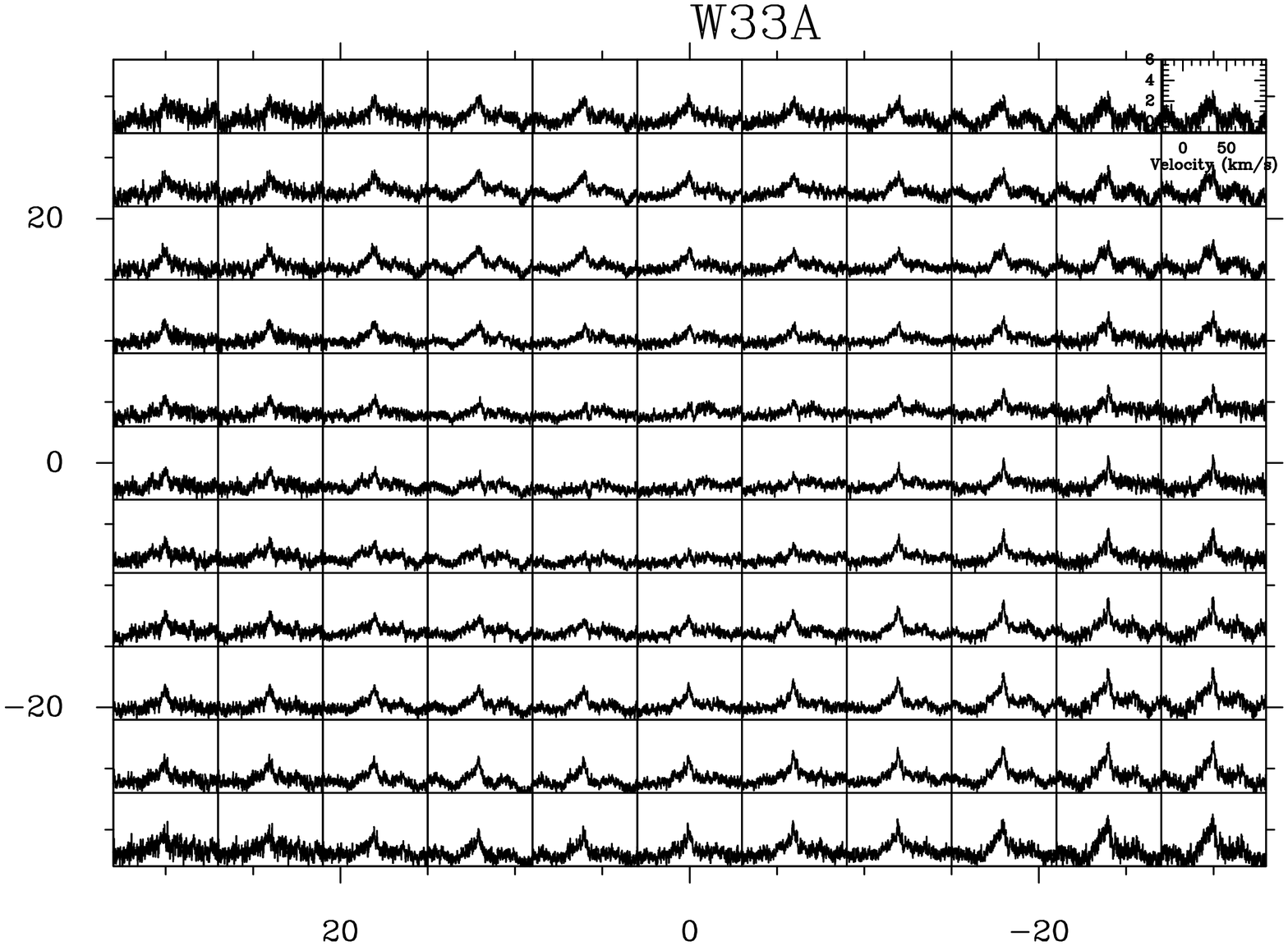}}
\caption{\label{fig:mapw33a} Montage of [\CII] spectra towards W33A. A 
  baseline has been subtracted from all spectra.  The horizontal axis is the LSR velocity in \kms, which runs from -23~\kms \ to 95~\kms, and the vertical axis the antenna temperature in
Kelvins, which runs from -1~\k \ to 6~\K.
 The x--axis shows the right ascension offset in arc-sec and the
  y--axis the declination offset in arc-sec, relative to the source position given in Table \ref{tab:sources}.
}
\end{figure}

\begin{figure}[h!]
\resizebox{7.cm}{!}{
\includegraphics{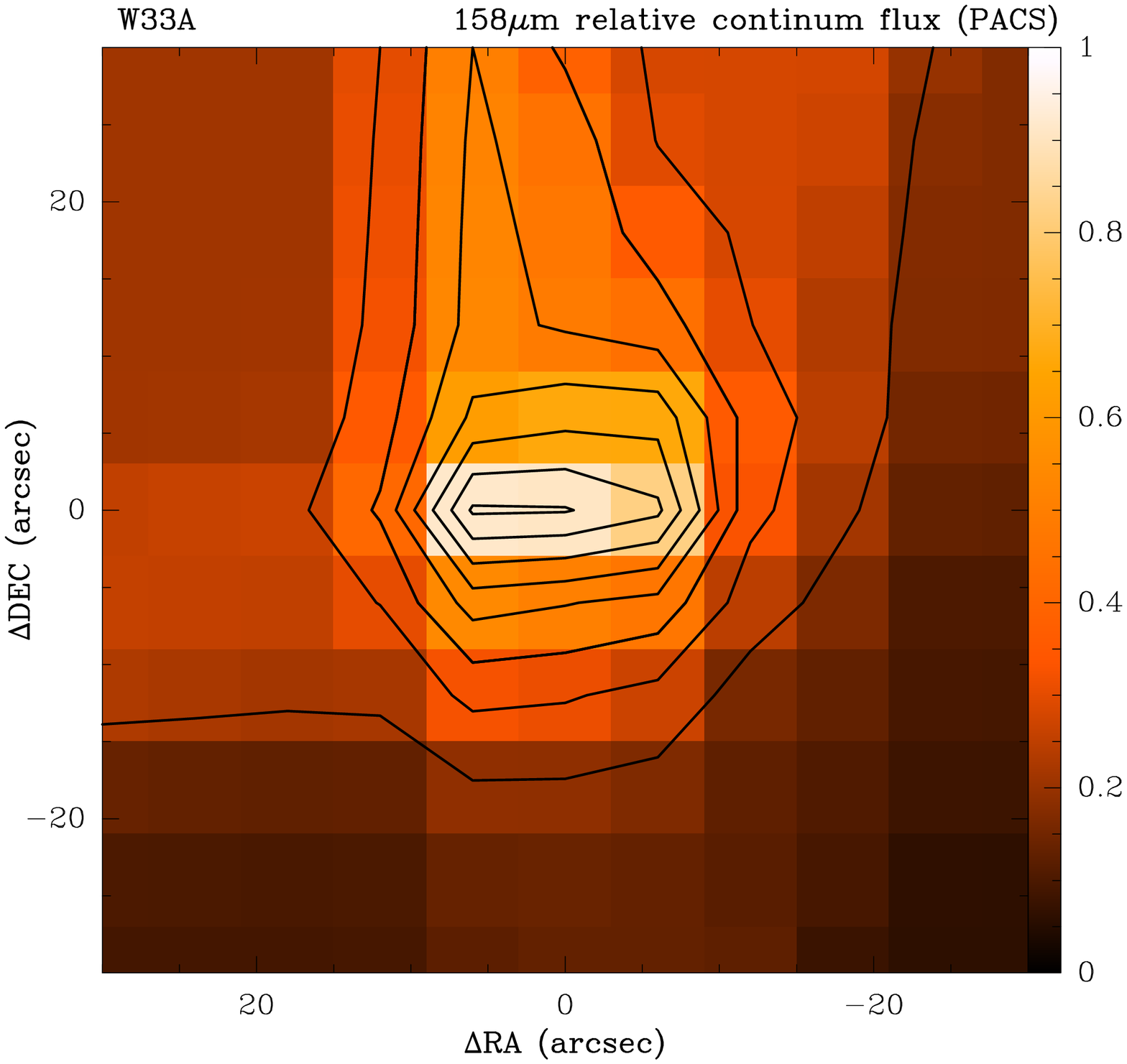}}
\\
\resizebox{7.cm}{!}{
\includegraphics{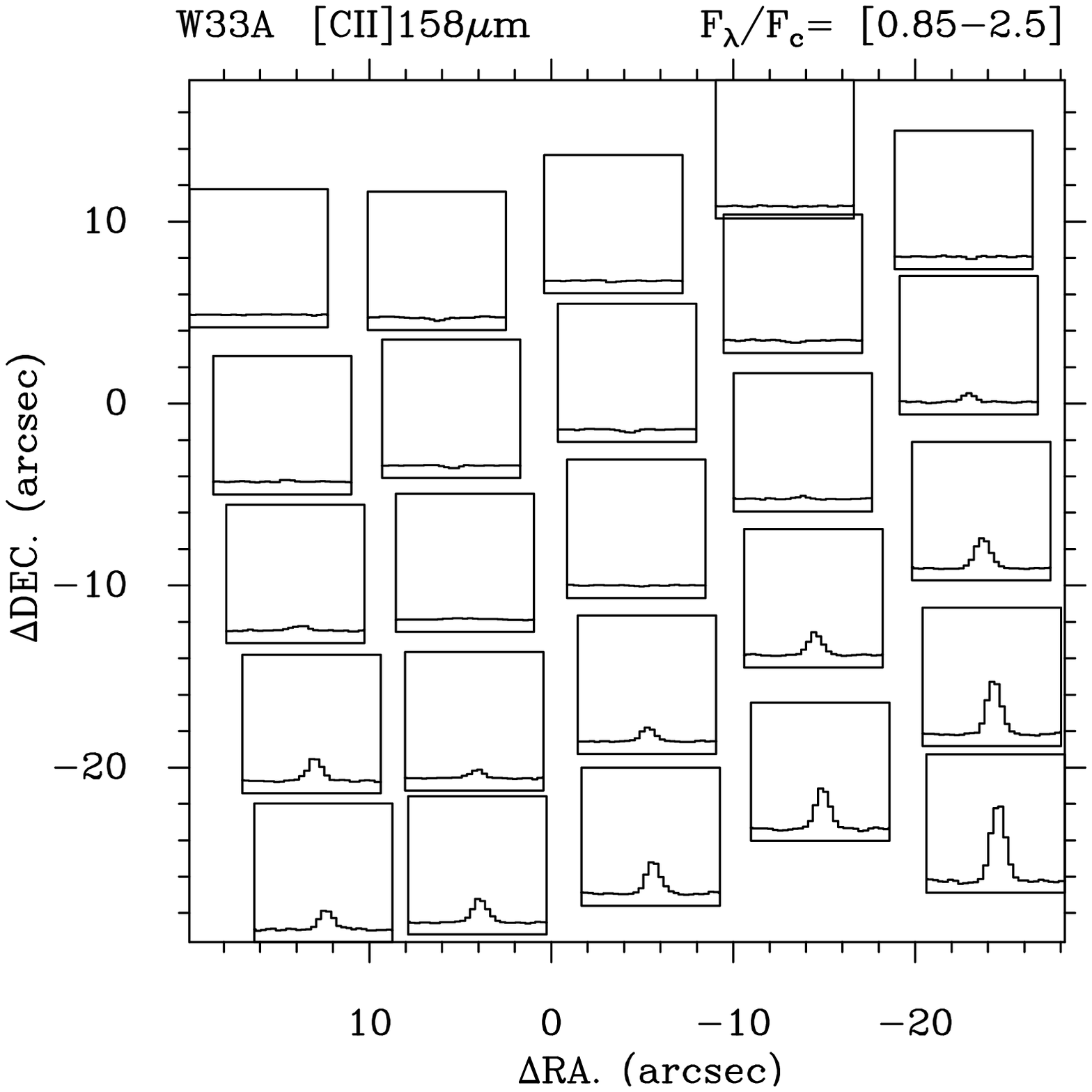}}
\caption{\label{fig:pacs-w33a} PACS data towards W33A. The
  offsets are given relative to the central position listed in Table
  \ref{tab:sources}. Top: Continuum
  emission at 158$\mu$m. Contour levels are 0.2, 0.3, 0.4 ... 0.9 relative
to the maximum.   Bottom: Map of the line to continuum emission/absorption in the 25
PACS spaxels. The vertical scale runs from 0.85 to 2.5. Note the very low
contrast in the central spaxels. The data have not been corrected for the
contamination in the OFF beam.}
\end{figure}

W33A(G12.9-0.3) is a highly embedded massive star at a distance of 2.4 \kpc \
\citep{immer}, with  a massive cold envelope.  A weak absorption from local
foreground gas has been
detected in CH$^+$ by \citet{godard:12} but the sensitivity of the present
data does not allow to detect the associated [\CII] feature. The HIFI spectra are shown in Figure \ref{fig:specw33a} and
\ref{fig:mapw33a}. The 
[\CII] emission from W33A is weak. It is barely detected at the continuum peak but increases as one moves away from
the continuum peak. Broad line wings are detected in [\CII], probably associated
with a molecular outflow. The [\CI] $^3P_1-^3P_0$ data show multiple weak emission
lines from line of sight diffuse gas, including features located further away from W33A.  The PACS spectral maps near 158$\mu$m are presented in Figure \ref{fig:pacs-w33a}.

\subsection{G34.3+0.15}

\begin{figure}[h!]
\resizebox{8cm}{!}{
\includegraphics{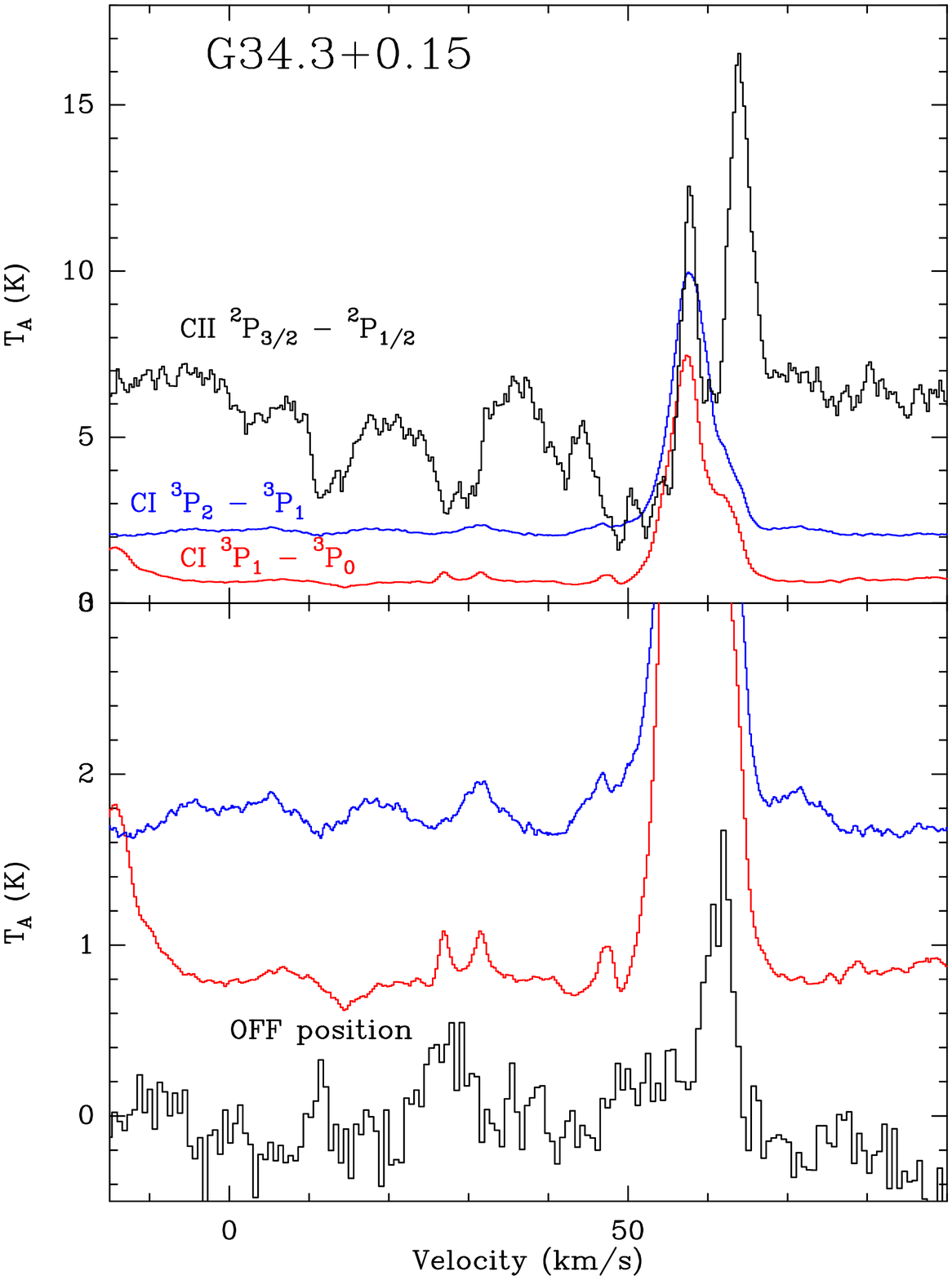}}
\caption{\label{fig:specg34} Top : Herschel/HIFI spectra towards G34.3+0.15.
The red line shows the [\CI]$^3P_1 - ^3P_0$ line at 492~GHz, the blue line shows
the [\CI]$^3P_2 - ^3P_1$ line at 809~GHz and the black line the
[\CII]$^2P_{3/2}-^2P_{1/2}$ line at 1.9~THz. The horizontal axis is the LSR
velocity in \kms \ and the vertical axis is the antenna temperature in
Kelvins. The continuum level for [\CII] corresponds to the SSB continuum level. 
Bottom : zoom on the [\CI] lines (red,blue as above), and average [\CII] 
spectrum of the OFF positions (black). The continuum levels have been shifted
for clarity in the bottom panel.  }
\end{figure}

\begin{figure}[h!]
\resizebox{8.5cm}{!}{
\includegraphics{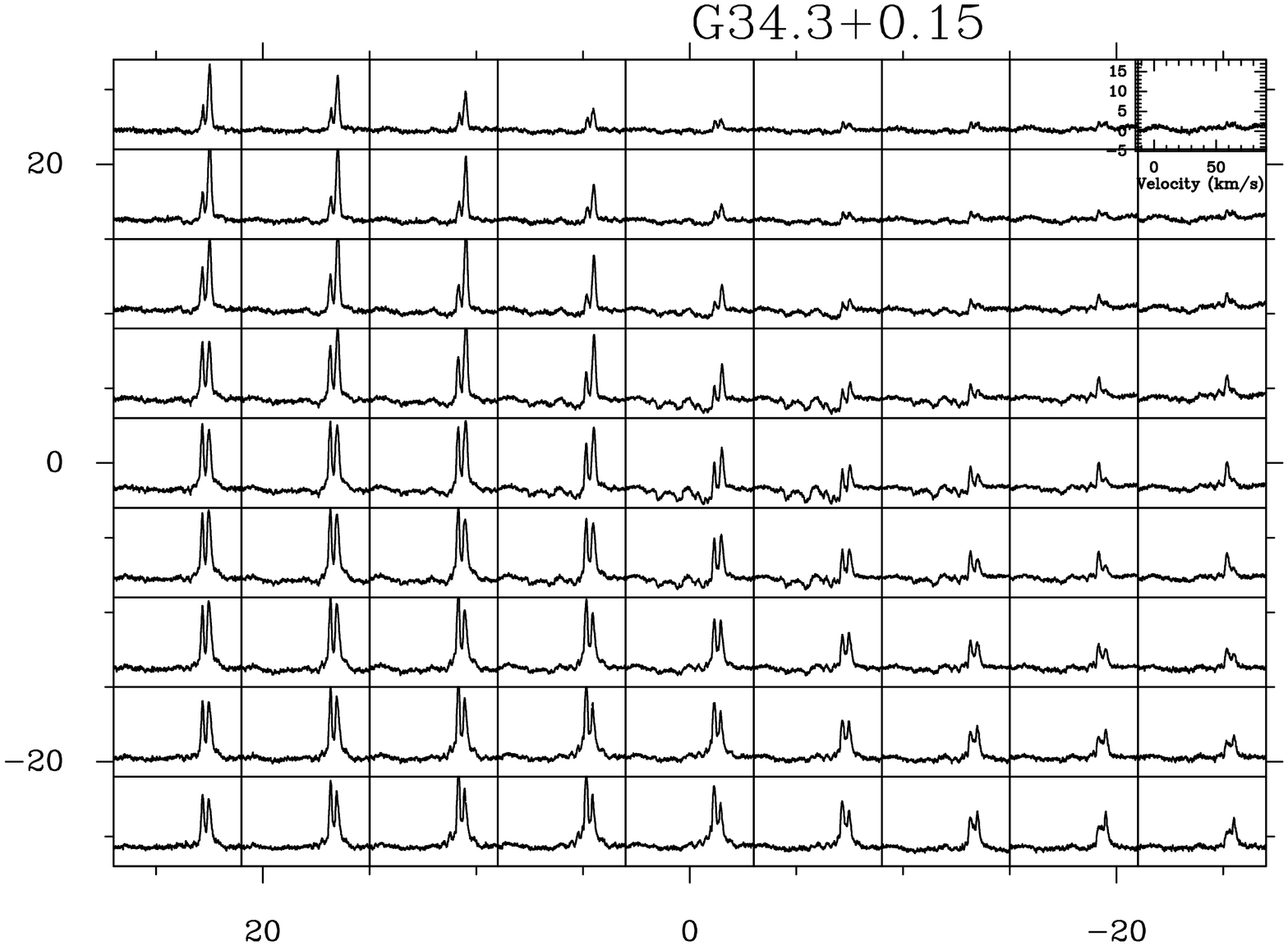}}
\caption{\label{fig:mapg34} Montage of [\CII] spectra towards G34.3+0.15. A 
  baseline has been subtracted from all spectra. The horizontal axis is the LSR
  velocity in  \kms, which runs from -15~\kms \ to 90~\kms,  and the  vertical axis is the antenna temperature in
  Kelvins, which runs from -5~\K \ to 18~\K.  The x--axis shows the right ascension offset in arc-sec and the  y--axis the declination offset in arc-sec, relative to the source position given
in Table \ref{tab:sources}.
}
\end{figure}

\begin{figure}[h!]
\resizebox{7.cm}{!}{
\includegraphics{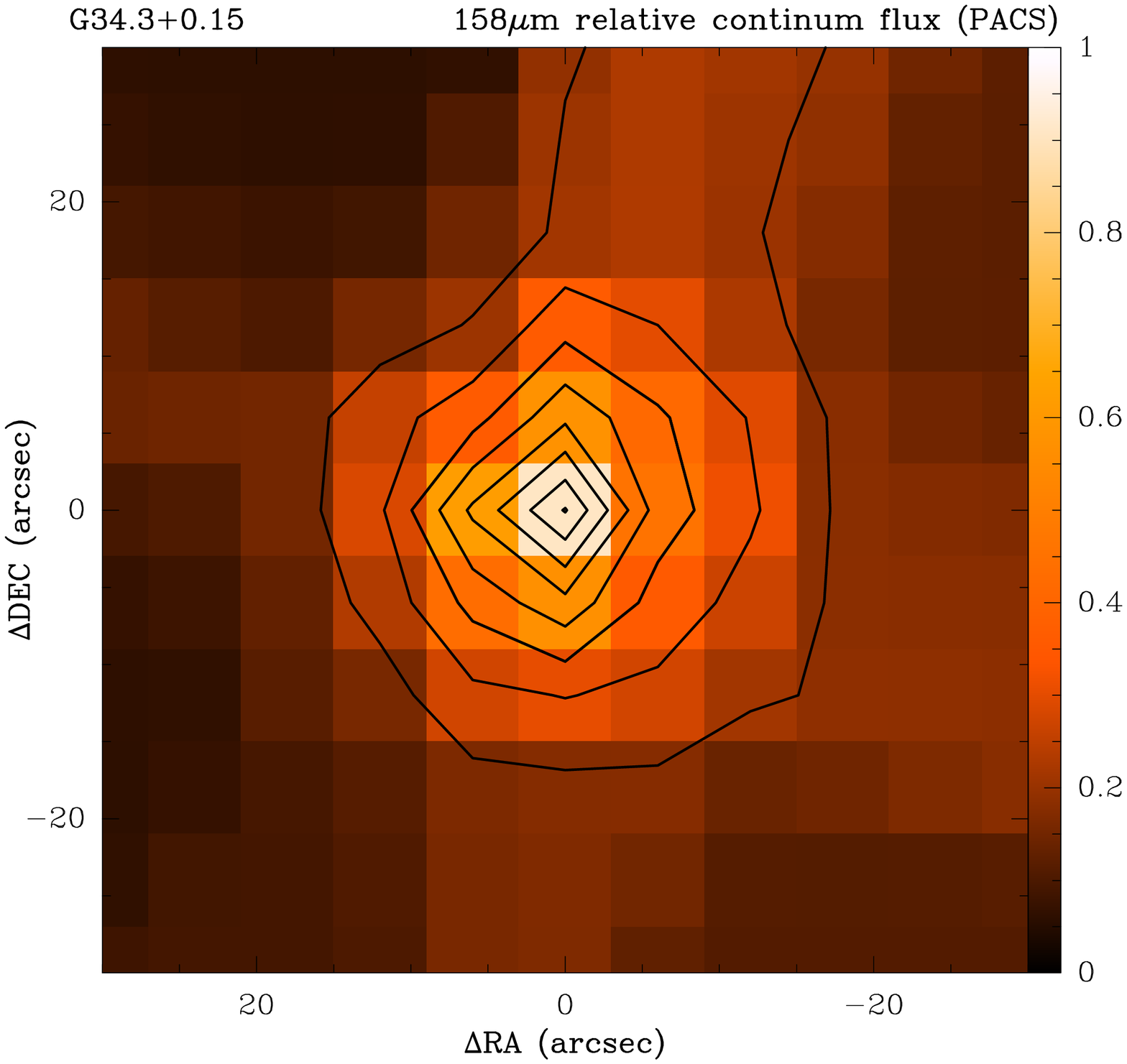}}
\\
\resizebox{7cm}{6cm}{
\includegraphics{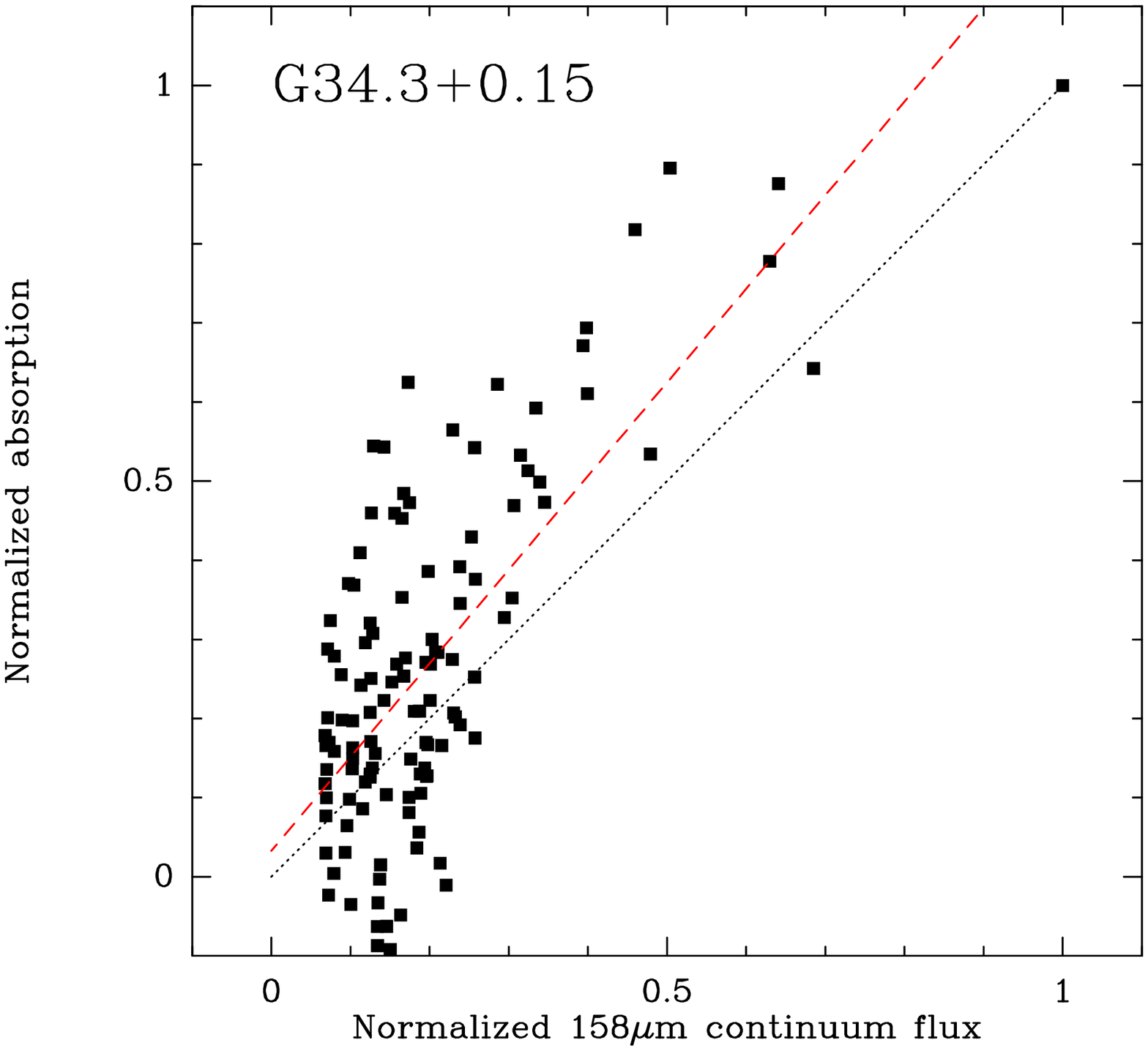}}
\\
\resizebox{7.cm}{6cm}{
\includegraphics{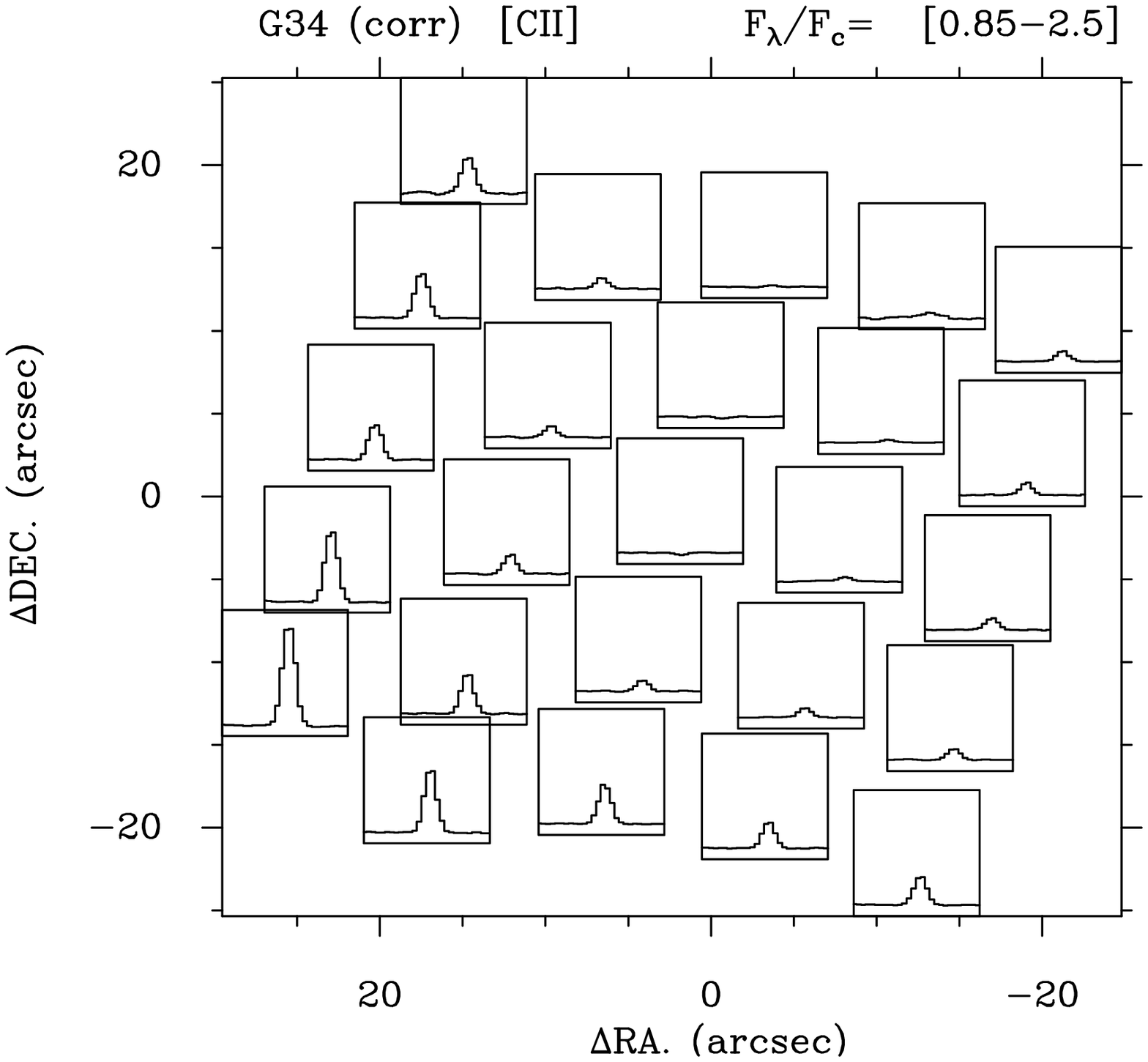}}
\caption{\label{fig:pacs-g34} PACS data towards G34.3+0.15. For all maps, the
  offsets are given relative to the central position listed in Tab
  \ref{tab:sources}. Top: Continuum
  emission at 158$\mu$m. Contour levels are at 0.2, 0.3, 0.4 ... 0.9 relative
to the maximum. Middle: Comparison of the integrated
absorption measured in the HIFI map relative to the absorption at the map
center, with the continuum flux measured in the PACS map relative to the
map center. The dashed red line shows the linear regression line 
and the dotted black line a 1:1 relationship. Bottom: [\CII]  Map of the line to continuum emission/absorption in the 25
PACS spaxels. The vertical scale runs from 0.85 to 2.5. The data have not been
corrected for possible contamination in the OFF position.}
\end{figure}

G34.3+0.15 is an ultra compact HII region located at about 3.3 \kpc \
\citep{kuchar}. The line of sight crosses the Sagittarius arm at velocities
between $\sim 20$ and $\sim 35$ \kms, and also samples
the local interstellar medium for LSR velocities lower than 20 \kms.  
The HIFI spectra are shown in Fig. \ref{fig:specg34} and
\ref{fig:mapg34}. The [\CII] line is strongly self-absorbed towards
G34.3+0.15 and presents an  asymmetric spatial distribution. The [\CI]
$^3P_1-^3P_0$ emission is weak in the foreground gas. The tiny absorption at
$\sim 15$ \kms \ is due to a low level contamination in the
reference position used in the Load Chop observations. The emission features
in the [\CI] $^3P_2 - ^3P_1$ spectra are excited molecular lines associated
with the  G34.3+0.15 hot cores.
The PACS spectral maps near 158$\mu$m are presented in fig \ref{fig:pacs-g34}. 

\subsection{W49N}
W49N is one of the most luminous star forming region in the Galaxy. It is located in the Perseus arm at a distance of 11.4~\kpc \ \citep{gwinn}. The line of sight crosses the Sagittarius arm twice along this very long line of sight.

\subsection{W51}

\begin{figure}[h!]
\resizebox{8cm}{!}{
\includegraphics{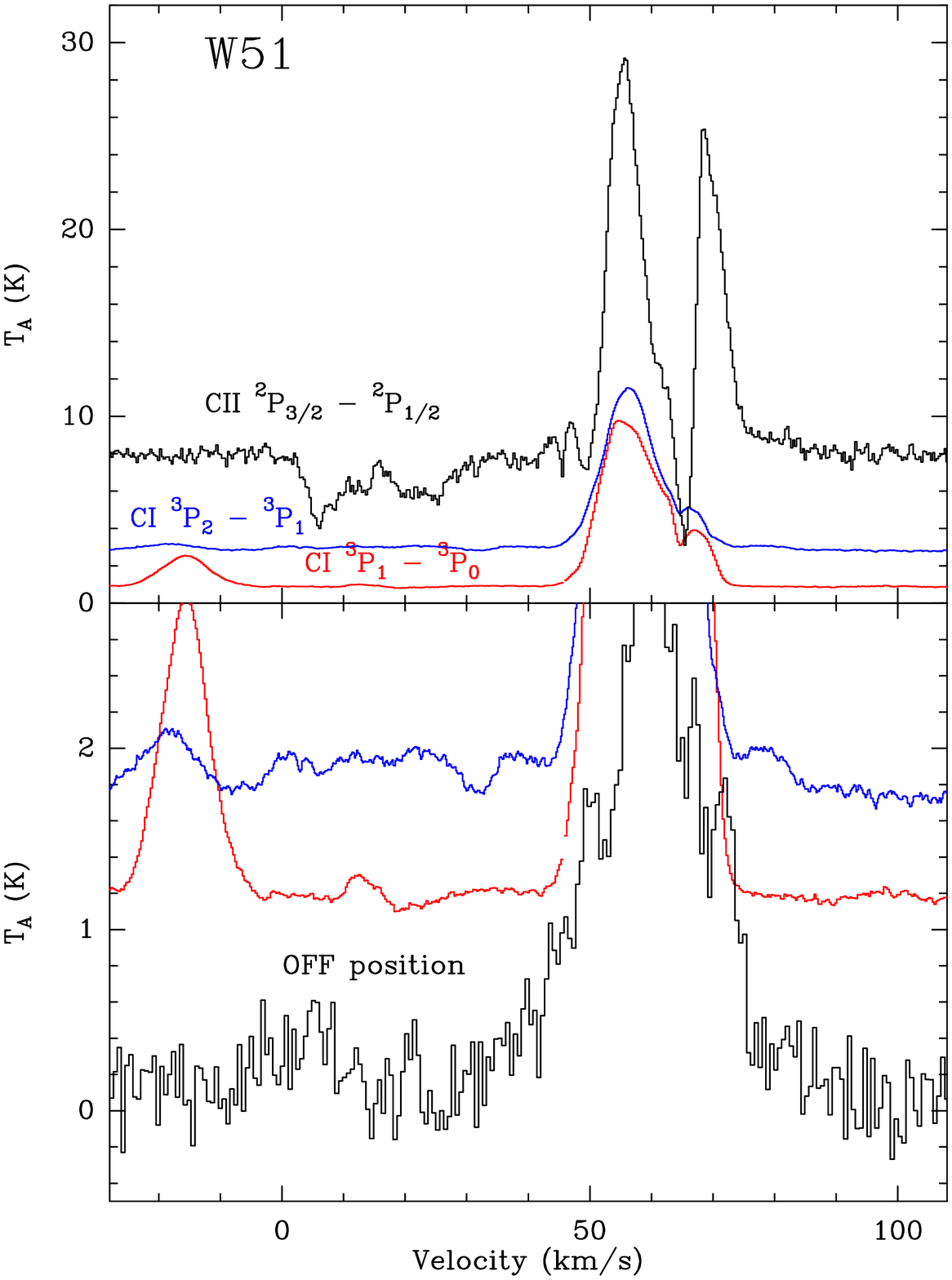}}
\caption{\label{fig:specw51} Top : Herschel/HIFI spectra towards W51.
The red line shows the [\CI]$^3P_1 - ^3P_0$ line at 492~GHz, the blue line shows
the [\CI]$^3P_2 - ^3P_1$ line at 809~GHz and the black line the
[\CII]$^2P_{3/2}-^2P_{1/2}$ line at 1.9~THz. The horizontal axis is the LSR
velocity in \kms and the vertical axis the antenna temperature in
Kelvins. The continuum level for [\CII] corresponds to the SSB continuum level. 
Bottom : zoom on the [\CI] lines (red,blue as above), and average [\CII] 
spectrum of the OFF positions (black). The continuum levels have been shifted
for clarity in the bottom panel.  }
\end{figure}

\begin{figure}[h!]
\resizebox{8.5cm}{!}{
\includegraphics{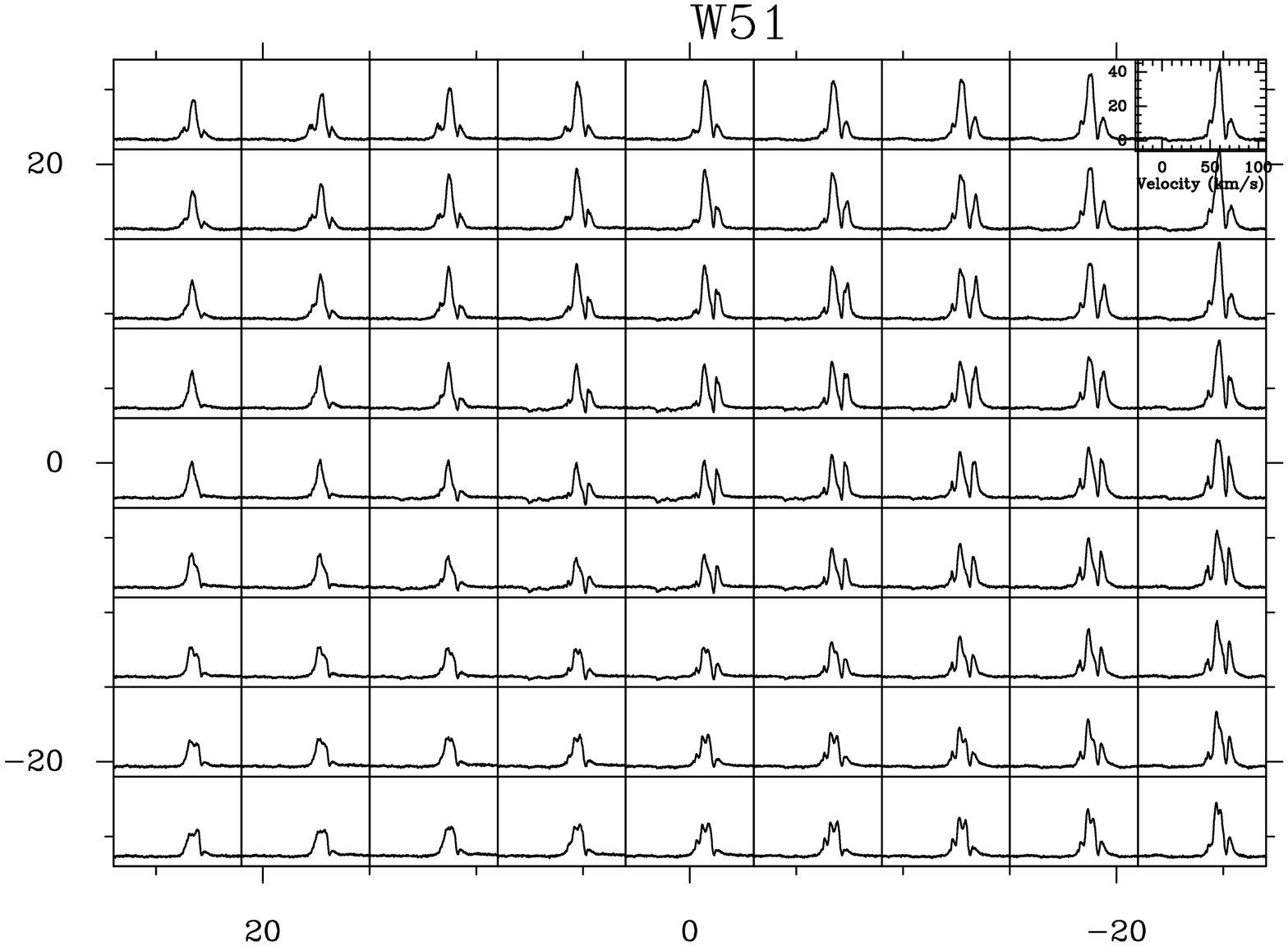}}
\caption{\label{fig:mapw51} Montage of [\CII] spectra towards W51. A 
  baseline has been subtracted from all spectra. The horizontal axis is the LSR
  velocity in  \kms,  which runs from -28~\kms \ to 108~\kms, and the  vertical axis is the antenna temperature in Kelvins, which runs from -6~\K \ to 47~\K.
 The x--axis shows the right ascension offset in arc-sec and the
  y--axis the declination offset in arc-sec, relative to the source position given
in Table \ref{tab:sources}. }
\end{figure}

\begin{figure*}[h!]
\resizebox{6cm}{!}{
\includegraphics{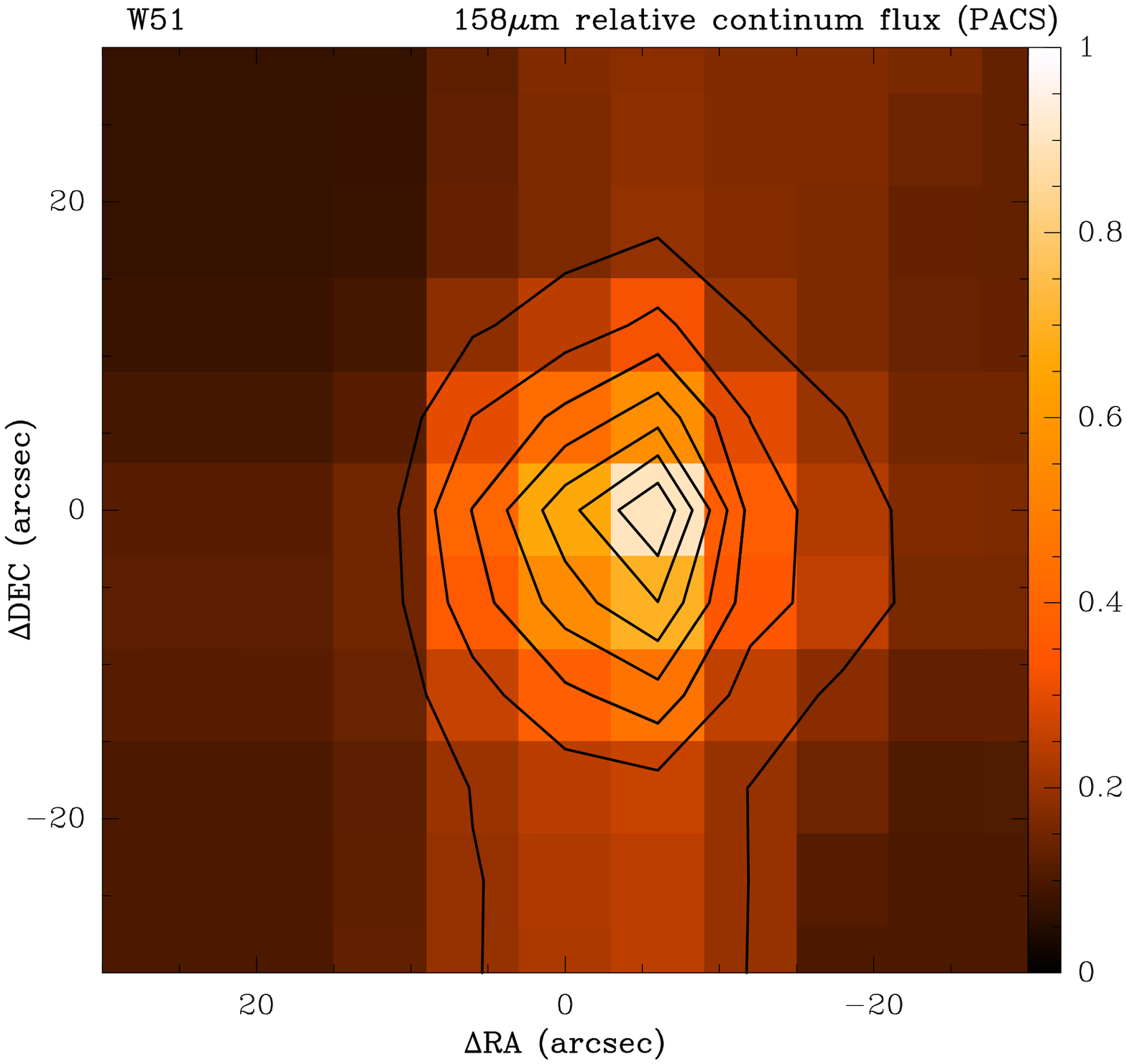}}
\resizebox{6cm}{!}{
\includegraphics{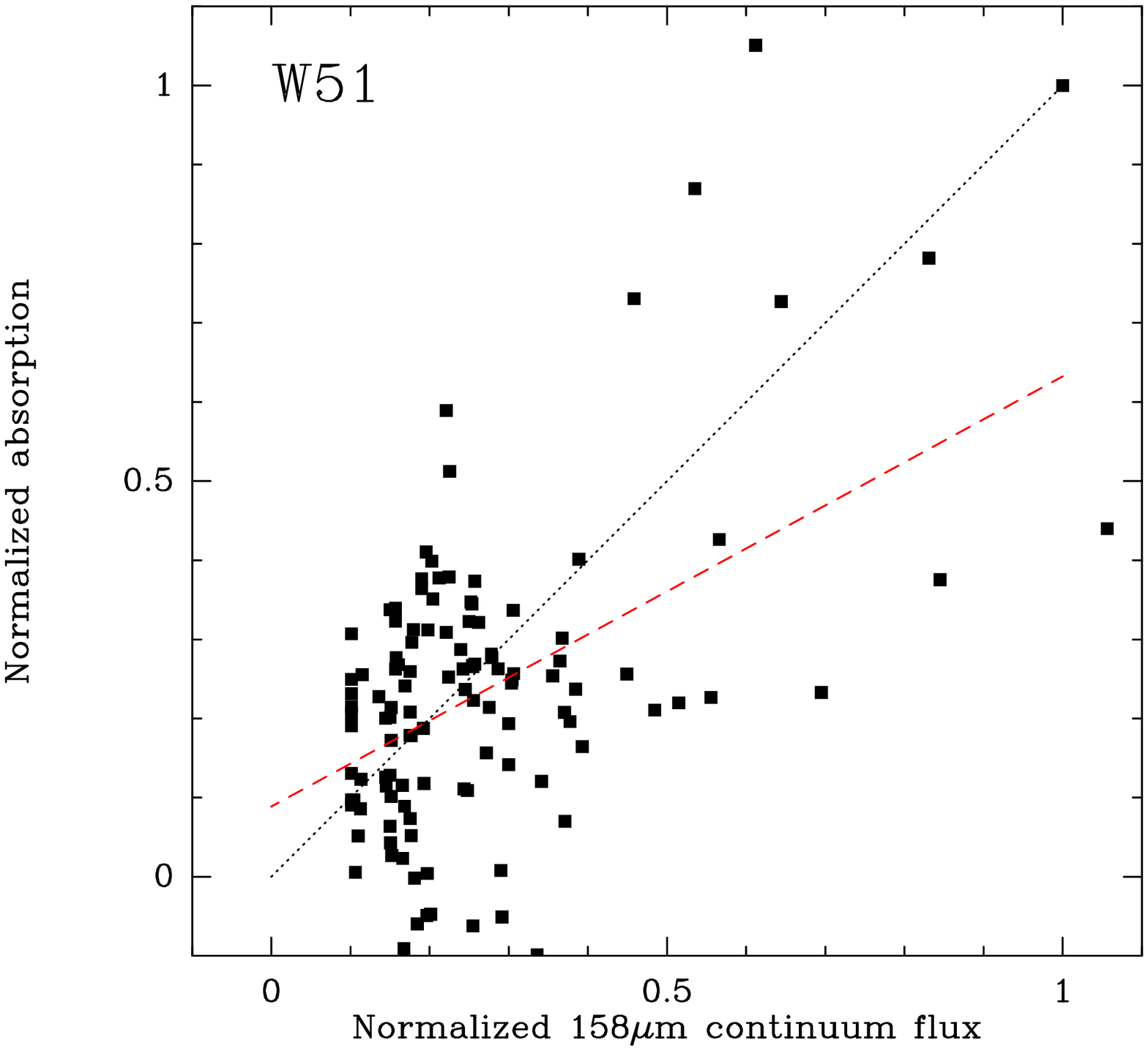}}
\\
\resizebox{6cm}{!}{
\includegraphics{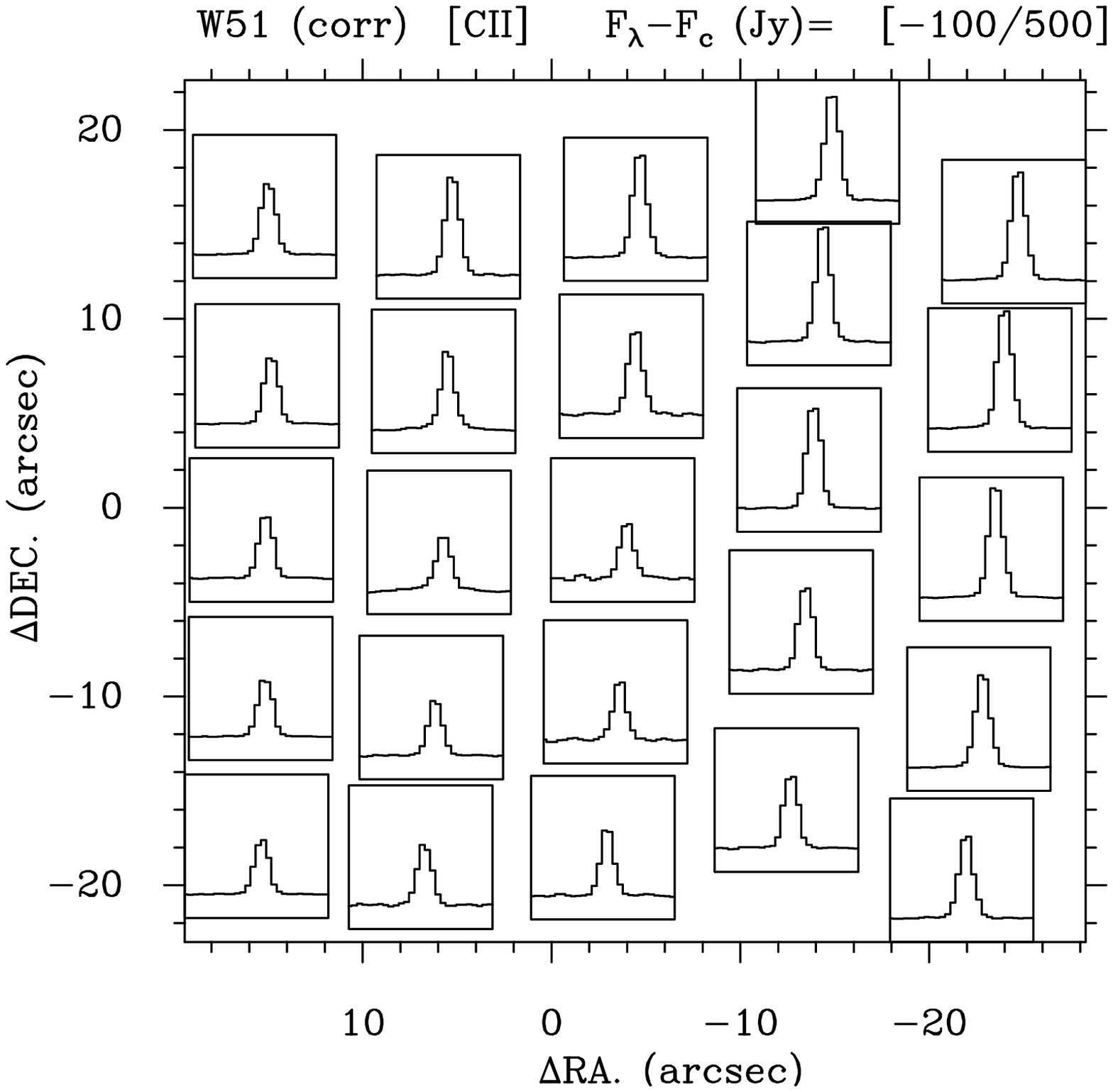}}
\resizebox{6cm}{!}{
\includegraphics{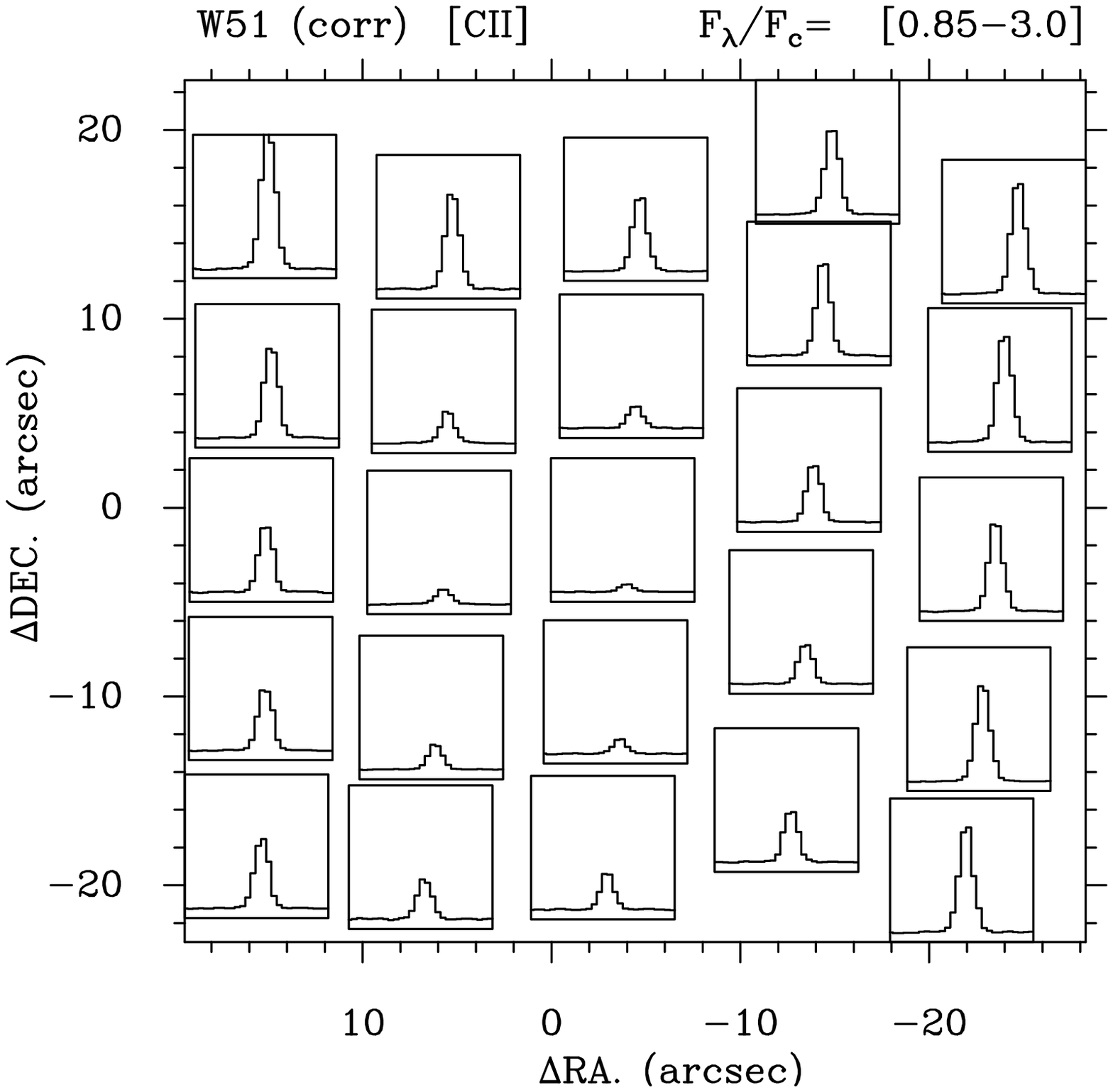}}
\resizebox{6cm}{!}{
\includegraphics{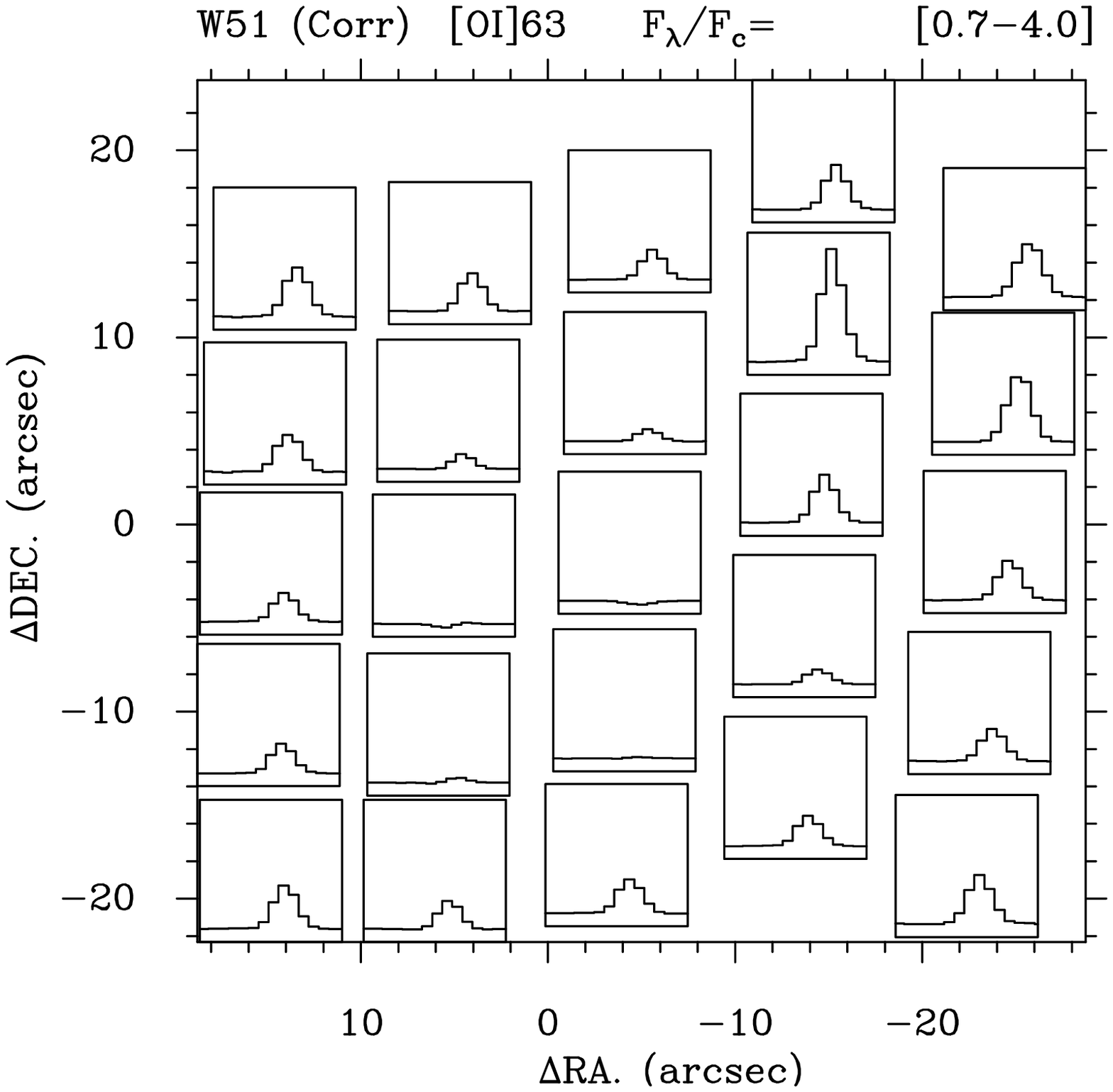}}
\caption{\label{fig:pacs-w51} PACS data towards W51. For all maps, the
  offsets are given relative to the central position listed in Tab
  \ref{tab:sources}. Top left: Continuum
  emission at 158$\mu$m. Contour levels are drawn at 0.2, 0.3, 0.4 ... 0.9 relative
to the maximum.  Top right: 
Comparison of the integrated
absorption measured in the HIFI map relative to the absorption at the map
center, with the continuum flux measured in the PACS map relative to the
map center. The dashed red line shows the linear regression line 
and the dotted black line a 1:1 relationship. Bottom left: [\CII] emission in the 25 PACS spaxels. The continuum
emission has been subtracted. The vertical scale runs from -100 to 500
Jy. Bottom middle: Map of the [\CII] 
line to continuum emission/absorption in the 25
PACS spaxels. The vertical scale runs from 0.85 to 3. Bottom right:  
Map of the [\OI] 63 $\mu$m line to continuum emission/absorption in the 25
PACS spaxels. The vertical scale runs from 0.7 to 4.
}
\end{figure*}

W51 is a very massive star formation complex in the Sagittarius arm at a
distance of $\sim 5.4$~\kpc \ \citep{sato:10}. The line of sight crosses the
local interstellar medium, and also probes relatively diffuse
material in the foreground of W51 at velocities $\sim 45$ and $\sim 65$ \kms.
The HIFI spectra are shown in Fig. \ref{fig:specw51} and
\ref{fig:mapw51}. The [\CII] spectra show extended wings associated with the
CO molecular outflow. The spectra are self reversed over a significant
fraction of the map. Despite the high sensitivity reached in these
data, we could not detect [\CI] either in absorption  nor in emission
associated with the prominent features at $\sim 6$ and $\sim 20$ \kms. The
weak emission features in the [\CI] $^3P_1 - ^3P_0$ and $^3P_2 - ^3P_1$ 
spectra are unrelated molecular lines
associated with the W51 E1/E2 hot cores. 
The PACS spectral maps near 158$\mu$m are presented in Figure \ref{fig:pacs-w51}.

\subsection{DR21(OH)}

\begin{figure}
\resizebox{8cm}{!}{
\includegraphics{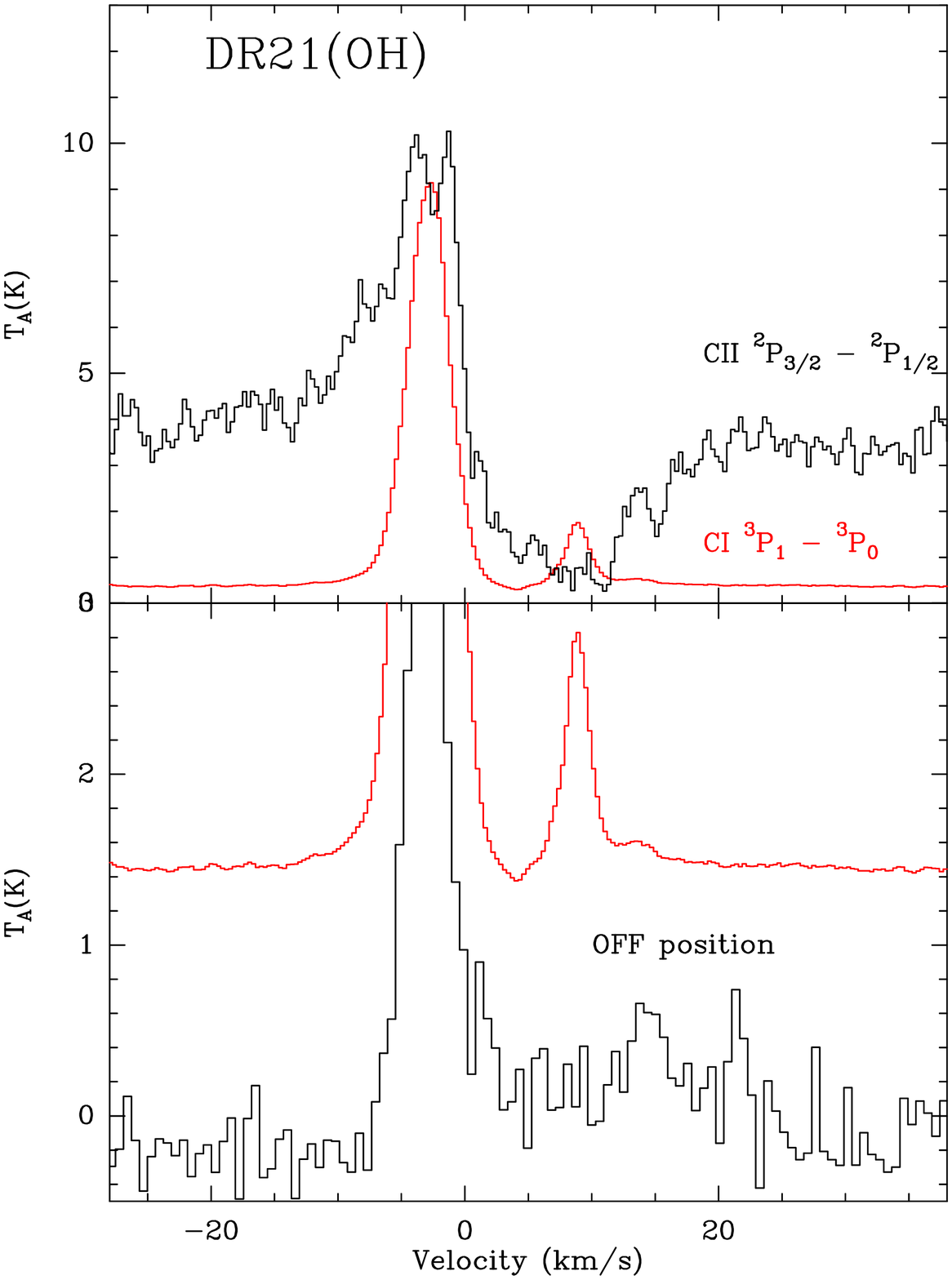}}
\caption{\label{fig:specdr21oh} Top : Herschel/HIFI spectra towards DR21(OH).
The red line shows the [\CI]$^3P_1 - ^3P_0$ line at 492~GHz, the blue line shows
the [\CI]$^3P_2 - ^3P_1$ line at 809~GHz and the black line the
[\CII]$^2P_{3/2}-^2P_{1/2}$ line at 1.9~THz. The horizontal axis is the LSR
velocity in \kms and the vertical axis the antenna temperature in
Kelvins. The continuum level for [\CII] corresponds to the SSB continuum level. 
Bottom : zoom on the [\CI] lines (red,blue as above), and average [\CII] 
spectrum of the OFF positions (black). The continuum levels have been shifted
for clarity in the bottom panel.  }
\end{figure}

\begin{figure}
\resizebox{8.5cm}{!}{
\includegraphics{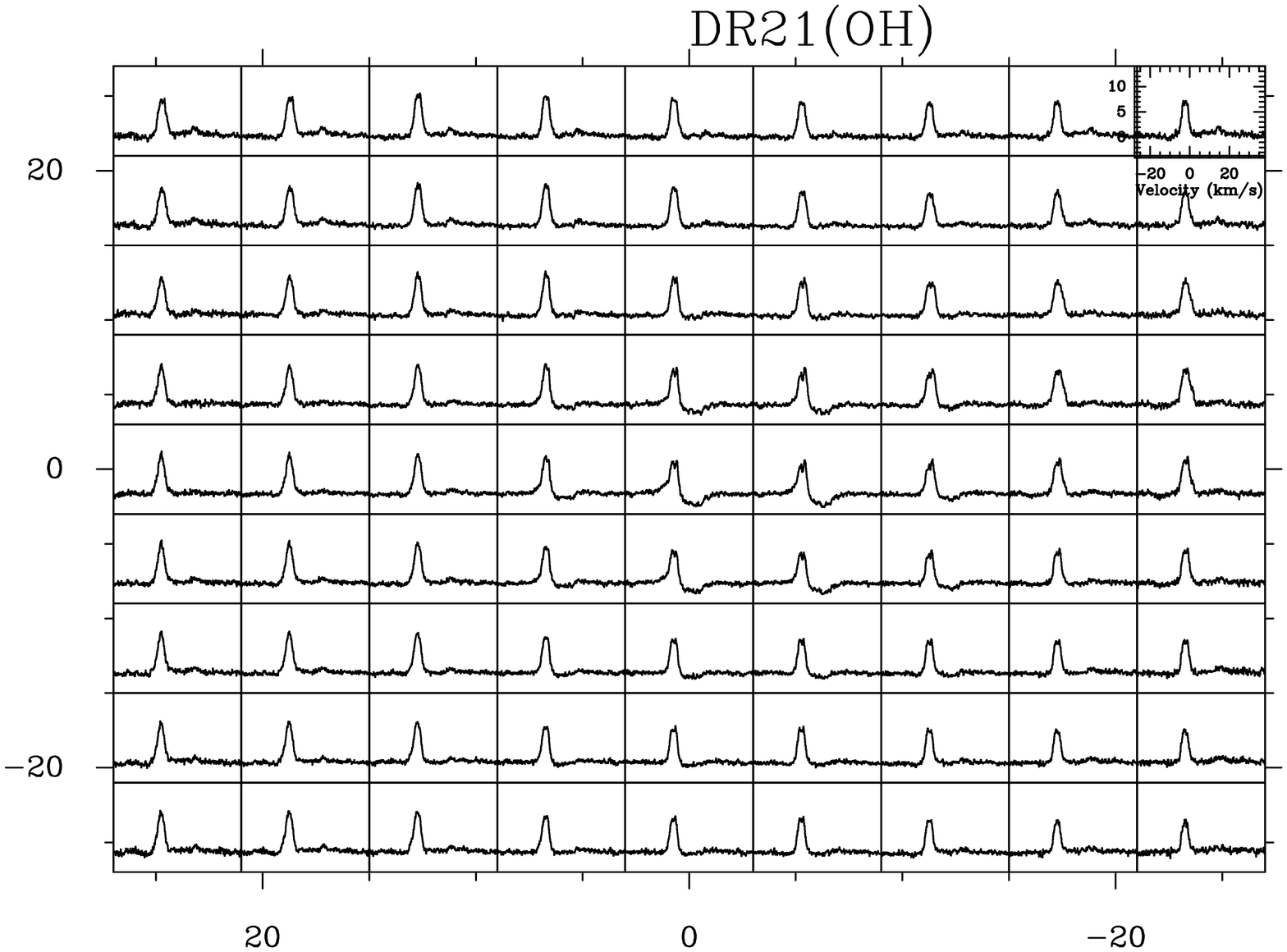}}
\caption{\label{fig:mapdr21oh} Montage of [\CII] spectra towards DR21(OH). A 
  baseline has been subtracted from all spectra. The horizontal axis is the LSR
  velocity in  \kms,  which runs from -28~\kms \ to 38~\kms, and  the  vertical axis is the antenna temperature in
  Kelvins, which runs from -4~\K \ to 14~\K.  The x--axis shows the right ascension offset in arc-sec and the
  y--axis the declination offset in arc-sec, relative to the source position given
in Table \ref{tab:sources}.  }
\end{figure}

\begin{figure}
\resizebox{7.cm}{!}{
\includegraphics{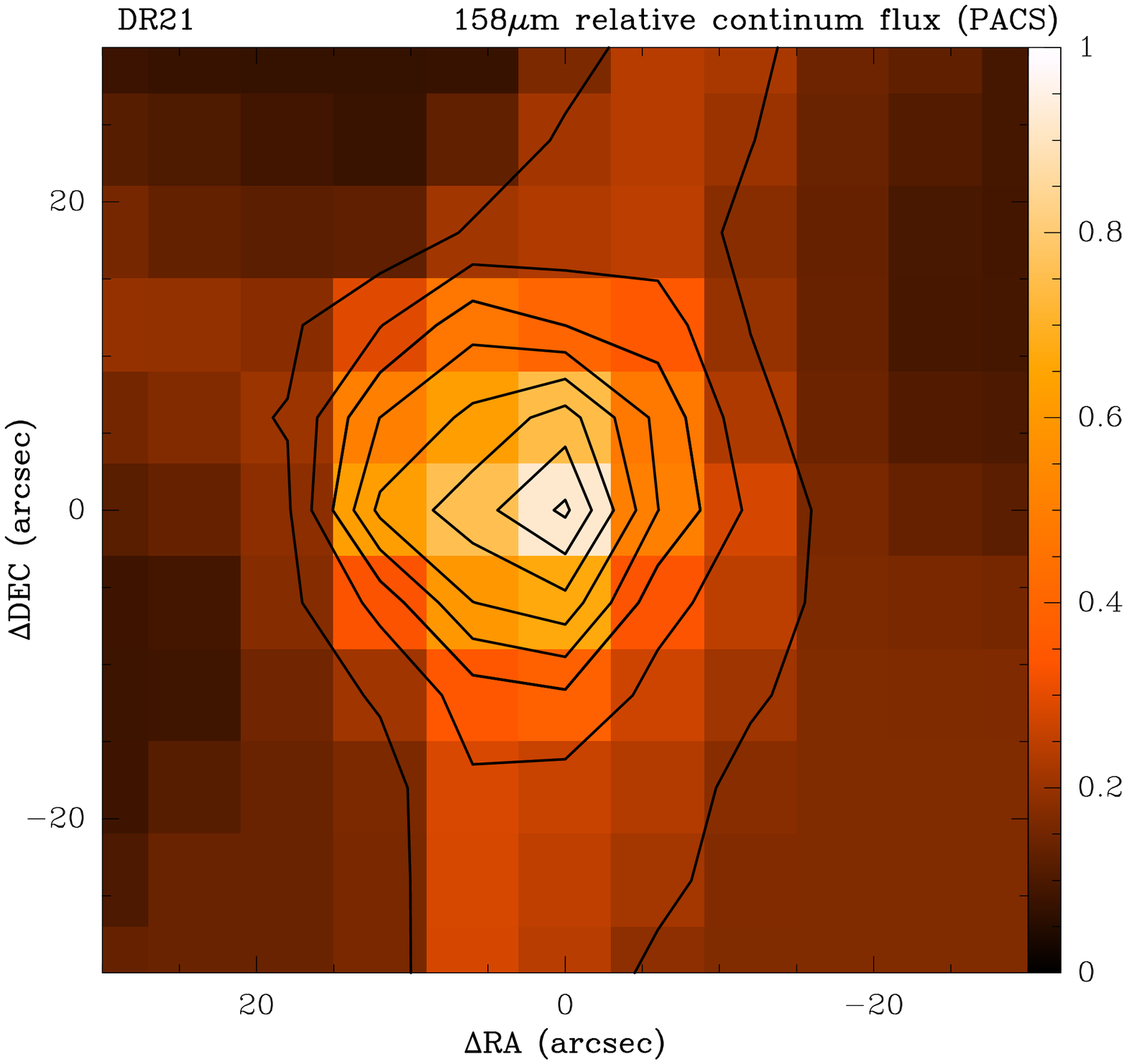}}
\\
\resizebox{7cm}{!}{
\includegraphics{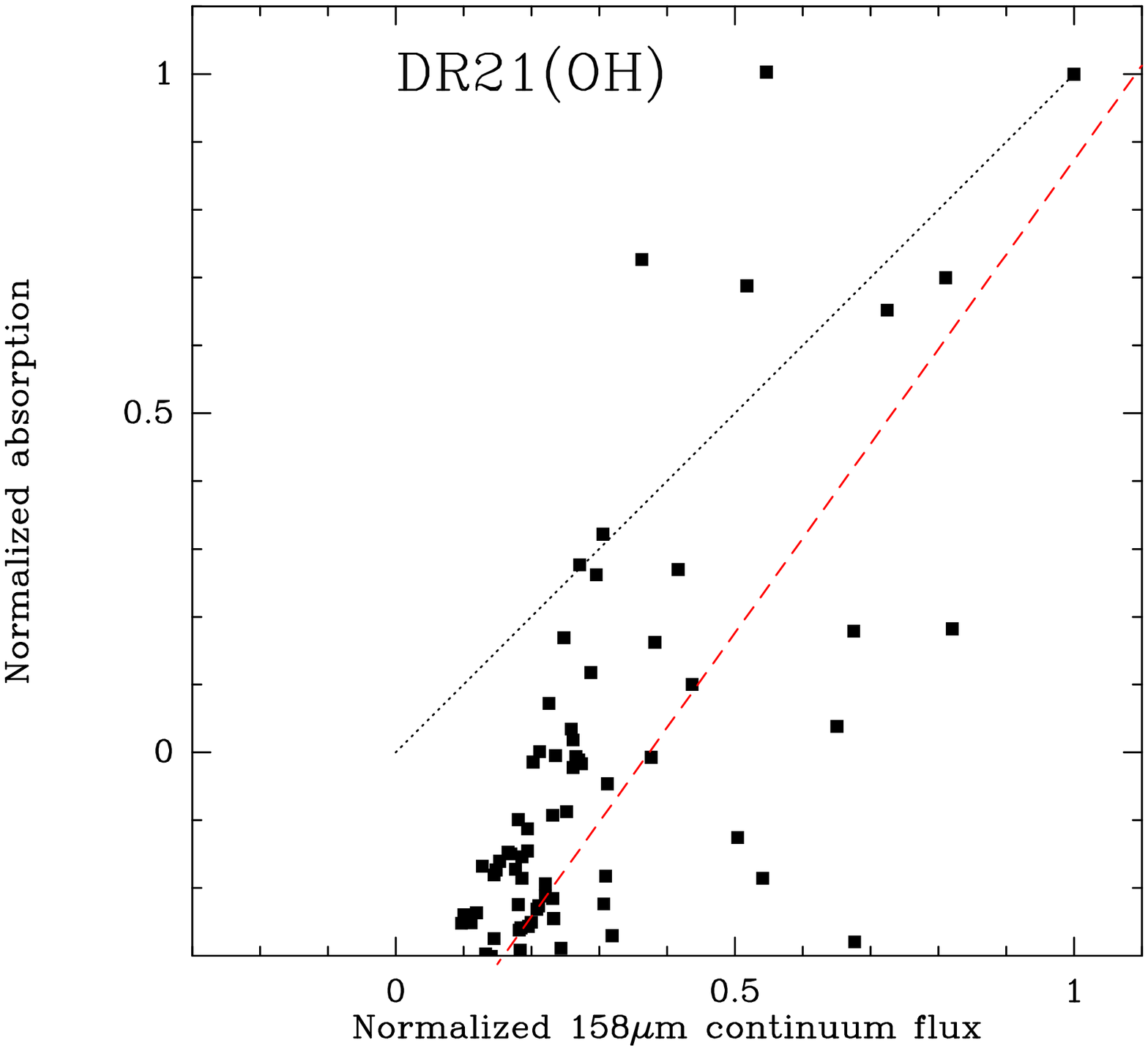}}
\\
\resizebox{7.cm}{!}{
\includegraphics{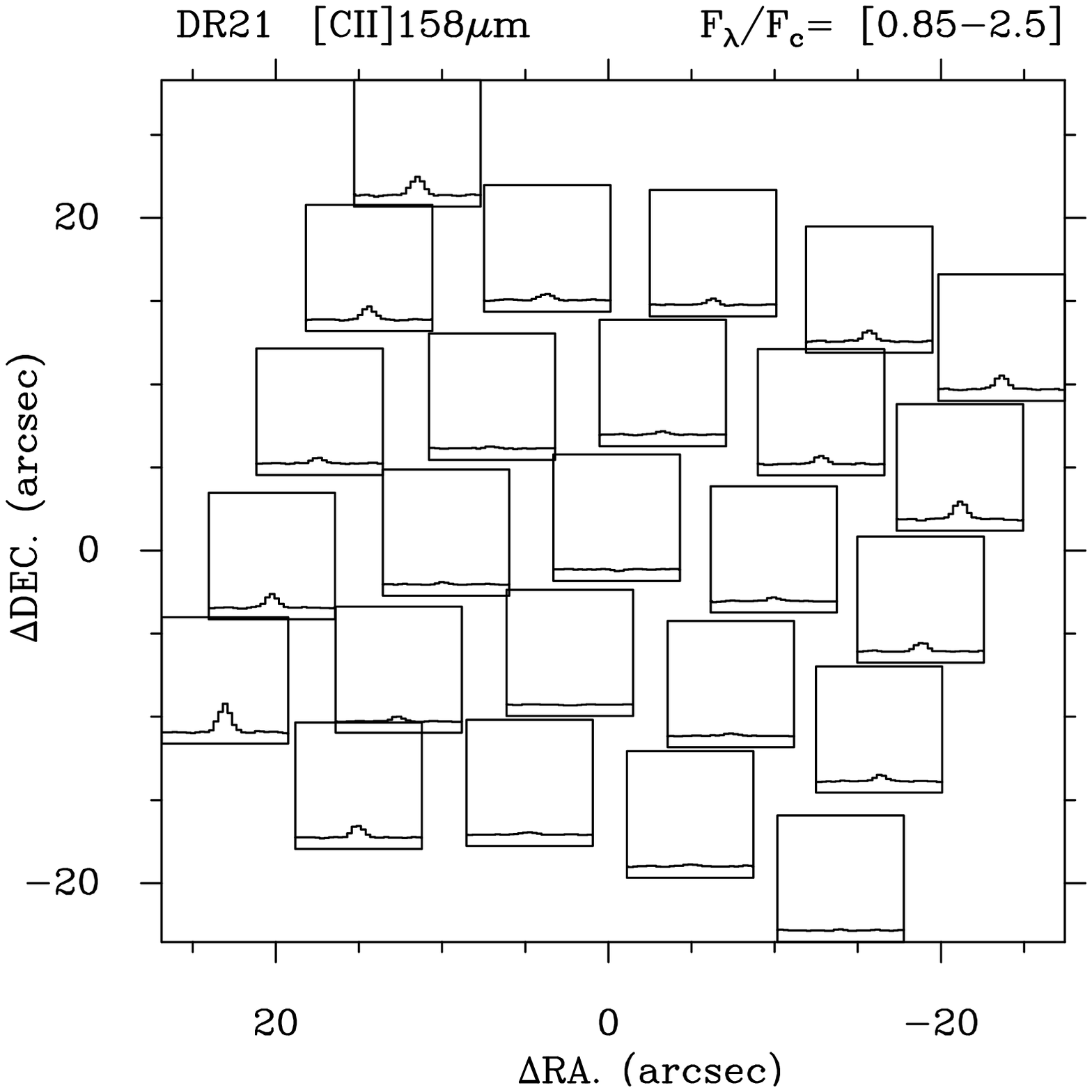}}
\caption{\label{fig:pacs-dr21} PACS data towards DR21(OH). For all maps, the
  offsets are given relative to the central position listed in Table
  \ref{tab:sources}.  Top: Continuum
  emission at 158$\mu$m. Contour levels are 0.2, 0.3, 0.4 ... 0.9 relative
to the maximum. Middle: Comparison of the integrated
absorption measured in the HIFI map relative to the absorption at the map
center, with the continuum flux measured in the PACS map relative to the
map center. The dashed red line shows the linear regression line 
and the dotted black line a 1:1 relationship.  Bottom:  Map of the line to continuum emission/absorption in the 25
PACS spaxels. The vertical scale runs from 0.85 to 2.5. The data have not been
corrected for contamination in the OFF position.}
\end{figure}

DR21(OH) is one of the most massive cores in the Cygnus molecular cloud
complex. At a distance of $\sim 1.5$ ~\kpc\ \citep{rygl}, the line of sight
mostly probes the material in the Cygnus region. While the material directly
associated with DR21(OH) appears at slightly negative LSR velocities, the
foreground gas is seen near 10 \kms and could be associated with W75N a few
arc-minutes North East.  The HIFI spectra are shown in Fig. \ref{fig:specdr21oh} and
\ref{fig:mapdr21oh}. The PACS spectral maps near 158$\mu$m are presented in Figure \ref{fig:pacs-dr21}.

As DR21(OH) is a very weak continuum source at 1.4~GHz, the HI data also
make use of the absorption towards the stronger continuum source DR~21, 3
arc-minutes South.

\subsection{W3-IRS5}

\begin{figure}[h!]
\resizebox{8cm}{!}{
\includegraphics{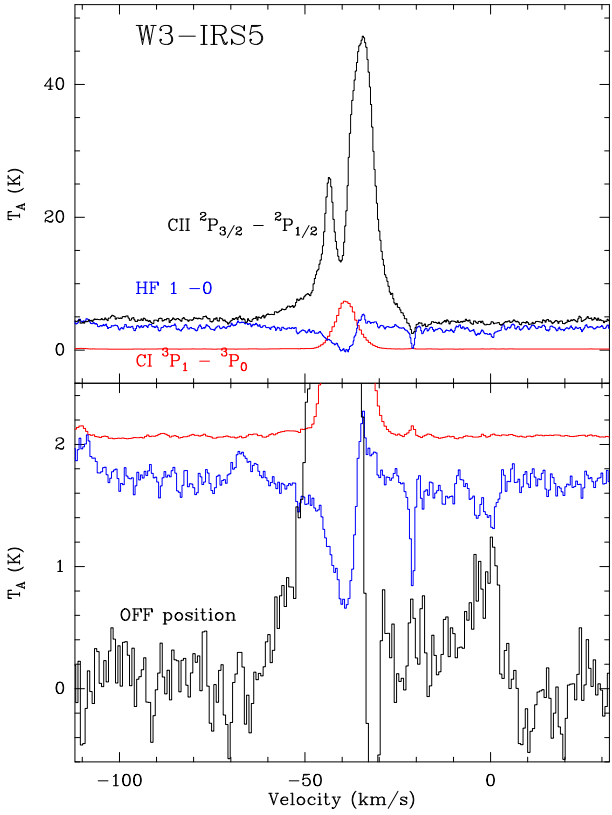}
}
\caption{\label{fig:specw3} Top : Herschel/HIFI spectra towards W3-IRS5.
The black line shows the [\CII]$^2P_{3/2}-^2P_{1/2}$ line at 1.9~THz, the red line shows the
[\CI]$^3P_1 - ^3P_0$ line at 492~GHz, and the blue line the ground state
transition of HF at 1.2~THz. The horizontal axis is the LSR
velocity in \kms \ and the vertical axis the antenna temperature in
Kelvins. The continuum level for [\CII] corresponds to the SSB continuum level. 
Bottom : zoom on the [\CI]  and HF lines (red and blue as above), and average [\CII] 
spectrum of the OFF positions (black). The continuum levels have been shifted
for clarity in the bottom panel.  }
\end{figure}

\begin{figure}
\resizebox{8.5cm}{!}{
\includegraphics{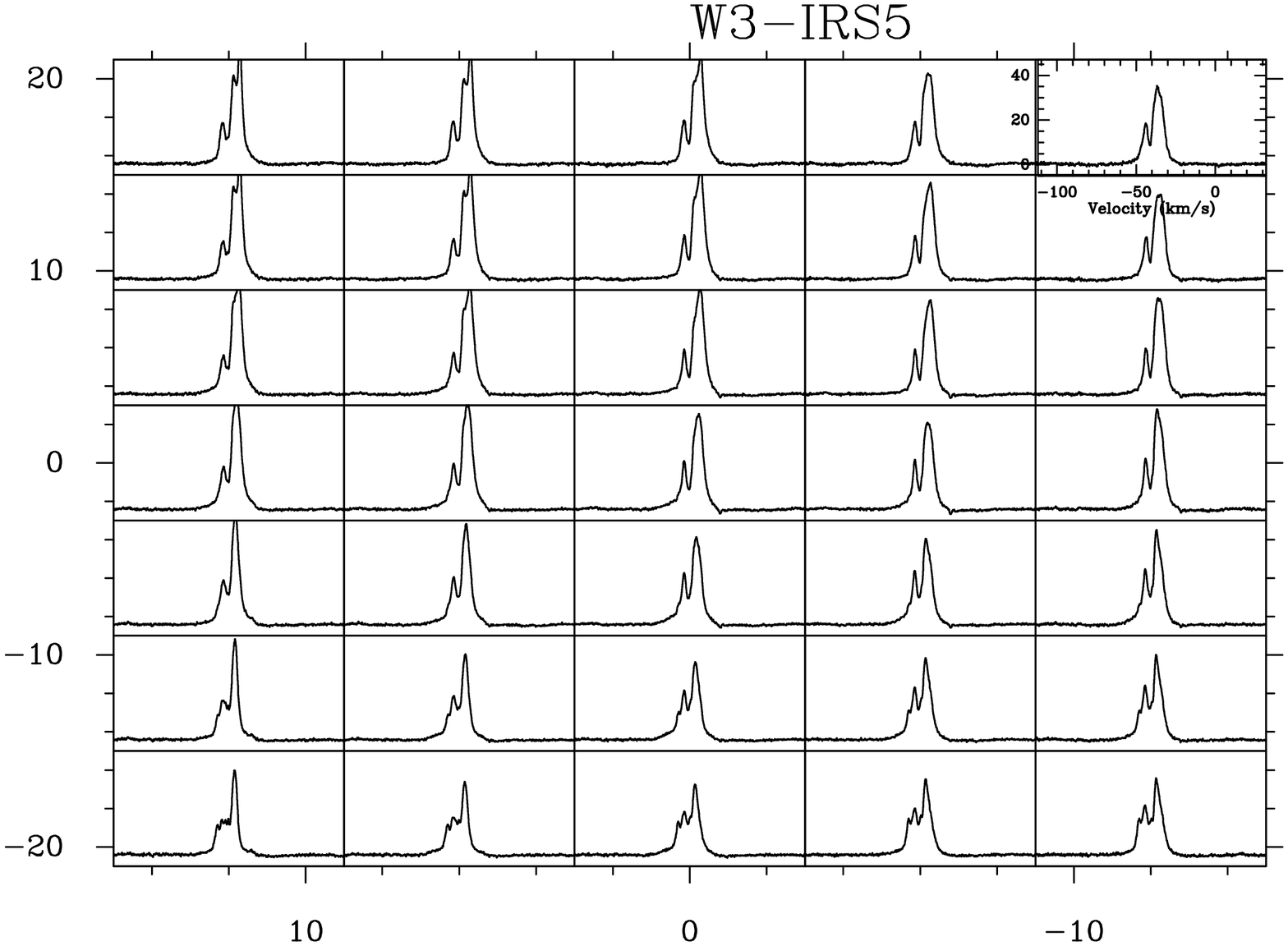}}
\caption{\label{fig:mapw3} Montage of [\CII] spectra towards W3-IRS5. A 
  baseline has been subtracted from all spectra. The horizontal axis is the LSR
  velocity in  \kms,  which runs from -112~\kms \ to 32~\kms, and   the  vertical axis is the antenna temperature in Kelvins, which runs from -5~\K \ to 
47~\K. The x--axis shows the right ascension offset in arc-sec and the
  y--axis the declination offset in arc-sec, relative to the source position given
in Table \ref{tab:sources}. }
\end{figure}

\begin{figure}
\resizebox{7.cm}{!}{
\includegraphics{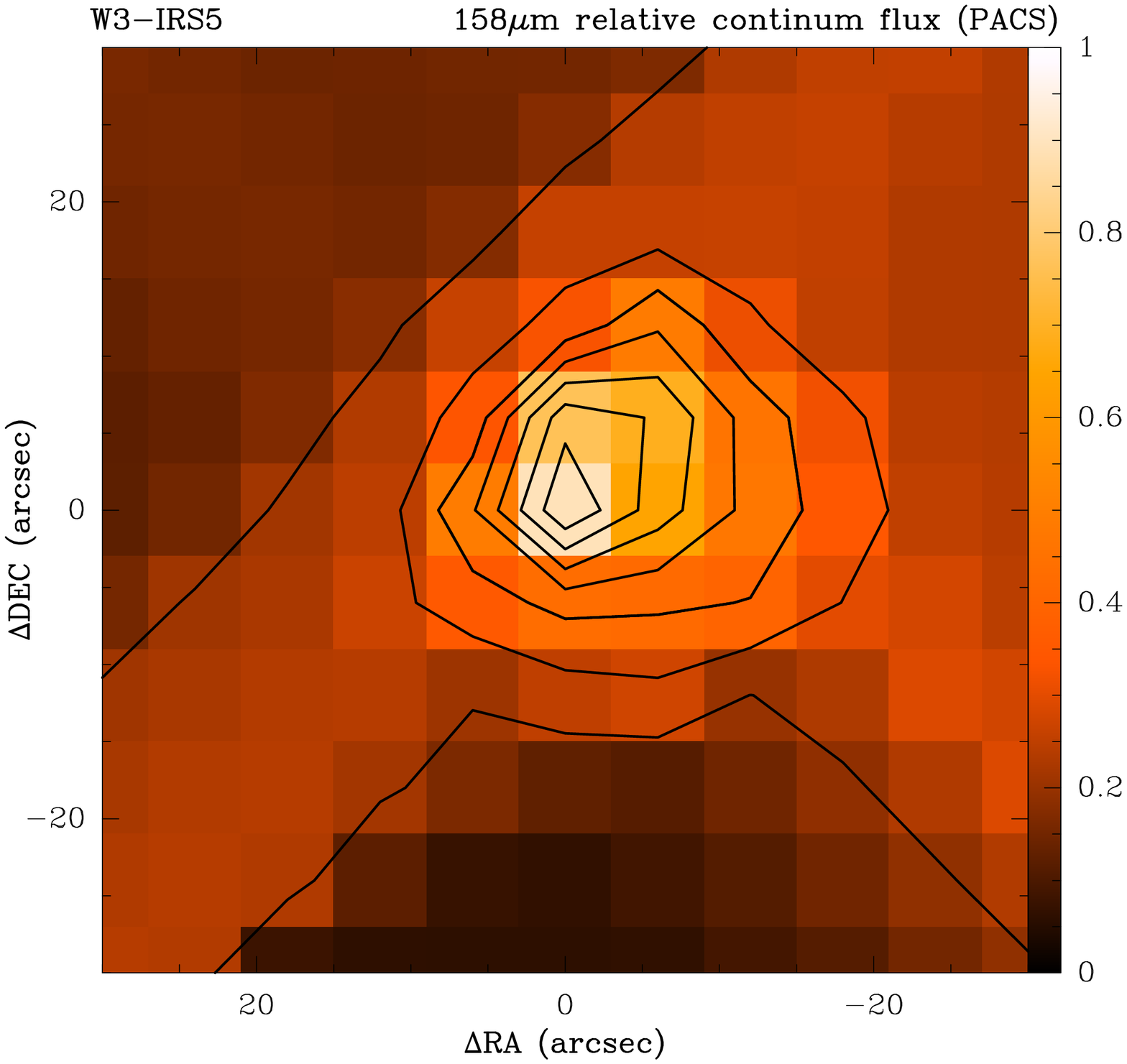}}
\resizebox{7cm}{!}{
\includegraphics{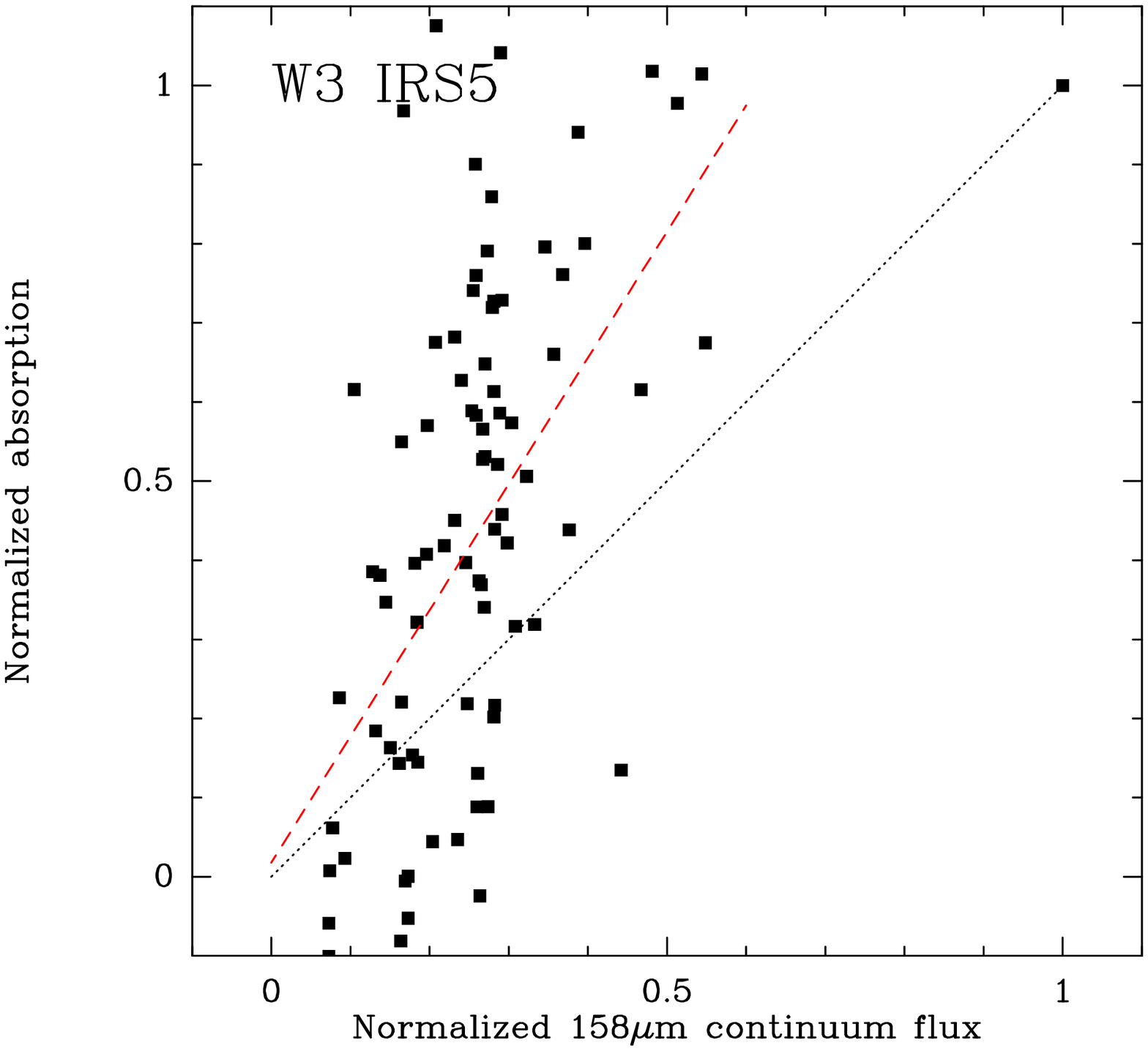}}
\resizebox{7.cm}{!}{
\includegraphics{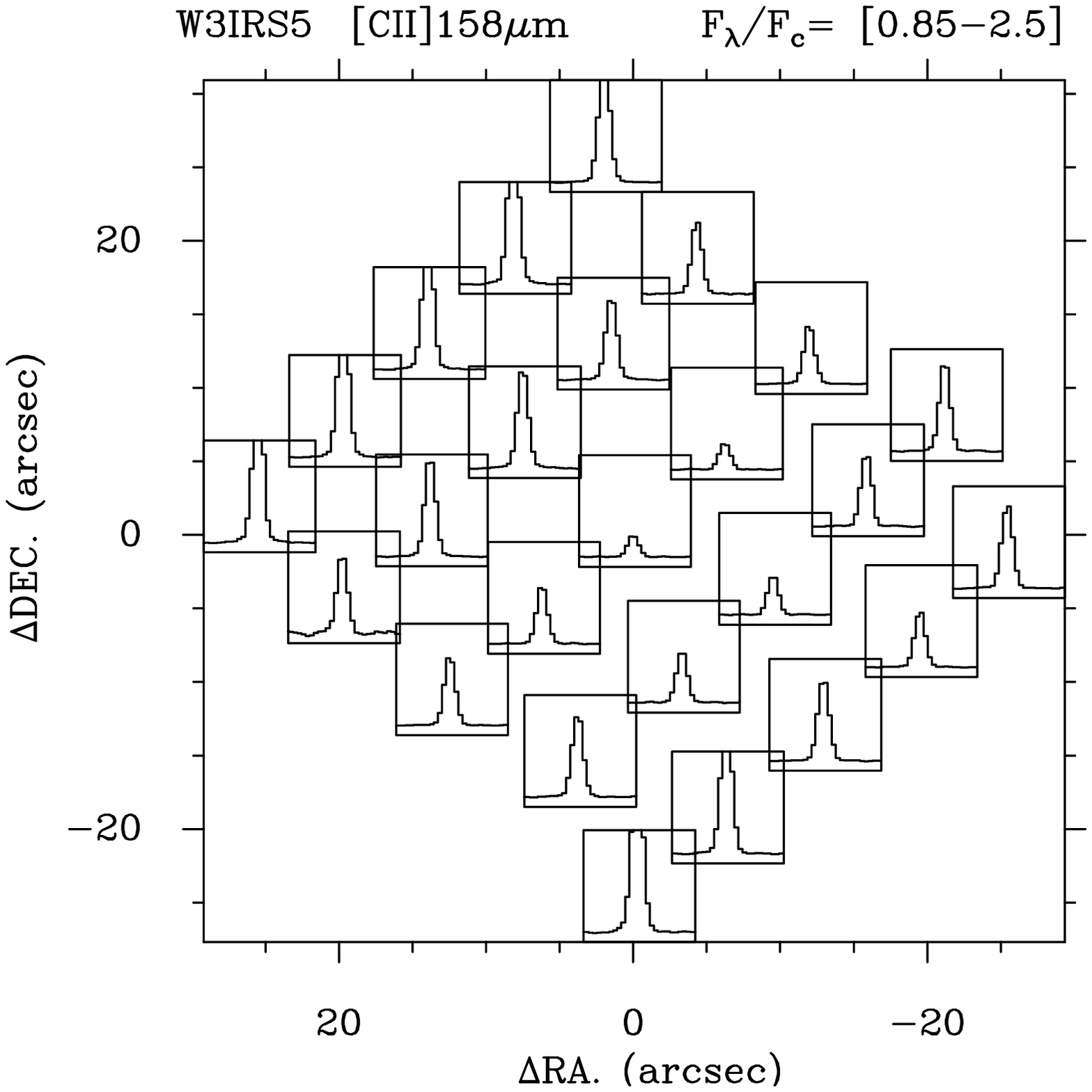}}
\caption{\label{fig:pacs-w3i} PACS data towards W3-IRS5. For all maps, the
  offsets are given relative to the central position listed in Table \ref{tab:sources}. 
Top: Continuum emission at 158$\mu$m. Contour levels are 0.2, 0.3, 0.4 ... 0.9 relative
to the maximum. Middle: Comparison of the integrated
absorption measured in the HIFI map relative to the absorption at the map
center, with the continuum flux measured in the PACS map relative to the
map center. The dashed red line shows the linear regression line 
and the dotted black line a 1:1 relationship. Bottom:  Map of the line to continuum emission/absorption in the 25
PACS spaxels. The vertical scale runs from 0.85 to 2.5. The data have not been
corrected for the contamination in the OFF position.
}
\end{figure}

W3-IRS5 is a massive protostar in the W3 star forming complex at 
a distance of $\sim 2$~\kpc \ \citep{xu:06}. Located in the second
Galactic quadrant, the line of sight towards  W3-IRS5 is  probing the outer Galaxy. The HIFI spectra are shown in Figs \ref{fig:specw3} and \ref{fig:mapw3}. 
They include the ground state transition of HF at 1.2~THz. 
A deep and narrow velocity component  is detected at $\sim -20$ \kms, and a
broader one near 0 \kms. The H$_2$ column densities have been derived from the HF data using [HF]/[H$_2$] = $1.3 \times 10^{-8}$ \citep{sonnentrucker:10}.

The PACS spectral maps near 158$\mu$m are presented in fig \ref{fig:pacs-w3i}.
The line of sight towards W3 has been studied in HI
\cite{normandeau:99}. Herschel-HIFI spectra exhibiting absorption from both
the W3 complex and foreground material have been presented by \citet{benz:10}.

\end{document}